\documentstyle[12pt,epsf,bezier]{article}
\def\photonatomright{\begin{picture}(3,1.5)(0,0)
                                \put(0,-0.75){\tencircw \symbol{2}}
                                \put(1.5,-0.75){\tencircw \symbol{1}}
                                \put(1.5,0.75){\tencircw \symbol{3}}
                                \put(3,0.75){\tencircw \symbol{0}}
                      \end{picture}
                     }

\def\photonatomup{\begin{picture}(1.5,3)(0,0)
                             \put(-0.75,3){\tencircw \symbol{3}}
                             \put(-0.75,1.5){\tencircw \symbol{2}}
                             \put(0.75,1.5){\tencircw \symbol{0}}
                             \put(0.75,0){\tencircw \symbol{1}}
                   \end{picture}
                  }

\def\photonright{\begin{picture}(30,1.5)(0,0)
                     \multiput(0,0)(3,0){10}{\photonatomright}
                  \end{picture}
                 }

\def\photonrighthalf{\begin{picture}(30,1.5)(0,0)
                     \multiput(0,0)(3,0){5}{\photonatomright}
                  \end{picture}
                 }

\def\photonup{\begin{picture}(1.5,30)(0,0)
                  \multiput(0,0)(0,3){10}{\photonatomup}
               \end{picture}
              }
\def\photonuphalf{\begin{picture}(1.5,15)(0,0)
                      \multiput(0,0)(0,3){5}{\photonatomup}
                   \end{picture}
                  }

\def\fermionurr{\begin{picture}(30,15)(0,0)
                        \put(-30,-15){\vector(2,1){15}}
                        \put(-15,-7.5){\line(2,1){15}}
                  \end{picture}
                 }
\def\fermionurrhalf{\begin{picture}(15,7.5)(0,0)
                        \put(-15,-7.5){\vector(2,1){7.5}}
                        \put(-7.5,-3.75){\line(2,1){7.5}}
                  \end{picture}
                 }
\def\fermiondrr{\begin{picture}(30,15)(0,0)
                        \put(0,0){\vector(2,-1){15}}
                        \put(15,-7.5){\line(2,-1){15}}
                  \end{picture}
                 }
\def\fermiondrrhalf{\begin{picture}(15,7.5)(0,0)
                        \put(0,0){\vector(2,-1){7.5}}
                        \put(7.5,-3.75){\line(2,-1){7.5}}
                  \end{picture}
                 }

\newenvironment{Feynman}[3]{\begin{center}
                            \setlength{\unitlength}{#3 mm}
                            \begin{picture}(#1)(#2)
                            \thicklines
                           }{\end{picture} \end{center}}

\def\theequation{\arabic{section}.\arabic{equation}}
\renewcommand{\cal}{\mathcal}
\newcommand{\ezero}{\setcounter{equation}{0}}

\newcommand{\mr}{\mathrm}
\newcommand{\as}{\alpha_s}

\newcommand{\nll}{\nonumber \\}

\newcommand {\Ql} {\mbox{$Q^2_{l}  $}}
\newcommand {\ql} {\mbox{$Q^2_{l}  $}}
\newcommand {\yl} {\mbox{$y  _{l}  $}}
\newcommand {\xl} {\mbox{$x   _{l}  $}}
\newcommand {\q}  {\mbox{$Q^2      $}}

\newcommand {\qm} {\mbox{$Q^2_{m}  $}}
\newcommand {\ym} {\mbox{$y  _{m}  $}}
\newcommand {\xm} {\mbox{$x   _{m}  $}}

\newcommand {\xh} {\mbox{$x  _{h}  $}}
\newcommand {\yh} {\mbox{$y  _{h}  $}}
\newcommand {\Qh} {\mbox{$Q^2_{h}  $}}
\newcommand {\qh} {\mbox{$Q^2_{h}  $}}
\newcommand {\Qt} {\mbox{$Q^2_{\tau}$}}
\newcommand {\xjb} {\mbox{$x_{_{\mathrm{JB}}}$}}
\newcommand {\yjb} {\mbox{$y_{_{\mathrm{JB}}}$}}
\newcommand {\yjbs} {\mbox{$y^2_{_{\mathrm{JB}}}$}}
\newcommand {\qjb} {\mbox{$Q^2_{_{\mathrm{JB}}}$}}
\newcommand {\qjbf} {\mbox{$Q^4_{_{\mathrm{JB}}}$}}

\newcommand {\LQZ} {\mbox{${\mr L}_{ \tau} $}}

\newcommand {\litwo} {\mbox{${\mr{ {Li}}}_{2} $}}
\newcommand {\lh} {\mbox{${\mr L}_{\mathrm h }$}}
\newcommand {\ljb} {\mbox{${\mr L}_{_{\mathrm{JB}}}$}}

\newcommand {\lhone}{\mbox{${\mr L}_{\mathrm{h1}}$}}

\newcommand {\Lb}{\mbox{${\mr L}_{\mathrm {\beta} }$}}

\newcommand {\oa}  {${\cal O}({\alpha})$}
\newcommand {\oaa} {${\cal O}({\alpha}^2)$}
\newcommand{\nn}{\noindent}

\newcommand{\bq}{\begin{equation}}
\newcommand{\eq}{\end{equation}}
\newcommand{\ba}{\begin{eqnarray}}
\newcommand{\ea}{\end{eqnarray}}

\newcommand {\he}  {{\tt HECTOR}}
\newcommand\order{{\cal O}}

\newcommand\GeV{\,\mbox{GeV}}

\newcommand {\xsi} {\mbox{$x_{_{\mathrm{\Sigma}}}$}}
\newcommand {\ysi} {\mbox{$y_{_{\mathrm{\Sigma}}}$}}
\newcommand {\qsi} {\mbox{$Q^2_{_{\mathrm{\Sigma}}}$}}

\newcommand {\yesi} {\mbox{$y_{_{\mathrm{e\Sigma}}}$}}
\newcommand {\xdo} {\mbox{$x_{_{\mathrm{DA}}}$}}
\newcommand {\ydo} {\mbox{$y_{_{\mathrm{DA}}}$}}
\newcommand {\qdo} {\mbox{$Q^2_{_{\mathrm{DA}}}$}}

\newcommand {\xan} {\mbox{$x_  {\theta \mathrm{y}}$}}
\newcommand {\yan} {\mbox{$y_  {\theta \mathrm{y}}$}}
\newcommand {\qan} {\mbox{$Q^2_{\theta \mathrm{y}}$}}

\begin{document}
\thispagestyle{empty}
\onecolumn
\begin{flushleft}
DESY 95--185
\\
November  1995
\end{flushleft}
\vspace*{.50cm}
\noindent
\LARGE
{{\tt{HECTOR~1.00}}
\\
 A program for the calculation of
QED, QCD and electroweak corrections
to $ep$ and $l^{\pm }\,N$ ~~deep  inelastic
neutral and charged current scattering~$^{*}$   \\   }

\vspace{2cm}
\nn
\large
A. Arbuzov$^{1\#}$,
D. Bardin$^{1,2}$,
J. Bl\"umlein$^{2}$,
L. Kalinovskaya$^{1}$,
T.~Riemann$^2$
\\

\vspace{1cm}
\noindent
\normalsize
{\it
$^1$Bogoliubov Laboratory for Theoretical Physics, JINR,
    ul. Joliot-Curie 6, RU-141980 Dubna,
\\
{}~Russia

\vspace{1mm}
\noindent
$^2$DESY -- Zeuthen,
          Platanenallee 6, D-15738 Zeuthen, Germany

\vspace{1mm}
\noindent
} 

\vspace{\fill}
\vfill
\thispagestyle{empty}
{\large
\centerline{ABSTRACT}
}
\vspace*{.2cm}
\small
\nn
A description of the Fortran program \he\ for a variety of
semi-analytical calculations of radiative QED, QCD, and electroweak
corrections to the double-differential cross sections of
$\cal NC$ and $\cal CC$ deep inelastic charged lepton proton (or
lepton deuteron) scattering is presented.
{\tt HECTOR} originates from the substantially improved and extended earlier
programs {\tt HELIOS} and {\tt TERAD91}.
It is mainly intended for
applications at HERA or LEP$\otimes$LHC,
but may be used also
for $\mu N$
scattering in fixed target experiments.
The QED corrections may be calculated in different sets of variables:
leptonic, hadronic, mixed, Jaquet-Blondel, double angle etc.
Besides the leading logarithmic approximation up to order \oaa,
exact \oa\ corrections and inclusive soft photon exponentiation
are
taken into account.
The photoproduction region is also covered.
\normalsize
\vfill
\vspace*{.5cm}

\bigskip
\vfill
\footnoterule
\nn
\small
$^{*}$~Supported by the EC network `Capital Humain et Mobilite' under
grant CHRX--CT92--0004.
\\
$^{\#}$~Supported by the Heisenberg-Landau fund.
\\
e--mail: hector@ifh.de.
\normalsize
\vfill \eject

\tableofcontents

\clearpage

\listoftables    

\listoffigures


%

\begin{minipage}[t]{8cm}{
\begin{sloppypar}
\vspace*{11.0cm}

{\sc Hector},
son of the Troyan King {\sc Priam}.
Troyan war leader.
{\sc Homer} describes {\sc Hector} to be {\it open, frank,
brave, cheerful in adversity, and tenderly compassionate} in the {\it Ilyad}
{}~\cite{wiw}.
Drawing by J. Flaxman~\cite{flaxman}.
\section*{PROGRAM SUMMARY}
   {\em Title of program\/}:  {\tt HECTOR}
\\ {\em Version\/}:           1.00 November 1995
\\ {\em Catalogue number\/}:
\\ {\em Program obtainable from\/}:
\\
http://www.ifh.de/theory or on
request from e-mail: hector@ifh.de
\\ {\em Licensing provisions\/}: non
\\ {\em Computers\/}: all
\\ {\em Operating system\/}:  all
\\ {\em Program language\/}:     {\tt FORTRAN-77}
\\ {\em Memory required to execute with typical data\/}: 4.2 Mb.
\\ {\em No. of bits per word\/}: 64;
\\
the branch
{\tt TERADLOW} uses 128
        bits per word for numerical quadric precision
\\ {\em No. of lines in distributed program\/}:  20.000
\end{sloppypar}
}
\end{minipage}
\hspace{.8cm}
\begin{minipage}[t]{8.cm}{
\vspace*{1.5cm}
\begin{sloppypar}
   {\em Other programs called\/}:
\\ libraries  {\tt KERNLIB} [1];
\\ {\tt FFREAD} -- part of {\tt PACKLIB} [2];
\\ {\tt DIZET} -- an electroweak library [3];
\\ optionally possible: {\tt PDFLIB} [4].
\\ {\em External files needed\/}:
\\ {\tt HECTOR.INP}     -- input cards to be read by the {\tt FFREAD} package.
\\ {\em Keywords\/}:  deep inelastic scattering, QED, electroweak and QCD
   radiative corrections, structure functions, higher order corrections.
\\
    {\em Nature of physical problem\/}:
\\  First and higher order QED radiative corrections to the
    lepton nucleon deep inelastic scattering; virtual
    electroweak and QCD corrections to the process;
    QCD corrections to structure functions.
   {\em Method of solution\/}:
\\ Numerical integration of analytical formulae.
\\ {\em Restrictions on complexity of the problem\/}:
\\ Only selective experimental cuts are possible.
   Results for full calculations in order \oa\
   are not available for all possible
   kinematical variables.
\\ {\em Typical running time\/}:
\\ The running time strongly depends upon the options used.
   One finds e.g.: LLA, leptonic variables, $\cal NC$, 25 points:
   about 25 sec. (CPU time),
   full $\order (\alpha)$ + $\order (\alpha^2)$ LLA + soft exponentiation
   + electroweak non-running couplings, leptonic variables, $\cal NC$,
   25 points:
   about 600 sec.
\\ {\em References\/}: \\  \vspace{-.8cm}
\begin{itemize}
\item[{[1]}]
CERN Program Library Z 001.
\item[{[2]}]
R. Brun, et al.,
FFREAD User Guide and Reference Manual,
CERN DD/US/71, CERN Program Library I 302, February 1987.
\item[{[3]}]
D. Bardin et al., {\it Comput. Phys. Commun.} {\bf 59} (1990) 303.
\item[{[4]}]
H. Plothow--Besch, {\it Comput. Phys. Commun.} {\bf 75} (1993) 396.
\end{itemize}
\end{sloppypar}
}
\end{minipage}


\clearpage

\section*{LONG WRITE-UP}
\vspace{.5cm}

\section{Introduction}

{\Huge {\tt HECTOR}
}
is a program to calculate radiative corrections
(RC) to the charged lepton nucleon neutral and charged
current deep inelastic scattering (DIS) $l^{\pm} N \to l' X$.
This requires to take into account the full set of virtual electroweak
(EW) corrections,
QCD  corrections, and the  complete
QED corrections in
lowest
order,
as well as
the leading higher order QED
corrections.
The  QCD radiative corrections are implemented up to
next-to-leading order (NLO).
They depend on the factorization scheme used.
The $\order (\alpha)$ QED radiative corrections due to
radiation from the lepton lines are known to be
large and can reach the order of $100\%$ in some regions of phase
space.
This makes, at least in parts of the phase space, the inclusion of
higher order corrections indispensable.
The aim of the project is to obtain a description at the percent level.

The program originates from two different previous codes:
{\tt HELIOS}~\cite{helios} and {\tt TERAD91}~\cite{terad}.
{\tt HELIOS} calculates QED RC in the leading logarithmic
approximation (LLA) in the $\order (\alpha)$ order.
The program is based on refs.~\cite{LLA1}--\cite{BLH}.
The
$\order (\alpha^2 L^2)$ corrections
and soft photon exponentiation have been
included recently~\cite{jblla2}.
{\tt TERAD91} calculates complete QED and the EW
corrections in the $\order (\alpha)$ order.
It is based on a series of papers~\cite{zfpc42}--\cite{dizet}.
The current Fortran program {\tt HECTOR} combines the two methods.

\section
{Basic notations
\label{basnot}}
\ezero

\subsection
{Born cross sections
\label{borsec}
}
\he\ allows the calculation of neutral current ($\cal NC$) and charged current
($\cal CC$) Born cross sections.

The $\cal NC$ cross section may be calculated in several  ways and
approximations: in the ultra-relativistic approximation or exact in all
masses; neglecting the longitudinal structure
functions
 or not; without or with electroweak one loop and higher order
corrections; the latter case is called improved Born cross section.
The Born cross section is calculated in subroutine {\tt sigbrn.f}:
\ba
\frac {d^2 \sigma_{\mr{NC}}^{\mr B}} {dx dy}
&=&
\frac{2 \pi \alpha^2(Q^2) S}{ Q^4}
\Biggl\{ \left[
2\left(1-y\right)-2xy\frac{M^2}{S}
+ \left(1-2\frac{m^2}{Q^2}\right)
\left(1+4x^2\frac{M^2}{Q^2} \right)
\frac{y^2}{1+R} \right]
{\cal F}_{2}(x,Q^2)
\nll
& &~+~x y(2-y){\cal F}_{3}(x,Q^2) \Biggr\}.
\label{born}
\ea
We also introduce a slightly modified notion for later use:
\ba
\frac {d^2 \sigma_{\mr{NC}}^{\mr B}} {dx dy}
&=&
\frac{2 \pi \alpha^2(Q^2)S}{ Q^4}
\Biggl\{ \left[ \frac{1}{S^2} {\cal S}_{2}^{\mr B}(y,Q^2)
+
\frac{1}{Q^2} {\cal S}_{1}^{\mr B}(y,Q^2)
\left(1+4x^2\frac{M^2}{Q^2} \right)
\frac{y^2}{1+R} \right]
{\cal F}_{2}(x,Q^2)
\nll
& & %
+~\frac{1}{2S^2} {\cal S}_{3}^{\mr B}(y,Q^2)
{\cal F}_{3}(x,Q^2) \Biggr\} ,
\label{bf12}
\ea
with
\ba
{\cal S}_{1}^{\mr B}(y,Q^2) &=& Q^2-2m^2,
\label{eqS1B}
\\
{\cal S}_{2}^{\mr B}(y,Q^2) &=& 2[(1-y)S^2-M^{2}Q^{2}],
\label{eqS2B}
\\
{\cal S}_{3}^{\mr B}(y,Q^2) &=& 2Q^{2} (2-y)S.
\label{eqS3B}
\ea
Here, $M$ and $m$ are the proton and lepton masses,
$s =-(k_1 + p_1)^2 = S+m^2+M^2 \approx 4E_eE_p$,
$x$ and $y$ denote
the Bjorken variables, and $q^2 = -Q^2$ is the four momentum
transfer squared. These variables
will be discussed in more detail in
appendix~A. The generalized structure functions are denoted by
${\cal F}_{i}(x,Q^2)$ and
$R$ is the ratio of the cross sections with virtual exchange of
longitudinal and transverse photons, respectively:
\ba
R(x,Q^2) = \frac{\sigma_L}{\sigma_T}
&=& \left(1+4x^2\frac{M^2}{Q^2}\right) \frac{{\cal F}_{2}(x,Q^2)}
{2x{\cal F}_{1}(x,Q^2)} - 1.
\label{rqcd}
\ea
These relations are thoroughly
used by {\tt HECTOR}.
We mention the  ultra-relativistic approximation of the
Born cross section:
\ba
\frac {d^2 \sigma^{\mr B}} {dx dy}
=
\frac{2 \pi \alpha^2(Q^2)S}{ Q^4}
\left[ Y_+
{\cal F}_{2}(x,Q^2)
+ Y_- x {\cal F}_{3}(x,Q^2)  -y^2 {\cal F}_{L}(x,Q^2)\right],
\label{xqBorn}
\ea
with
\ba
Y_{\pm}(y) = 1 \pm (1-y)^2.
\label{ypm}
\ea

For the $\cal CC$ reaction, a similar cross section formula holds:
\ba
\frac {d^2 \sigma_{CC}^{\mr B}} {dx dy}
&=&
\frac{G_{\mu}^2S}{4\pi}
\left[\frac{M_W^2}{\q+M_W^2}\right]^2
\Biggl\{
Y_+ {\cal W}_{2}(x,Q^2) +
Y_- {\cal W}_{3}(x,Q^2) -y^2 {\cal W}_L(x,Q^2) \Biggr\}.
\label{bf14}
\ea

The generalized structure functions for the $\cal NC$ case (${\cal F}$) and
for the $\cal CC$ case ($\cal W$) are defined in section~\ref{genfun}, while
the relations of the basic input parameters follow in the next
section,
 and the description of the running of $\alpha$ in
section~\ref{alfarun}.

In a lowest order description one may assume the validity of
Callan--Gross relations~\cite{callan}
 for ${\cal F}_{1,2}$ and ${\cal W}_{1,2}$:
\ba
{\cal F}_{L}(x,Q^2) &\equiv& {\cal F}_{2}(x,Q^2)
- 2 x {\cal F}_{1}(x,Q^2) = 0,
\nonumber\\
{\cal W}_{L}(x,Q^2) &\equiv& {\cal W}_{2}(x,Q^2)
- 2 x {\cal W}_{1}(x,Q^2) = 0.
\label{bjli}
\ea

\subsubsection
{Input parameters
\label{inpt}
}
\he\ takes into account that not all the physical parameters
are independent of each other in the Standard Model.
There are three sets of input values foreseen.
The default set is chosen in file {\tt hecset.f} with flag
setting {\tt IMOMS}=1.
For this, the input is expressed by the following parameters:
\begin{itemize}
\item
     fine structure constant $\alpha= 1/137.0359895$;
     its running is discussed in section~\ref{alfarun};
\item
     muon decay constant $G_{\mu} = 1.166388\cdot 10^{-5}$ GeV$^{-2}$;
\item
     $Z$ boson mass $M_Z=91.175$ GeV;
\item
     fermion masses, including the top quark mass $m_t=180$ GeV;
\item
     Higgs boson mass $M_H$, for which the default value of
     $300 \GeV$ is assumed;
\item
     QCD parameter $\Lambda_{\overline{\rm MS}}$,
which is chosen in
     file {\tt alpqcd.f} in accordance with the
pre-selected {\tt PDF} set (set of parton distribution functions).
\end{itemize}
The default values of constants are set in file {\tt
  hecset.f}.
The strong coupling constant $\alpha_s$ is calculated from
$\Lambda_{\overline{\rm MS}}$ as described in section~\ref{alsms}.

Several additional parameters may be calculated from the above input.
The most important ones are the $W$ boson mass:
\ba
M_W &=& M_Z \left\{
\frac{1}{2} + \frac{1}{2} \left[
1-\left(
\frac{74.562}{M_Z}
\right) ^2\frac{1}{1-\Delta r}\right]^{1/2}\right\}^{1/2},
\label{mw}
\ea
and the weak mixing angle:
\ba
\sin^2 \theta_W &=& 1-\frac{M_W^2}{M_Z^2}.
\label{sw}
\ea
Details on the definition of $\Delta r $ within \he\ may be found
in~\cite{dizet,YR95}. It depends on the chosen electroweak parameters
and on the fermion masses.

If one wants to use the $W$ mass as an input parameter instead of
$M_Z$, the default setting is:
\begin{itemize}
\item
     $M_W=80.22$ GeV.
\end{itemize}
Then, one has to determine $M_Z$ by iteratively solving the equation:
\ba
M_Z &=& M_W \left[
1-\left(\frac{74.562}{M_W}\right)^2 \frac{1}{1-\Delta r}
\right]^{-1/2}.
\label{mz}
\ea
This is done when using {\tt IMOMS}=2.

A third possibility,  {\tt IMOMS}=3, is to take both $M_Z$
and $M_W$ together with $\alpha$ as an
input and to calculate $G_{\mu}$:
\ba
G_{\mu}
&=&
\frac{\pi \alpha}{\sqrt{2}M_W^2\sin^2\theta_W(1-\Delta r)}.
\label{azw}
\ea

For any setting of the
flag {\tt IMOMS}, it is also possible to use an
effective weak mixing angle\footnote{
For more details, see the discussion in~\cite{YR95} on the definition of
$\sin^2 \theta_W^{\mr{eff}}$.
}
{\em instead} of the definition~(\ref{sw}).
This is done by choosing the flag  {\tt IWEA}=0, which
switches off the calculation of virtual weak corrections.
The weak mixing angle as defined in~(\ref{sw}) is not used for
the calculation of weak neutral couplings in this case:
\ba
\sin^2\theta_W &\rightarrow& \sin^2 \theta_W^{\mr{eff}} .
\ea
In file {\tt hecset.f} a default value
is given:
\ba
\sin^2 \theta_W^{\mr{eff}}
&=& 0.232.
\label{sw2effdef}
\ea
\subsubsection{%
\label{alfarun}%
The running electromagnetic coupling $\alpha(Q^2)$
}
The running of the QED coupling is taken into account if {\tt IVPL}=1.
Then, instead of $\alpha$, the value
\ba
\alpha(Q^2)
&=&
\frac{\alpha}
{1-\Delta \alpha}
\label{ddalf}
\ea
is used.
The correction consists of the pieces:
\ba
\Delta \alpha &=&
\Delta \alpha_l + \Delta \alpha_{udcsb} + \Delta \alpha_t
\label{dalf}
\ea
corresponding to the charged leptons, the lighter quarks ($u$-$b$),
and
the top quark.

The value of
$\Delta \alpha$ is calculated in function {\tt XFOTF1} of the
package {\tt DIZET}.
The contribution due to
leptons is:
\ba
\Delta \alpha_l &=& \sum_{f=e,\mu,\tau}
Q_f^2 N_f
\Delta F_f(Q^2),
\label{vac0}
\\
\Delta F_{f}(Q^2)
&=&
\frac{\alpha}{\pi}
\left\{-\frac{5}{9}+\frac{4}{3}\frac{m_f^2}{Q^2}+\frac{1}{3}
\beta_f
\left(1-\frac{2m_f^2}{Q^2}\right)
\ln \frac{\beta_f+1}{\beta_f-1} \right\},
\label{vac1}
\nll
\\
\beta_f
&=&
\sqrt{1+\frac{4m_f^2}{Q^2}}.
\label{vac2}
\ea
The color factor, $N_f$, and the
charge squared, $Q_f^2$, are unity for leptons.
In the weak library,
$\Delta F_f$ is calculated by the
function {\tt XI3}:
$\Delta F_f = 2$~{\tt XI3}.
The contribution
from the light quarks has been parametrised in two different
ways.
It may be
calculated by~(\ref{vac0})
setting {\tt IHVP}=2. In this case,
we use effective quark masses~\cite{x21,x24}:
$m_u=m_d=0.041, m_s=0.15, m_c=1.5, m_b=4.5$~GeV.
The preferred, default approach (with {\tt IHVP}=1) uses a
parameterization of the
hadronic vacuum polarization~\cite{jeger5}, which is contained in
file {\tt hadr5.f}. An older parameterization~\cite{x19} may also
be
accessed choosing {\tt IHVP}=3.

Finally, the $t$ quark corrections are:
\ba
\Delta \alpha_t &=&
Q_t^2 N_t
\Delta F_t(Q^2) + \Delta \alpha
^{\mathrm{2loop},\alpha\alpha_s}.
\label{vact}
\ea
The last term contains higher order corrections and is calculated by the
functions {\tt ALQCDS} (approximate) or {\tt ALQCD} (exact):
\ba
\Delta \alpha
^{\mathrm{2loop},\alpha\alpha_s}
&=&
\frac{\alpha\alpha_s}{3\pi^2}Q_t^2\frac{m_t^2}{Q^2}
\Biggl\{
{\cal R}e
\Pi_t^{VF}(Q^2) + \frac{45}{4}\frac{Q^2}{m_t^2} \Biggr\} .
\label{daqcd}
\ea
$\Pi_t^{VF}(Q^2)$ is a two loop self energy function~\cite{x22,x23}.
For {\tt IQCD}=0 (default value), this tiny correction is
neglected.
For {\tt IQCD}=1 approximate and for {\tt IQCD}=2 exact
calculations according to~(\ref{daqcd}) are made.
 The first one is a crude approximation only, the second,
on the other hand, results in a very time consuming computation.
\subsection{Structure functions
\label{sf}
}
\subsubsection
{Generalized structure functions
\label{genfun}
}
The generalized structure functions describe the electroweak interactions
of leptons with beam particle
charge $Q_l$
and lepton beam polarization $\xi$
via the exchange of a photon,  $Z$ boson, or $W$ boson
 with unpolarized nucleons.

The generalized structure functions are calculated for the $\cal NC$ cross
sections in file {\tt genstf.f} or, for the low
$Q^2$ branch {\tt TERADLOW},
in file {\tt gsflow.f},
 and for the $\cal CC$
cross sections in file {\tt gccstf.f}.
The exact treatment of virtual electroweak corrections (for the choice
{\tt IWEAK}=1) will be described in section~\ref{impsec}.
Here we assume {\tt IWEAK}=0.
Then, the generalized $\cal NC$ structure functions ${\cal F}_{i}(x,Q^2)$
are expressed in terms of the
usual {\it hadronic} structure functions
$F_i$, $G_i$, $H_i$ by the following equations:
\begin{eqnarray}
{\cal F}_{1,2}(x,Q^2)
&=& F_{1,2}(x,Q^2) + 2 |Q_{e}| \left( v_{e} + \lambda a_e \right)
\chi(Q^2) G_{1,2}(x,Q^2)
\nll & &+~4 \left( v_{e}^{2} + a_{e}^{2} + 2 \lambda v_e a_e \right)
\chi^2(Q^2) H_{1,2}(x,Q^2),
\label{f112}
\\
x {\cal F}_3(x,Q^2)
&=& -2~\mbox{\rm{sign}}(Q_l) \Biggl\{
 |Q_{e}|
\left( a_{e} + \lambda v_e \right) \chi(Q^2) xG_{3}(x,Q^2)
\nll
& &
+
\left[2v_{e} a_e + \lambda \left(v_e^2 + a_e^2 \right) \right]
\chi^2(Q^2) xH_{3}(x,Q^2)\Biggr\},
\label{f123}
\end{eqnarray}
with $Q_e=-1$ and
\ba
\lambda&=&\xi \, \mbox{\rm{sign}}(Q_l),
\label{laxi}
\\
v_e&=&1-4 \sin^{2}\theta_{W}^{\mr{eff}} ,
\\
a_{e}&=&1,
\ea
and
\ba
\chi (Q^2) =
{G_\mu \over\sqrt{2}}{M_{Z}^{2} \over{8\pi\alpha(Q^2)}}{Q^2 \over
{Q^2+M_{Z}^{2}}}.
\label{chiq}
\ea
The structure functions
$F_{1,2}, G_{1,2}, H_{1,2}$ and $G_3, H_3$ are defined in the next
section.

The charged current generalized structure functions
for {\tt IWEAK}=0 are:
\ba
{\cal W}_2(x,\q)
&=&
\frac{1+\lambda}{2} W_2^{Q_l}(x,\q),
\\
{\cal W}_3(x,\q)
&=&
- {\mr{sign}} (Q_l) \frac{1+ \lambda}{2} xW_3^{Q_l}(x,\q).
\label{ccgsf}
\ea
They are as well
defined in the subsequent section.

For the case of an exact inclusion of the electroweak virtual
corrections, we refer to section~\ref{impsec}.
There we will see that
the notion of the generalized structure functions holds further on, but that
of hadronic structure functions will be lost.

\bigskip

The photon exchange parts of the generalized structure functions have
to vanish at low $Q^2$.
This is ensured by non-perturbative
modifications of the corresponding structure
functions as described in section~\ref{lowq2}.
\subsubsection
{Structure functions
\label{strfun}
}
There are several ways to choose structure functions in {\he}.
This will be described in detail in section~\ref{sotc}.

One option is to derive structure functions from parton distribution
functions (PDF).
The different structure functions are in a lowest order
approach\footnote{NLO corrections are delt with
in section~\ref{pardis}.}
\ba
\vphantom{\int\limits_t^t}
\label{lf2}
F_2(x,Q^2) &=&
x \sum_q |Q_q|^2 \, [ q(x,Q^2) + {\bar q}(x,Q^2) ],
\\
\vphantom{\int\limits_t^t}
G_2(x,Q^2) &=&
x \sum_q |Q_q| \, v_q \, [ q(x,Q^2) + {\bar q}(x,Q^2) ],
\\
\vphantom{\int\limits_t^t}
H_2(x,Q^2) &=&
x \sum_q \frac{1}{4} \left(v_q^2 + a_q^2 \right)
[ q(x,Q^2) + {\bar q}(x,Q^2) ],
\\
\vphantom{\int\limits_t^t}
x G_3(x,Q^2) &=& x \sum_q |Q_q| \, a_q \,
[ q(x,Q^2) - {\bar q}(x,Q^2) ],
\\
\label{lxh3}
\vphantom{\int\limits_t^t}
x H_3(x,Q^2) &=& x \sum_q \frac{1}{2} \, v_q a_q \,
[ q(x,Q^2) - {\bar q}(x,Q^2) ] ,
\label{qpdis}
\ea
where $q(x,Q^2)~(\overline{q}(x,Q^2))$ denote the quark (antiquark)
distributions.
The neutral current couplings are in our notations:
\ba
v_{f}&=&1-4 |Q_{f}| \sin^{2}\theta_{W}^{\mr{eff}},
\\
a_{f}&=&1.
\label{veae}
\ea
Here, $\theta_{W}^{\mr{eff}}$ is the effective weak mixing angle
introduced in section~\ref{inpt}
and $s_e |Q_e| = Q_{e} = -1$ is the electron charge.

The $\cal CC$ structure functions are:
\ba
\label{lw2p}
W_2^{+}(x,\q) &=&
2x\sum_i\left[d_i(x,\q)+{\bar u}_i(x,\q)\right],
\label{w2pd}
\\ 
W_2^{-}(x,\q) &=&
2x\sum_i\left[u_i(x,\q)+{\bar d}_i(x,\q)\right],
\label{w2md}
\\
x W_3^{+}(x,\q)
&=&
2x\sum_i\left[d_i(x,\q)-{\bar u}_i(x,\q)\right],
\label{w3pd}
\\ 
x W_3^{-}(x,\q)
&=& 2x\sum_i\left[u_i(x,\q)-{\bar d}_i(x,\q)\right] ,
\label{w3md}
\ea
with $u_i$ and $d_i$ denoting the up ($u,c,t$) and down quark
$(d,s,b)$ densities, respectively.

The structure functions ${\cal S}_1(x,Q^2) =
F_1(x,Q^2), G_1(x,Q^2), H_1(x,Q^2),
W_1^{-}(x,\q)$ are calculated in \he\ by
\ba
2x {\cal S}_1
&=&
{\cal S}_2 - {\cal S}_L,
\label{l2xw1m}
\ea
and analogously for the other distributions.
The longitudinal structure functions vanish in lowest order of QCD.
\subsubsection
{Generalized structure functions for improved Born cross sections
\label{impsec}
}
For the introduction of the generalized structure functions in the
 presence of
virtual weak interactions, we use a slightly different
notation compared to that of
section~\ref{genfun} in order to
make the relation between \he\ and~\cite{zfpc42} and~\cite{zfpc44}
sufficiently transparent.
To these references (and for later improvements also to~\cite{YR95}) we
refer for all the details which cannot     be described here.

In the $\cal NC$ case the structure functions are given by:
\ba
{\cal F}_2^{\cal{SM}}(x,\q)
&=&
\sum_{B=\gamma,I,Z} \, \sum_{i=1}^3 \, \sum_{q=Q,{\bar Q}} {\cal K}(B,p)
\, {\cal V}(B,p) \, q_i(x,\q),
\label{calft2}
\\ 
{\cal F}_3^{\cal{SM}}(x,\q)
&=&
\sum_{B=\gamma,I,Z} \, \sum_{i=1}^3 \, \sum_{q=Q,{\bar Q}} p \, {\cal K}(B,p)
\, {\cal A}(B,p) \, q_i(x,\q).
\label{calft3}
\ea
Here, index $i$ denotes the quark generation.
$B=\gamma$ stands for the pure photon exchange contribution, $Z$
for the $Z$ exchange and $I$ for their interference.
The parameter $p$ depends on the relative
charge assignments of the beam particle and the struck quark:
\bq
\begin{array}{rlcll}
p&=&1&\hspace{1.cm}&{\mr{for:}} \,\, e^-Q_i, \, \, e^+{\bar Q}_i,
\\
p&=&-1&\hspace{1.cm}&{\mr{for:}} \,\, e^+Q_i, \, \, e^-{\bar Q}_i.
\end{array}
\label{p}
\eq
The function $\cal K$ contains overall normalizations:
\ba
{\cal K}(\gamma,p) &=&
\left( Q_e \, Q_Q \right)^2,
\\ 
{\cal K}(I,p) &=&
2 \left| Q_e \, Q_Q \right|  \chi(\q) \, \rho_Z(p),
\\ 
{\cal K}(Z,p) &=&
\left[ \chi(\q) \rho_Z(p) \right]^2.
\label{calk}
\ea
The function
$\chi(\q)$ has been defined in~(\ref{chiq}).
The real-valued weak form factor $\rho_Z(p)$ was  introduced as
$\rho_{(I)}^Z(p)$ in~\cite{zfpc42}.
The form factors depend on a variety of variables:
\ba
\rho_Z(+1) &=& \rho_Z(xS, \q, xU; e,q),
\nll
\rho_Z(-1) &=& \rho_Z(xU, \q, xS; e,q),
\\
U &=& Q^2 - S.
\label{rhoz}
\ea
An explicit expression for
$\rho_Z(S,\q,U;a,b)$ in
 \oa\, including the $ZZ,WW$ box terms,
 which are
crucial here, is given by
eq.~(A.1) in~\cite{zfpc42}.

The dependence on the weak mixing angle and its radiative corrections,
but also that on a longitudinal beam polarization $\xi$,
 is contained
in the functions ${\cal V}$ and ${\cal A}$:
\ba
{\cal V}(\gamma,p)  &=& 1,
\nll 
{\cal A}(\gamma,p)  &=& 0,
\nll  
{\cal V}(I,p)       &=& v_{eq} + \lambda a_e v_q,
\nll  
{\cal A}(I,p)       &=& \left(a_{e} + \lambda v_e\right) a_q,
\nll  
{\cal V}(Z,p)       &=& \left(v_{e}^2 + a_e^2\right)a_q^2 a_e^2v_q^2 + v_{eq}^2
+ 2 \lambda a_e \left(v_ea_q^2+v_{eq}v_q\right),
\nll  
{\cal A}(Z,p)       &=& 2 \left[ a_e a_q\left(v_{e}v_q + v_{eq} \right)
+ \lambda a_q \left(v_e v_{eq}+a_{e}^2 v_q\right)\right].
\label{va}
\ea
The vector and axial vector couplings
contain finite radiative corrections from
the real-valued weak form factors $\kappa_a(p)$:
\ba
a_e &=& 1,
\nll
a_q &=& 1,
\nll
v_e &=& 1 - 4 \sin^2 \theta_W |Q_e| \kappa_e(p),
\nll
v_q &=& 1 - 4 \sin^2 \theta_W |Q_q| \kappa_q(p),
\nll
v_{eq} &=& v_e + v_q -1 +16 \sin^4 \theta_W |Q_e Q_q| \kappa_{eq}(p).
\label{kappa}
\ea
The weak form factors $\kappa_a(p)$, $a=e,q,eq$, are defined in
eqs.~(A.2) and~(A.3) of~\cite{zfpc42}:
\ba
\kappa_a(+1) &=& \kappa_a (xS, \q, xU; e,q),
\nll
\kappa_a(-1) &=& \kappa_a (xU, \q, xS; e,q).
\label{kappa2}
\ea
Some universal higher order corrections to $\rho_Z$ and $\kappa_a$ may
be found in~\cite{YR95}.
The form factors are calculated in {\tt DIZET}, file {\tt
  rokap.f}, and transferred to other parts of the package as elements of
a vector {\tt XFF}.

An approximation to the exact electroweak expressions in terms of an
effective weak mixing angle may be introduced as follows:
\ba
\rho                      &\rightarrow& 1,
\nll
\kappa_{eq}               &\rightarrow& \kappa_e \kappa_q,
\nll
\kappa_e \approx \kappa_q &\rightarrow& \kappa,
\ea
together with the definition
\ba
\sin^{2}\theta_{W}^{\mr{eff}}
 =\kappa \sin^2 \theta_W.
\label{swtwo}
\ea
(cf.~sections~\ref{inpt}, \ref{genfun}, and~\ref{strfun}).
In \he\, a numerical value may be chosen for
$\sin^{2}\theta_{W}^{\mr{eff}}$.
This value may be identified by that used
in the description of the $Z$ resonance physics at LEP~1:
\ba
\sin^{2}\theta_{W}^{\mr{eff}}
=
\kappa_e
\sin^2 \theta_W,
\label{swtwe}
\ea
where the form factor $\kappa_e$ describes the decay of the $Z$ boson
into two charged leptons.
Thus, a certain scale has been chosen ($\q = - M_Z^2$), there are
no contributions from box diagrams included etc.
For details see~\cite{YR95,abr,zfitter}.

For the $e^- p$~~$\cal CC$ scattering, the corresponding formulae are:
\ba
{\cal W}_2^{-\cal{SM}}(x,\q)
&=&
\frac{1-\xi}{2} x \sum_{i=1}^3 \left[
\rho_C^2(1) {u}_i(x,\q)
+ \rho_C^2(-1) {\bar d}_i(x,\q) \right],
\label{calfc}
\\ 
{x\cal W}_3^{-\cal{SM}}(x,\q)
&=&
\frac{1-\xi}{2} x \sum_{i=1}^3 \left[
\rho_C^2(1) {u}_i(x,\q)
- \rho_C^2(-1) {\bar d}_i(x,\q) \right],
\label{calcc}
\ea
whereas, for the $e^+ p$ scattering:
\ba
{\cal W}_2^{+\cal{SM}}(x,\q)
&=&
\frac{1+\xi}{2} x \sum_{i=1}^3 \left[
\rho_C^2(-1) {d}_i(x,\q)
+ \rho_C^2(1) {\bar u}_i(x,\q) \right],
\label{calf3}
\\ 
{x\cal W}_3^{+\cal{SM}}(x,\q)
&=&
\frac{1+\xi}{2} x \sum_{i=1}^3 \left[
\rho_C^2(-1) {d}_i(x,\q)
- \rho_C^2(1) {\bar u}_i(x,\q) \right].
\label{calc3}
\ea
For {\tt IMOMS}=1,2,
the real-valued weak form factors $\rho_C$ are defined as follows:
\ba
\rho_C(+1) &=& 1+\delta \rho_W(xS,\q,xU;e,q),
\nll
\rho_C(-1) &=& 1+\delta \rho_W(xU,\q,xS;e,q),
\label{rhoc12}
\ea
where the arguments $e,q$ indicate the isospin and charge contents of the
scattering particles.
For {\tt IMOMS}=3, the definition is instead:
\ba
\rho_C(+1) &=& \frac{ \left[1+\delta
  \rho_W(xS,\q,xU;e,q)\right]}{1-\Delta r},
\nll
\rho_C(-1) &=& \frac{ \left[1+\delta
  \rho_W(xU,\q,xS;e,q)\right]}{1-\Delta r}.
\label{rhoc}
\ea

The weak correction $\delta \rho_W$ is a part of the electroweak
virtual $\cal CC$ corrections, which include also electromagnetic
contributions.
The latter have to be combined with real photonic bremsstrahlung in order to
cancel infrared divergent intermediary terms.
As is well-known, the separation of a weak and an electromagnetic part of the
virtual corrections,
\ba
\rho_W = 1 + \delta \rho_W +\delta \rho_W^{\mr{QED}},
\label{rwq}
\ea
is not unique for the charged current.
For practical purposes, a recipe was proposed in~\cite{zfpc44}, which
allows a reasonable definition of both parts.
In this reference, further details and explicit expressions for the
form factors may be found in appendix~A.

\bigskip

For all the weak form factors $\rho$ and $\kappa$, the masses of the
light leptons and quarks may be neglected while the $t$ quark mass has
to be taken into account.
The exception is $\Delta \alpha$, which depends on the
light fermion masses logarithmically.
To the QED part of the corrections, which is discussed later,
terms of the order $\ln \left [
\{Q^2,S,U\}/m_f^2 \right ]$ will also contribute.
For details we again refer to the quoted literature.


\subsection
{Parton distributions and QCD corrections to structure functions
\label{pardis}
}
Nucleons and hadrons will be assumed to be built out of quarks.
The quark distributions are labelled as follows:
\ba
q_i &=& (u,d,s,c,b,t), \hspace{1.0cm} \mathrm{for}~~i=1,\ldots,6,
\\
{\bar q}_i &=& ({\bar u}, {\bar d},{\bar s},{\bar c},{\bar b},{\bar
  t}), \hspace{1.0cm} \mathrm{for}~~i=1,\ldots,6.
\label{qdistr}
\ea
These quark distributions contain both the valence and sea quark
contributions.
The explicit expression for $xW_3^-$, which was introduced
in~(\ref{w3md}) may serve as an example:
\ba
x W_3^{-}(x,\q) &=& 2 \left[ {q}_{12}^{-}(x,Q^2) +
{q}_{43}^{-}(x,Q^2)+{q}_{65}^{-}(x,Q^2) \right] .
\label{w3mod}
\ea
Here, we use the following notations for the parton densities:
\ba
{q}_{ij}^{\pm}(x,Q^2) &=&
x \left[ q_i(x,Q^2) \pm {\bar q}_j(x,Q^2) \right].
\label{qij}
\ea
The combinations ${q}_{ij}^L(x,Q^2)$ are defined in eq.~(\ref{mslon}).
The quark densities $q_i$
 are functions of $x$ and $Q^2$.
In \he, there are three approaches to the parton distributions available:
\begin{itemize}
\item
       Leading order (LO) distributions;
\item
       Next to leading order (NLO)
distributions in the ${\overline {\rm MS}}$
       scheme;
\item
       Next to leading order distributions in the ${\rm DIS}$
       scheme.
\end{itemize}
The NLO corrections are only active if the flag {\tt IWEA = 0} is
set. Otherwise only LO distributions are accessed.
\subsubsection
{The strong coupling constant}
\label{alsms}
For the renormalization of the strong coupling constant, $\alpha_s$,
 the
$\overline{\rm MS}$ scheme is choosen.
The expressions for $\alpha_s$ in leading and next to leading order are
\ba
\as^{\rm LO}(\q) &=& \frac{4\pi}{\beta_0 \ln \left(\q/\Lambda^2\right)},
\\
\as^{\rm NLO}(\q) &=& \as^{\rm LO}(\q)
\left[ 1 - \frac{\beta_1}{\beta_0^2}
\frac{\ln\ln\left(\q/\Lambda^2\right)}
{\left[\ln\left(\q/\Lambda^2\right)\right]^2}
\right],
\ea
with
\ba
\beta_0 &=& 11 - \frac{2}{3} N_f,
\\
\beta_1 &=& 102 - \frac{38}{3} N_f.
\label{cf}
\ea
Here, $\Lambda \equiv
\Lambda_{\overline{\rm MS}}$ denotes the QCD parameter.
$\Lambda_{\overline{\rm MS }}$ depends on the number of active
flavours, $N_f$. On passing the respective flavour thresholds the value
of $\Lambda$ is changed leaving $\alpha_s$ continuous. In the case
of distributions contained in {\tt PDFACT} this treatment of
$\alpha_s$ is
provided. For other cases the corresponding function {\tt ALPQCD(ISSET,Q)}
has to be added    by the user. The mass thresholds in the case of the
distributions~\cite{MRS} were choosen as in~\cite{GRV95}.
%
\subsubsection
{Parton distributions in the $\overline{\rm MS}$ factorization
scheme}
\label{msb}
Several types of parton densities emerge in the structure
functions and were defined in~(\ref{qij}):
${q}_{ij}^{+}(x,Q^2)$, ${q}_{ij}^{-}(x,Q^2)$, and ${q}_{ij}^L(x,Q^2)$.

The distribution $q_{ij}^+$ reads
\ba
\label{msplu}
{q}_{ij}^+(x,Q^2)
&=& {q}_{ij}^+(x,Q^2)_{\overline{{\rm MS}}}
+ \frac{\as(\q)}{2 \pi}
\Biggl\{
\int_x^1 \frac{dy}{y}
\Biggl[ \frac{x}{y}
C_q\left(\frac{x}{y}\right)
\left[
 {q}_{ij}^+(y,Q^2)_{\overline{{\rm MS}}}
- {q}_{ij}^+(x,Q^2)_{\overline{{\rm MS}}} \right]
\nll &&+~
 2\frac{x}{y} C_g\left(\frac{x}{y}\right)
{\cal G}(y,\q)_{\overline{{\rm MS}}}
\Biggr]
-{q}_{ij}^+(x,Q^2)_{\overline{{\rm MS}}} \int_0^x dy C_q(y,Q^2)
\Biggr\},
\label{qmsb}
\ea
where
\ba
{\cal G} &\equiv& x G,
\\
C_q(z) &=& C_F \left[ \frac{1+z^2}{1-z}
\left( \ln\frac{1-z}{z} -\frac{3}{4}\right) + \frac{1}{4} (9+5z)\right],
\label{cq}
\\
C_g(z) &=& \frac{1}{2} \left\{
\left[z^2+(1-z)^2\right]\ln\frac{1-z}{z}+8z(1-z)-1
\right\},
\label{cg}
\ea
with
$C_F = (N_c^2-1)/(2N_c) \equiv 4/3$.
\label{c-f}

Correspondingly $q_{ij}^-$ is:
\ba
{q}_{ij}^-(x,Q^2) = x \left[ q_i(x,\q) - {\bar q}_j(x,\q) \right].
\label{qhat}
\ea
For this combination one has:
\ba
\label{msmin}
{q}_{ij}^-(x,Q^2)
&=& {q}_{ij}^-(x)_{\overline{{\rm MS}}}
+ \frac{\as(\q)}{2 \pi}
\Biggl\{
\int_x^1 \frac{dy}{y} \frac{x}{y}
\left[
C_3\left(\frac{x}{y}\right)
{q}_{ij}^-(y,Q^2)_{\overline{{\rm MS}}}
- C_q\left(\frac{x}{y}\right) {q}_{ij}^-(x,Q^2)_{\overline{{\rm MS}}}
\right]
\nll
& &-~
{q}_{ij}^-(x,Q^2)_{\overline{{\rm MS}}} \int_0^x {dy} C_q(y)
\Biggr\},
\label{cqh}
\ea
with:
\ba
C_3(z) = C_q(z) - (1+z) C_F.
\label{c3}
\ea
Finally, for the longitudinal structure function, the correction reads:
\ba
\label{mslon}
{q}_{ij}^L(x,Q^2)
&=&
\frac{\as(\q)}{2\pi}
\int_x^1 \frac{dy}{y} \left(\frac{x}{y}\right)^2
\left\{
2C_F {q}_{ij}^+(y,Q^2)_{\overline{{\rm MS}}} + 4\left(1 - \frac{x}{y}
\right )
{\cal G}(y,\q)_{\overline{{\rm MS}}}
\right\} .
\label{lsf}
\ea
In a series of parameterizations all active quark flavours are delt with
as massless. In these cases the above relations (and those given in
section~\ref{diss}) yield the complete description. On the other
hand, some parameterizations account for mass effects explicitely
(e.g.~\cite{GRV95}) leading to  coefficient functions
different from~(\ref{msplu}, \ref{msmin}, \ref{mslon})
for
heavy flavour contributions $q = c,b,t$.
In general these contributions are process dependent.
{\tt HECTOR} foresees a simple approach for parameterizations contained
in {\tt PDFACT}
in the present version:
the light flavours $(u,d,s)$
 are parametrized according to~\cite{GRV95},
while the heavy flavours $(c,b)$
 are taken from~\cite{GRV92}, which is
some suitable approximation (cf.~\cite{GRV95}).

In all other cases (including the use of {\tt PDFLIB})
the user has to provide the correct mappings her/himself.

The distributions
$[q_{i}(x,\q)\pm {\bar{q}}_{i}(x,\q)]_{\overline{{\rm MS}}}$
and $G(y,\q)_{\overline{\rm  MS}}$ have to be
taken from one of the corresponding
libraries.

In the above we have chosen $Q^2$ as the factorization scale always.
\subsubsection
{Parton distributions in the DIS factorization
scheme
\label{diss}
}

The $+$ distributions~(\ref{qij}) are associated with the coefficient
function
\ba
\delta(1 - z)
\ea
in the DIS scheme. Thus the
generalized structure functions ${\cal F}_2, {\cal G}_2,
{\cal H}_2, {\cal W}_2$, are just linear combinations of the quark
quark distributions
in the DIS scheme.
The generalized structure functions ${\cal G}_3, {\cal H}_3,
{\cal W}_3$, on the other hand, constitute of linear combinations of
\ba
{q}_{ij}^-(x,Q^2)
&=&
{q}_{ij}^-(x,Q^2)_{\rm DIS}
+ \frac{\as(\q)}{2 \pi}
\int_x^1 \frac{dy}{y}
\frac{x}{y}
C_3^{\rm DIS}\left(\frac{x}{y}\right)
{q}_{ij}^-(y,Q^2)_{\rm DIS}.
\label{dish}
\ea
with
\ba
C_3^{\rm DIS}(z) = - C_F (1 +z).
\ea
The expression for the
longitudinal structure functions are identical in the $\overline{\rm MS}$
and DIS schemes.

Again
the distributions
$[q_{i}(x,\q)\pm {\bar{q}}_{i}(x,\q)]_{\rm DIS}$
and $G(y,\q)_{\rm DIS}$ have to be
taken from one of the corresponding
libraries.
The heavy flavour distributions are delt with as in section~\ref{msb}.

\subsubsection
{Structure functions in different schemes
\label{sfids}
}
The physical expressions of the structure
functions~(\ref{lf2})--(\ref{lxh3}), (\ref{lw2p})--(\ref{l2xw1m})
in the different schemes are obtained
by linear combinations of the functions:
\ba
{q}_{ij,\overline{\rm MS}({\rm DIS})}^+(x,Q^2),  \\
{q}_{ij,\overline{\rm MS}({\rm DIS})}^-(x,Q^2),  \\
{q}_{ij,\overline{\rm MS}({\rm DIS})}^L(x,Q^2),
\ea
replacing the
 corresponding leading order expressions there.

%
\subsection
{Low $Q^2$ modifications of structure functions and parton distributions
\label{lowq2}
}
In order to ensure the vanishing of the photonic parts of the
generalized structure functions at low $Q^2$, dedicated modifications
of the photonic structure functions  are
foreseen in \he.
They are realized by a global suppression factor {\tt FVAR} in the
files       {\tt disepm.f},
{\tt  genstf.f}, {\tt prodis.f}, and {\tt gsflow.f}.
The default setting is
{\tt IVAR}=0 corresponding to~\cite{Prokh}.
The structure functions are
\ba
{\cal F}_{1,2}(x,Q^2)
&\rightarrow&
F_{var} \, {\cal F}_{1,2}(x,Q^2),
\\
F_{var}
&=&
[1 - \exp(-aQ^2)], ~~~~~~
a=3.37 \,  {\mr GeV}^{-2}.
\label{prokh}
\ea
Other approaches have a similar form:
\ba
F_{var}(Q^2)
&=&
\left[1 - W_2^{el}(Q^2)\right] .
\label{w2el}
\ea
Parameterizations of the elastic form factor $W_2^{el}$,
\ba
\label{wgegm}
W_2^{el}(Q^2)
&=&
\frac{G_e^2 + \tau G_m^2}{1+\tau},
\\
\tau &=& \frac{Q^2}{4M^2},
\ea
may be chosen (cf. e.g.~\cite{brasse})
as is listed in section~\ref{sfp}. The other
structure functions are continued by their value at $Q_0^2$ to values
in the range $Q^2 < Q^2_0$ at a given value of $x$.

\begin{figure}[hbtp]
\begin{center}
\vspace*{2.3cm}
\mbox{
\epsfysize=20.0cm
\epsffile{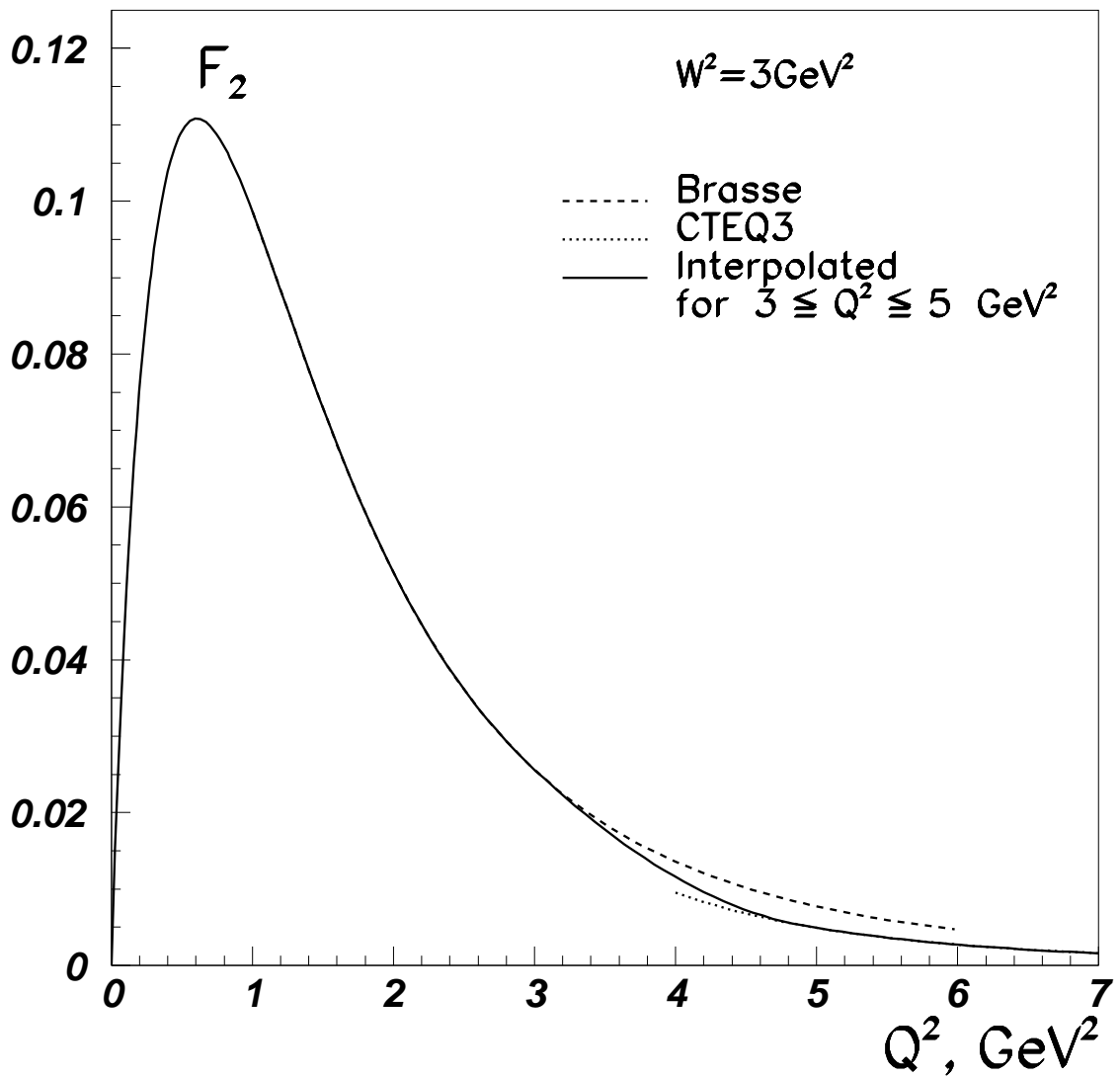}
}
\end{center}
\vspace*{-2.2cm}
\caption[\it
Low $Q^2$ interpolation for the structure function $F_2(x,Q^2)$
]
{\it
Low $Q^2$ interpolation for the structure function $F_2(x,Q^2)$
between $F_2 = F_2^{LO}(CTEQ3)$ for $Q^2 \geq 5 {\mr GeV^2}$
({\tt ISCH = 0,
ISSE =1}), and the low $Q^2$ extension seleced by the flag {\tt IVAR=2}
for $Q^2 \leq 3 {\mr GeV^2}$.
\label{fcc4}
}
\end{figure}

In figure~\ref{fcc4} we show as an example the combination of the
structure function $F_2$ from~\cite{cteq3} with a low
$Q^2$ parameterization as proposed in~\cite{stein,brasse}.

%
%

\clearpage
\newpage
%
\subsection
{QED corrections. Introduction
\label{qedcor}
}
\begin{figure}[tbhp]
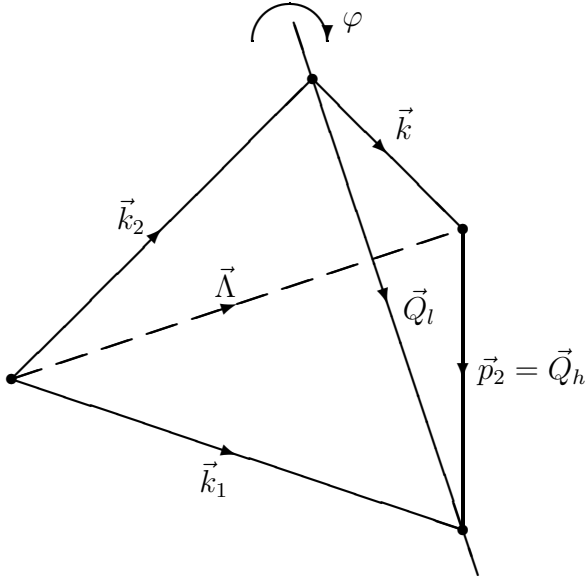

\begin{minipage}[bht]{12.cm}{
\begin{center}
\begin{Feynman}{70,75}{0,0}{1.}
%
\thicklines 
\put(00,20){\line(1, 1){40}} 
\put(00,20){\line(3,-1){60}} 
\put(60,00){\line(0, 1){40}} 
\put(40,60){\line(1,-1){20}} 
\put(40,60){\line(1,-3){20}} 
%
\multiput(00,20)(6,2){10}{\line(3,1){4}}  
\put(60,00){\line(1,-3){2.0}}  
\put(40,60){\line(-1,3){2.5}}  
\put(37,65){\oval(10,10)[t]}
\put(42,65){\vector(0,-1){0}}
\put(44,67){$\varphi$}
%
\put(20,40){\vector(1, 1){0}}
\put(14,40){${\vec k}_2$}
\put(30,10){\vector(3,-1){0}}
\put(25,05){${\vec k}_1$}
\put(60,20){\vector(0,-1){0}}
\put(62,20){${\vec p}_2={\vec Q}_h$}
\put(50,50){\vector(1,-1){0}}
\put(51,52){${\vec k}$}
\put(50,30){\vector(1,-3){0}}
\put(52,28){${\vec Q}_l$}
\put(30,30){\vector(3, 1){0}}
\put(27,31){${\vec \Lambda}$}
\put(00,20){\circle*{1.5}}
\put(60,00){\circle*{1.5}}
\put(60,40){\circle*{1.5}}
\put(40,60){\circle*{1.5}}
%
\end{Feynman}
\end{center}
}\end{minipage}
\vspace*{1.cm}
\caption[\it
Configuration of 3--momenta for deep inelastic scattering in the
proton rest system
]{\it
Configuration of the 3--momenta in the reaction $e(k_1) + p(p_1)
\rightarrow l'(k_2) + X(p_2) + \gamma(k)$ in the proton rest system,
${\vec p}_1=0$.
}
\label{tetra}
\end{figure}

The inclusion of photonic corrections,
\bq
e(k_1) + p(p_1) \rightarrow e(k_2) + X(p_2) + n\gamma(k),
\label{eqdeep}
\eq
is crucial for a correct analysis of deep inelastic scattering.
The radiative corrections may {\it differ considerably}
comparing different sets of outer kinematical variables.
If Born kinematics is valid, the scaling variable $y$, e.g., may be
defined from momentum measurements of the leptons or from the hadrons
with no difference:
\ba
 \yl &=& \frac{p_1(k_1-k_2)}{p_1k_1},
\\
 \yh &=& \frac{ p_1(p_2-p_1)}{p_1k_1}.
\ea
As may be seen from figure~\ref{tetra}~\cite{MI}, the two definitions are quite
different if a photon with non-vanishing momentum $k$ is emitted.
The same holds true for the variables
$x$ and $Q^2$, or $y$.

In fixed target experiments mostly leptonic variables are used.
At $ep$ colliders, a lot of different sets of variables may be measured:
the scattering angles of the lepton or the hadronic system, the
energies of the scattered lepton or the hadrons, the transverse
hadronic momentum, the kinetic energy of the hadrons
etc.~(cf.~refs.~\cite{KIN0}-\cite{KIN3}).
Those definitions of $x, y, Q^2$, which may be used in \he\, are
introduced in appendix~\ref{appvardef}.

The {\tt HELIOS} part of {\tt HECTOR} allows the calculation of
leading logarithmic approximations (LLA) including higher order
corrections to $\cal NC$ and $\cal CC$ scattering in a large variety
of different
variable sets.

With the {\tt TERAD} part one may calculate complete \oa\ QED
corrections, including soft photon exponentiation, in a model
independent approach.
\hfill
The structure functions are  not

%

\clearpage

\noindent
necessarily calculated in the quark
parton model and the kinematics has been worked out exact in the
proton mass and ultra-relativistic in the electron mass.

The {\tt TERAD} part of \he\ may also be used for the calculation of
complete \oa\ QED corrections, including soft photon exponentiation,
in the quark parton model.
This makes use of the {\tt DISEP} branch.
In the quark parton model, one may, in addition to the leptonic
corrections, also
calculate  the lepton quark interference terms
and treat the
corrections from the quark lines.

The complete calculations are available for both $\cal NC$ and $\cal
CC$ scattering
for a series of sets of kinematical
variables.

For $\cal NC$ scattering at very low $Q^2$,
the complete leptonic corrections
described in leptonic variables are available from the {\tt TERADLOW}
branch.
Here, the exact kinematics both in the electron and proton masses is
taken into account.

The {\it combined} use of both parts {\tt HELIOS} and {\tt TERAD} is
also possible.
Then, the complete \oa\ corrections of the {\tt TERAD} part, without
its soft photon exponentiation, are used together with the complete
higher order corrections of the {\tt HELIOS} part.

Details on the general
structure of \he\ may be seen in figure~\ref{hecflow}
and on the availability of variables in figure~\ref{branches}.
\subsection{{\tt HELIOS}: The leading logarithmic
  approximation
\label{lla}
}
\begin{figure}[t]
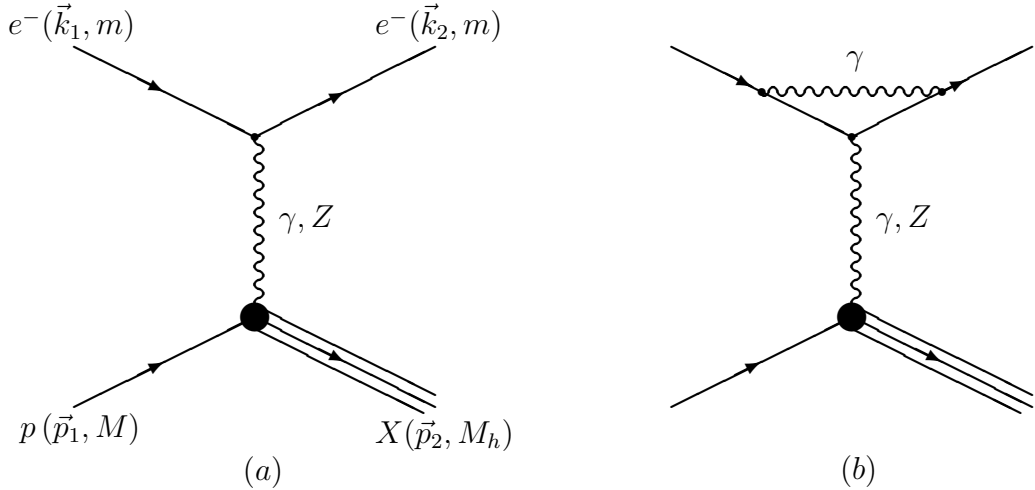

\vspace*{1.cm}
\begin{minipage}[tbhp]{7.8cm}{
\begin{center}
\begin{Feynman}{60,60}{0,0}{0.8}
%
\put(00,60){\fermiondrr}
\put(-11,62){$e^-(\vec k_1,m)$}  
\put( 50,62){$e^-(\vec k_2,m)$}
\put( 50,-06){$X(\vec p_2,M_h)$}
\put( 28,-12){$(a)$}
\put(-09,-05){$p \, (\vec p_1,M)$}
\put(34,30){$\gamma, Z$}
\put(60,60){\fermionurr}
\put(30.5,15){\photonup}
\put(30,15){\circle*{5}}
\put(30,17){\line(2,-1){30}}
\put(30,13){\line(2,-1){28}}
\put(30,45){\circle*{1.5}}
\put(30,15){\fermionurr}
\put(30,15){\fermiondrr}
\end{Feynman}
\end{center}
}\end{minipage}
\begin{minipage}[tbhp]{7.8cm} {
\begin{center}
\begin{Feynman}{60,60}{0,0}{0.8}
%
\put(34,30){$\gamma, Z$}
\put(29,57){$\gamma$}
\put(00,60){\vector(2,-1){13.5}}
\put(11,54.5){\line(2,-1){19.}}
%
\put(45,52.5){\line(2,1){15}}
\put(30,45){\vector(2,1){19.}}
\put(14.5,52.4){\photonright}
\put(45.0,52.4){\circle*{1.5}}
\put(15.0,52.4){\circle*{1.5}}
\put( 28,-12){$(b)$}
\put(30,45){\circle*{1.5}}
\put(30.5,15){\photonup}
\put(30,15){\circle*{5}}
\put(30,17){\line(2,-1){30}}
\put(30,13){\line(2,-1){28}}
\put(30,15){\fermionurr}
\put(30,15){\fermiondrr}
\end{Feynman}
\end{center}
}\end{minipage}
\vspace*{1.cm}
%
\caption[\it
$\cal NC$ deep inelastic scattering of electrons off protons:
Born diagram and leptonic QED vertex correction
]{\it
NC deep inelastic scattering of electrons off protons:
(a) Born diagram, (b) leptonic QED
vertex correction
\label{fig1}
}
\end{figure}

\begin{figure}[tbhp]
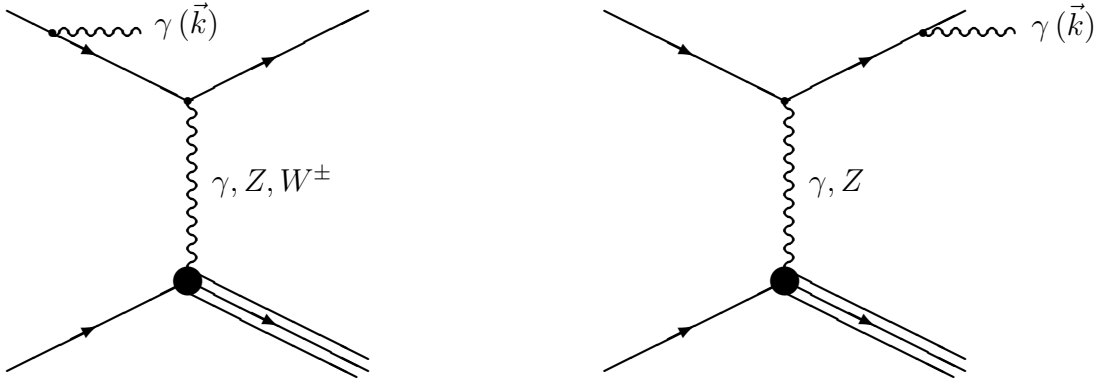

\vspace*{1.cm}
\begin{minipage}[tbh]{7.8cm}{
\begin{center}
\begin{Feynman}{60,60}{0,0}{0.8}
%
\put(00,60){\fermiondrr}
\put(60,60){\fermionurr}
\put(30.5,15){\photonup}
\put(30,15){\circle*{5}}
\put(24.5,56.25){$\gamma \, (\vec k)$}
\put(34,30){$\gamma, Z, W^{\pm}$}
\put(30,17){\line(2,-1){30}}
\put(30,13){\line(2,-1){28}}
\put(30,45){\circle*{1.5}}
\put(30,15){\fermionurr}
\put(30,15){\fermiondrr}
\put(7.5,56.25){\circle*{1.5}}
\put(7.0,56.25){\photonrighthalf}
\end{Feynman}
\end{center}
}\end{minipage}
\begin{minipage}[tbh]{7.8cm} {
\begin{center}
\begin{Feynman}{60,60}{0,0}{0.8}
%
\put(00,60){\fermiondrr}
\put(60,60){\fermionurr}
\put(30.5,15){\photonup}
\put(71,56){$\gamma \, (\vec k)$}
\put(34,30){$\gamma, Z $}
\put(30,15){\circle*{5}}
\put(30,17){\line(2,-1){30}}
\put(30,13){\line(2,-1){28}}
\put(30,45){\circle*{1.5}}
\put(30,15){\fermionurr}
\put(30,15){\fermiondrr}
\put(53.0,56.25){\photonrighthalf}
\put(53.0,56.25){\circle*{1.5}}
\end{Feynman}
\end{center}
}\end{minipage}
\vspace*{.5cm}
%
\caption
{\it
Leptonic QED
bremsstrahlung diagrams
\label{fig4}
}
\end{figure}

\newcommand{\CALL}{{\cal L}}
\newcommand{\CALH}{$\cal H$}
The LLA corrections are calculated  by the {\tt HELIOS} branch of
\he.
They are related to logarithmic mass singularities, which
distinguish  them
 from the remaining corrections within a given
order of perturbation theory.
In leading logarithmic approximation, one may separate the terms
due to
initial and final state radiation.
The QED corrections are composed out of several contributions.
In {\tt HECTOR}, the LLA cross section is\footnote{
There are also leading logarithmic corrections related to the emission of
photons from the initial or final state quarks.
They dependend on $\ln (Q^2/m_q^2)$ iff calculated
in the on-mass-shell scheme. These terms deserve a
special treatment.
One can absorb these corrections into parton
distributions~\cite{RUJU,LLA1,KP,LLAmix,hsllaq}
or fragmentation functions~\cite{LLA1} leading to modified expressions
of the scaling violations of these quantities.}:
\ba
\frac{d^2 \sigma^{LLA}}{dxdy}
&=&
\frac{d^2 \sigma^{0}}{dx dy}
+
\frac{d^2 \sigma^{ini,1loop}}{dxdy}
+
\frac{d^2 \sigma^{ini,2loop}}{dxdy}
+
\frac{d^2 \sigma^{ini,>2,soft}}{dx dy}
+
\frac{d^2 \sigma^{ini,e^- \rightarrow e^+}}{dx dy}
\nll &&+~
\frac{d^2 \sigma^{fin,1loop}}{dxdy}
+
\frac{d^2 \sigma^{C}}{dxdy} .
\label{lla1}
\ea
The Born contribution has been discussed in section~\ref{borsec}.
It is calculated in {\tt sigbrn.f} and may or may not include weak
virtual corrections depending on flag {\tt IWEAK} and QCD corrections
to the parton distributions depending on flag {\tt ISCH}.
For neutral current scattering both initial and final state LLA
corrections may contribute, while for charged current scattering there
is only initial state radiation.
For the $\cal NC$ case, the corresponding Born diagram is shown in
figure~\ref{fig1}(a).
The virtual corrections to order \oa\ arise from the diagram in
figure~\ref{fig1}(b) for the $\cal NC$ case.
For $\cal CC$ scattering, it is replaced by a diagram with a photon exchange
between incoming electron and virtual $W$ boson.
The \oa\ diagrams with real photon emission are shown in
figure~\ref{fig4}.

\bigskip

The two first order LLA corrections have a common generic structure:
\ba
\frac{d^2 \sigma^{ini(fin),1loop}}{dxdy}
&=&
\frac{\alpha}{2\pi} L_e
\int\limits_{0}^{1}dz
P_{ee}^{(1)}
\left\{
\theta(z-z_0)
{\cal J}(x,y,Q^2)
\left.
\frac{d^2 \sigma^{0}}{dxdy}\right|_{x={\hat x}, y={\hat y}, S={\hat S}}
-
\frac{d^2 \sigma^{0}}{dxdy}
\right\} ,
\label{lla2}
\nll
\\
P_{ee}^{(1)} &=&  \frac{1+z^2}{1-z} .
\label{pee1}
\ea
Here, we introduced the notion
\ba
L_e
= {\ln}  \frac{Q^2}{m_e^2} -1.
\label{ln1}
\ea
The definition~(\ref{ln1}) reproduces the soft photon terms of complete
calculations in leptonic variables (see e.g.~\cite{MI}).

The
expressions~(\ref{lla2}) for initial and final state radiation
differ by the scaling properties of
the kinematical variables, the resulting Jacobian, $\cal J$,
and the integration
bounds.
In table~\ref{tabLLAini}, for the initial state corrections
these parameters
are listed for a variety of different sets of variables.
Table~\ref{tabLLAfin} contains the same for final state corrections if
they exist.
For some variables, e.g. the
Jaquet-Blondel variables, the final state is treated
completely
inclusive. Then, in accordance with the Kinoshita-Lee-Nauenberg
theorem~\cite{KNL} there is no LLA correction.

Furthermore, there are certain second order corrections from the initial
state  which may be non-negligible~\cite{jblla2}:
\ba
\frac{d^2 \sigma^{ini,2loop}}{dx dy} &=&
 \left[ \frac{\alpha}{2 \pi}
L_e\right]^2
\int_0^1 dz
 P_{ee}^{(2,1)}(z)
\left \{
\theta(z - z_0) {\cal J}(x,y,z)
\left.\frac{d^2 \sigma^{0}}{dx dy} \right|_{x=\hat{x},
y=\hat{y},S=\hat{S}}  -  \frac{d^2 \sigma^{0}}{dx dy} \right \}
\nonumber
\end{eqnarray}
\begin{equation}
+
\left( \frac{\alpha}{2 \pi} \right )^2
\int_{z_0}^1 dz
 \left \{
L_e^2
P_{ee}^{(2,2)}(z)
+  L_e
\sum_{f=l,q} \ln\frac{Q^2}{m^2_f} \,
P_{ee,f}^{(2,3)}(z) \right \}
{\cal J}(x,y,z)
\left.\frac{d^2 \sigma^{0}}{dx dy} \right|_{x=\hat{x},
y=\hat{y},S=\hat{S}}.
\end{equation}

\newpage

\bigskip

\noindent
The second order splitting functions are:
\begin{eqnarray}
\label{mellin1}
P_{ee}^{(2,1)}(z) &=& \frac{1}{2}
\left [P_{ee}^{(1)} \otimes P_{ee}^{(1)} \right ](z)
\nonumber\\
&=& \frac{1 + z^2}{1 - z} \left [ 2\ln(1-z) - \ln z + \frac{3}{2} \right ]
+ \frac{1}{2}(1 + z) \ln z - (1 -z),
\\
\label{mellin2}
P_{ee}^{(2,2)}(z) &=& \frac{1}{2}
\left [
P_{e \gamma}^{(1)} \otimes P_{\gamma e}^{(1)}
\right ](z)  \nonumber\\
&\equiv&
(1 + z) \ln z +
\frac{1}{2}(1 - z) + \frac{2}{3} \frac{1}{z} (1 - z^3),
\\
\label{eq-p23}
P_{ee,f}^{(2,3)}(z) &=&
N_c(f) Q_f^2
\frac{1}{3} P_{ee}^{(1)}(z)
                        \theta \left( 1 - z - \frac{2 m_f}{E_e}
                        \right).
\ea
The symbol
$\otimes$ in~(\ref{mellin1}) and~(\ref{mellin2}) denotes the Mellin
convolution:
\ba
\label{eqmellin}
A(x) \otimes B(x) = \int_0^1 dx_1 \int_0^1 dx_2 \delta(x-x_1x_2)
A(x_1) B(x_2).
\ea
$Q_f$ is the fermion charge, and
$N_c(f) = 3$ for quarks, $N_c(f) = 1$ for leptons, respectively.
In~(\ref{eq-p23}),
we parametrize $\alpha_{_{QED}}(s, m_{f_i})$
in terms of {effective} quark masses
and
use the parameters
$m_u =  62$ MeV, $m_d =  83$ MeV, $m_s = 215$ MeV,
$m_c = 1.5$ GeV, and $m_b = 4.5$ GeV as obtained in \cite{x21,x24}.

The {\tt HELIOS} part of \he\ allows to include
higher order soft photon corrections
into cross section
calculations:
\begin{equation}
\frac{d^2 \sigma^{(>2,soft)}}{dx dy} =
 \int_0^1  dz
P_{ee}^{(>2)}(z,Q^2) \left \{
\theta(z - z_0) {\cal J}(x,y,z)
\left. \frac{d^2 \sigma^{(0)}}{dx dy} \right|_{x=\hat{x},
y=\hat{y},S=\hat{S}}  -  \frac{d^2 \sigma^{(0)}}{dx dy}
\right\} ,
\end{equation}
with~\cite{jblla2}:
\ba
\label{eqP3}
P_{ee}^{>2}(z,Q^2) &=& D_{NS}(z,Q^2) -
\frac{\alpha}{2 \pi} L_e \,
\frac{2}{1 - z} \left \{ 1 +
\frac{\alpha}{2\pi} L_e \,
\left [ \frac{11}{6} + 2 \ln(1 - z) \right ] \right \},
\nll
\\
\label{eqNS}
D_{NS}(z,Q^2) &=& \zeta (1 - z)^{\zeta - 1}
\frac{\exp\left [ \frac{1}{2} \zeta \left (
\frac{3}{2} - 2 \gamma_E \right ) \right ]} {\Gamma(1 + \zeta)},
\\
\label{eqzet}
\zeta &=& -3 \ln \left [ 1 - (\alpha/3\pi) L_e
\right ].
\ea

Equation~(\ref{eqNS}) was derived in~\cite{GL} already.
If the charge of the final state electron is not recorded,
also the conversion cross section:
\begin{equation}
\label{conv}
\frac{d^2 \sigma^{(2,e^- \rightarrow e^+)}}{dx dy} =
\int_{z_0}^1 dz
P(z,Q^2; e^- \rightarrow e^+)
 {\cal J}(x,y,z)
\left.\frac{d^2 \sigma^{(0)}}{dx dy} \right|_{x=\hat{x},
y=\hat{y},S=\hat{S}} ,
\end{equation}
with the conversion rate:
\begin{equation}
P(z,Q^2; e^- \rightarrow e^+) = \left(\frac{\alpha}{2 \pi} \right )^2
L_e^2 P_{ee}^{(2,2)}(z)
\end{equation}
has to be included into the radiative corrections.

The explicit expressions for the splitting functions
have been taken from~\cite{jblla2} where additional useful comments on
the higher order corrections may be found.

Finally, we come to the Compton correction.
It has to be taken into account in the case of leptonic variables:
\ba
\frac{d^2 \sigma^{C1}}{d\xl d\yl}
&=&
\frac{\alpha^3}{\xl S}
\left[1+(1-\yl)^2\right]
\ln \left( \frac{\ql}{M^2}  \right)
\int \limits_{x_l}^{1} \frac{dz}{z^2}
\frac{z^2+(\xl-z)^2}{\xl(1-\yl)}
\sum_f \left[q_f(z,\ql) + {\bar q}_f(z,\ql)\right].
\nll
\label{compt1}
\ea
This expression may be extracted from the complete calculation of the
\oa\ corrections~\cite{LLA1,MI}
and is accessed setting the  flag  {\tt ICMP}=1.
An alternative description of the Compton peak has been derived
in~\cite{BLH}~({\tt ICMP}=2):
\ba
\label{compt2}
\frac{d^2 \sigma^{C_2}}{d\xl d\yl}
&=& \int_0^1 \frac{dz}{z} D_{\gamma/p}(z,Q^2_l) \left .
\frac{d^2 \hat{\sigma}
(e \gamma \rightarrow e \gamma)}{d \hat{x} d y_l}
\right |_{\hat{s} = zs, \hat{x} = x_l/z} ,
\ea
with:
\ba
\label{compt3}
\frac{d^2 \hat{\sigma}
(e \gamma \rightarrow e \gamma)}{d \hat{x} d y_l}
&=& \frac{2 \pi \alpha^2}{\hat{s}} \frac{1 + (1 -y_l)^2}{1 - y_l}
\delta(1 - \hat{x}) ,
\\
D_{\gamma/p}(x_l,Q^2_l) &=& \frac{\alpha}{2 \pi} \int_{x_l}^{1} dz
\int_{Q^2_{h,min}}^{Q^2_l} \frac{dQ^2_h}{Q^2_h} \frac{z}{x_l}
\left [ \frac{1 + (1-z)^2}{z^2} F_2 \left (\frac{x_l}{z}, Q^2_h \right )
\right.
\nonumber\\
& &~- \left.
F_L \left (\frac{x_l}{z}, Q^2_h \right ) \right ].
\ea
In principle, the Compton correction has to be taken into account also
in the case of
mixed variables.
But, although being logarithmically enhanced, it is not
larger than the rest of the \oa\ corrections in this case~\cite{MI}.

\begin{table}[thbp]
{\small
\begin{center}
\begin{tabular}[]{|c|c|c|c|c|c|}
\hline
      \multicolumn{5}{|c}{}& \\
      \multicolumn{5}{|c}{\hspace{3cm}Initial state radiation}& \\
      \multicolumn{5}{|c}{}& \\
\hline
          &&&&&              \\
          &  $\hat{S}  $
          &  $\hat{Q}^2$
          &  $\hat{y}  $
          &  $z_0      $
          &  ${\cal J}(x,y,z) $     \\
          &&&&&              \\
\hline
          &&&&&                                              \\
leptonic  variables
    &  $ S z                                            $
    &  $ Q^2_l z                                        $
    &  $\frac{{\displaystyle z+\yl-1}}{\displaystyle z} $
    &  $\frac{{\displaystyle 1-y_l}}
             {{\displaystyle 1-x_ly_l}}                 $
    &  $\frac{{\displaystyle  y_l}}
             {{\displaystyle z+y_l-1}}                  $    \\
          &&&&&                                              \\
\hline
          &&&&&                                              \\
mixed     variables
    &  $ S z                                            $
    &  $\Ql z                                           $
    &  $\frac{{\displaystyle\yjb}}{{\displaystyle z}}    $
    &  $                    \yjb                         $
    &  $1                                               $    \\
          &&&&&                                              \\
\hline
          &&&&&                                              \\
hadronic  variables
    &  $ S z                                            $
    &  $ \Qh                                            $
    &  $ \frac{{\displaystyle \yh}}{\displaystyle z}    $
    &  $ \yh                                            $
    &  $ \frac{\displaystyle 1}{\displaystyle z}        $    \\
          &&&&&                                              \\
\hline
          &&&&&                                              \\
JB        variables
    &  $ S z                                            $
    &  $\frac{\displaystyle \qjb(1-\yjb)z}
             {\displaystyle z-\yjb}                      $
    &  $\frac{\displaystyle \yjb}{\displaystyle z}       $
    &  $\frac{\displaystyle \yjb}
             {\displaystyle 1-\xjb(1-\yjb)}              $
    &  $\frac {\displaystyle 1-\yjb}
              {\displaystyle z-\yjb}                     $    \\
          &&&&&                                              \\
\hline
          &&&&&                                              \\
double angle method
    &  $ S z                                            $
    &  $\qdo z^2                                        $
    &  $\ydo                                            $
    &  $ 0                                              $
    &  $ z                                              $    \\
          &&&&&                                              \\
\hline
          &&&&&                                              \\
$\theta_l, \yjb$
    &  $ Sz                                             $
    &  $ \qan \frac{\displaystyle z(z-\yjb)}
                   {\displaystyle    1-\yjb}             $
    &  $\frac{\displaystyle\yjb}{\displaystyle z}        $
    &  $\yjb                                             $
    &  $\frac{\displaystyle z-\yjb}{\displaystyle 1-\yjb} $    \\
          &&&&&                                              \\
\hline
          &&&&&                                              \\
$\Sigma $   method
    &  $ S z                                            $
    &  $\qsi                                            $
    &  $\ysi                                            $
    &  $\xsi                                            $
    &  $ \frac{\displaystyle 1}{\displaystyle z }       $    \\
          &&&&&                                              \\
\hline
          &&&&&                                              \\
$e\Sigma $   method
    &  $ S z                                            $
    &  $\ql z                                           $
    &  $\yesi z                                         $
    &  $\xsi                                            $
    &  $ 1                                              $    \\
          &&&&&                                              \\
\hline
 \end{tabular}
\end{center}
} 
\caption[
\it
Scaling properties of kinematical
variables for leptonic initial state radiation
]{
\it
Scaling properties of various sets of kinematical
variables for leptonic initial state radiation.
The different parameters
are defined in appendix~A.
\label{tabLLAini}
}
\end{table}

\begin{table}[thbp]
{\small
\begin{center}
\begin{tabular}[]{|c|c|c|c|c|c|}
\hline
      \multicolumn{5}{|c}{}& \\
      \multicolumn{5}{|c}{\hspace{3cm}Final state radiation}& \\
      \multicolumn{5}{|c}{}& \\
\hline
          &&&&&              \\
          &  $\hat{S}  $
          &  $\hat{Q}^2$
          &  $\hat{y}  $
          &  $z_0      $
          &  ${\cal J}(x,y,z) $     \\
          &&&&&              \\
\hline
          &&&&&                                              \\
leptonic  variables
    & $  S                                               $
    & $\frac{\displaystyle \Ql}{\displaystyle z}         $
    & $\frac{\displaystyle z+\yl-1}{\displaystyle z}     $
    & ${\displaystyle 1-\yl(1-\xl) }                     $
    & $\frac{\displaystyle \yl}{\displaystyle z(z+\yl-1)}$   \\
          &&&&&                                              \\
\hline
          &&&&&                                              \\
mixed     variables
   & $ S                                                   $
   & $ \frac{\displaystyle \Ql}{\displaystyle z}           $
   & $ \yjb                                                $
   & $ \xm                                                 $
   & $ \frac{\displaystyle 1}{\displaystyle z}             $ \\
          &&&&&                                              \\
\hline
          &&&&&                                              \\
$ \Sigma $   method
   &  $ S                                                  $
   &  $ \qsi \frac{\displaystyle 1-\ysi(1-z)}
                  {\displaystyle z^2}                      $
   &  $ \frac{\displaystyle  \ysi z}
             {\displaystyle 1-\ysi(1-z)}                   $
   &  $z_0^{\Sigma,f}$
   &  $ \frac{\displaystyle 1}{\displaystyle z^2}          $ \\
          &&&&&                                              \\
\hline
          &&&&&                                              \\
$e \Sigma $   method
   &  $ S                                                 $
   &  $ \frac{\displaystyle \ql}
                    {\displaystyle z}                     $
   &  $ \frac{\displaystyle \yesi z^2}
             {\displaystyle [1-\yesi (1 - z)]^2 }         $
   &  $z_0^{\Sigma,f}$
   &  $ \frac{\displaystyle 1+\yesi (1 - z)     }
             {\displaystyle [1-\yesi(1-z)] z }            $ \\
          &&&&&                                             \\
\hline
\end{tabular}
\end{center}
} 
\caption[
\it
Scaling properties of kinematical
variables for leptonic final state radiation
]
{
\it
Scaling properties of various sets of kinematical
variables for leptonic final state radiation.
The different parameters
are defined in appendix~A.
\hfill 
\label{tabLLAfin}
}
\end{table}

\subsubsection{Cuts
\label{zcut}
}
In the LLA approach, it is possible to apply a cut on the
reconstructed electron energy by cutting the integration variable $z$.
The variable {\tt ZCUT} is defined by:
\ba
    \left [ E_J(1 - \cos \theta_J)
+ E_{e'}(1 - \cos \theta_{e'})\right ]
     \geq {\tt ZCUT}.
\ea
It is also possible to reject the production of final states with a
$Q_h^2$ and/or an invariant hadronic mass $W_h^2$ below some
bound.
See also the description of flag {\tt IHCU} in section~\ref{smtc}.

\subsection{{\tt TERAD}:
Complete \oa\ leptonic corrections in the
model independent approach for $\cal NC$ scattering
\label{secMI}
}
If one wants to improve the LLA approximation of QED corrections, a
natural step is to calculate the contributions from figures~\ref{fig1}(b)
and~\ref{fig4}, which lead to the leptonic corrections, exact in order
\oa.
This is much more involved than the LLA calculation alone
 and has been done
so far for a rather limited set of variables only.
A comprehensive presentation of the calculation and of the numerical
details may be found in~\cite{MI} and in references quoted therein.
Here
we restrict ourselves to a short collection of formulae, which
are realized in \he.

Let us remind the Born cross section as given in~(\ref{bf12}):
\ba
\frac {d^2 \sigma^{\mr B}} {dy dQ^2} =
        \frac{2 \pi \alpha^{2}}{{\lambda_S}} \frac{S}{Q^4}\;
        \sum_{i=1}^3 {\cal A}_i(x,Q^2) {\cal S}_{i}^{\mr B} (y,Q^2),
\label{eqBorn}
\ea
with $\lambda_S=S^2-4m^2M^2$.
The factorized kinematical functions ${\cal S}_i^{\mr B}$ are
introduced in~(\ref{eqS1B})--(\ref{eqS3B}).
The functions ${\cal A}_i$ are related to the generalized
structure functions by:
\begin{eqnarray}
\begin{array}{rclcl}
{\cal A}_{1}(x,Q^2)
&=&2  {\cal F}_{1}(x,Q^2),
  \\
{\cal A}_{2}(x,Q^2)
&=&\frac{\displaystyle 1}{
\displaystyle
yS}
{\cal F}_{2}(x,Q^2),
\\
{\cal A}_{3}(x,Q^2)
&=&\frac{
\displaystyle
1}{
\displaystyle
2 Q^2}
{\cal F}_{3}(x,Q^2).
\end{array}
\label{deffi}
\end{eqnarray}

In the {\tt TERAD} part of \he, the QED corrections have a structure
quite similar to that of the Born cross section~(\ref{eqBorn}):
\begin{eqnarray}
\frac{d^2 \sigma^{QED}}{d{\cal E}} &=&
 {2 \alpha^{3}S^2 \over { \pi\lambda_S } } \int d^2{\cal I} \,
\frac{1} {Q_h^{4}} \,
 S({\cal E,I}),
\\
 S({\cal E,I})
&=&
\sum_{i=1}^3 {\cal A}_i(x,Q^2)\, {\cal S}_{i}({\cal E,I}).
\label{eq3.3}
\end{eqnarray}
Here,
${\cal E}$
is a set of two variables on which $d^2 \sigma$
finally depends. In the
case of leptonic, mixed, hadronic
variables, e.g.:\\
{\centerline{
${\cal E}= \{\yl,\ql \}, \{y_h,\ql \}, \{y_h,Q^2_h\}   $.
}}
The
${\cal I}$ is a corresponding set of two variables to be integrated
over: \\
{\centerline{
${\cal I}=\{y_h, Q_h^2 \}, \{\yl,Q^2_h\}, \{\yl,\ql \}     $.
}}

The functions $S_i({\cal E,I})$ in~(\ref{eq3.3}) mark the difference
to the Born cross section and have to be calculated.
They are results of an
 one-dimensional analytical integration.
In some of the variable sets additional integrations have been
performed analytically.

We only mention here that in the exact calculations the correct
treatment of the kinematically allowed regions of the integration
variables may be crucial.
A careful and exhaustive analysis of this may be found in appendix~B
of~\cite{MI}.

The cross sections may be represented in the following generic form:
\ba
\frac{d^{2}{\sigma}^{QED}}{d{\cal E}}
&=&
\left[1+\frac{\alpha}{\pi}\;\delta_{{VS}}^{expon}({\cal E})\right]
\frac{d^{2} {\sigma}^{B}}    {d{\cal E}}
+\frac{d^{2}{\sigma}^{brems}}    {d{\cal E}}.
\label{62b}
\ea
The term:
\ba
\frac{\alpha}{\pi} \;\delta_{{VS}}^{expon}({\cal E})
&=&
\exp
\left[\;\frac{\alpha}{\pi} \delta_{{inf}}({\cal E}) \right]
-1+\frac{\alpha}{\pi}
\Bigl[\delta_{{VR}} ({\cal E}) -\delta_{{inf}}({\cal E})\Bigr] \;
\label{611}
\ea
 contains the (exponentiated) virtual
and soft photon part of the corrections and $d^{2}{\sigma}^{brems}$
the finite part of real bremsstrahlung.
Exponentiation may be taken into account with flag {\tt IEXP}.

The Born cross section is called from subroutine {\tt sigbrn.f}, while the
non-factorizing part of the QED corrections directly uses file
{\tt genstf.f}.
The corrections may or may not include weak
virtual corrections depending on the
flag {\tt IWEAK} and QCD corrections
to the parton distributions depending on the flag {\tt ISCH}.
These two types of corrections are not yet
thoroughly
compatible in the present version
of \he\ as mentioned in section~\ref{pardis}.
%
\subsubsection{Leptonic variables
\label{MIlepv}
}
%

In {\em leptonic} variables one has:
\ba
\delta_{{VR}} ({\cal E}_l)
 &=& \delta_{{inf}}({\cal E}_l)
 -\frac{1}{2}\mbox{ln}^{2} {\left[
   \frac{1-\yl(1-\xl)}{1-\yl\xl}\right]}
 + {\mr {Li}}_{2}{\left[
   \frac{1-\yl}{(1-\yl\xl)[1-\yl(1-\xl)]}\right]}
\nll & &
 +~\frac{3}{2} \mbox{ln}\frac{\Ql}{m^2} - {\mr {Li}}_2 (1)-2 \; ,
\label{65}
\ea
where
\ba
\delta_{{inf}} ({\cal E}_l)=
\left( \mbox{ln}\frac{\Ql}{m^2}-1 \right)
     \mbox{ln} {\left[ \frac{y_l^2(1-\xl)^2}
     {(1- \yl\xl)[1-\yl(1-\xl)]}\right]} .
\label{66}
\ea
Here, ${\rm Li}_2(x)$ denotes the Euler dilogarithmic function
\ba
   {\rm Li}_2(x) = - \int_0^1 dz \frac{\ln(1 - xz)}{z}.
\label{diloga}
\ea
The finite bremsstrahlung correction has a factorizing and a
non-factorizing part:
\ba
\frac {d^2 \sigma^{brems}}{d \yl d \ql}
&=& \frac {2\alpha^3 S^2}{\lambda_S}
     \int d\yh d\qh  \sum_{i=1}^{3} \Biggl[ {\cal A}_{i}(\xh,\qh)
     \frac{1}{Q^4_h} {\cal S}_{i}(\yl,\ql,\yh,\qh)
\nll
& &-~{\cal A}_{i}(\xl,\ql)\frac{1}{Q_{l}^4} \;
     {\cal S}_{i}^{B}(\yl,\ql)\;
     {\cal L}^{\mr {IR}}(\yl,\ql,\yh,\Qh) \Biggr] ,
\label{*052}
\ea
The radiators ${\cal S}_i({\cal E,\cal I})$ of the non-factorizing
part are:
\begin{eqnarray}
\label{eq313}
{\cal S}_{1}( {\cal{E, I}})
&=& \Biggl\{ \frac{1}{\sqrt{C_{2}}}
\label{eq314}
\left[2 m^2 - \frac{1}{2}
\left( \Qh+\Ql \right) + \frac{Q_h^4-4m^{4}}{\Qh-\Ql}\right]
\nll
& &-~m^{2}(\Qh-2m^2)\frac{B_{2}}{C_{2}^{3/2}} \Biggr\}
 +\frac{1}{\sqrt{A_2}}
 - \Biggl\{  (1)  \leftrightarrow - (1-\yl)  \Biggr\},
\\
{\cal S}_{2}( {\cal{E, I}})
&=& \Biggl\{\frac{1}{\sqrt{C_{2}}} \Bigl[ M^{2}(\Qh + \Ql)-
  \yh (1-\yl) S^2 \Bigr]
  \nll
& & +~
\frac{1}{(\Qh-\Ql)\sqrt{C_{2}}}
 \Biggl[ \Qh \Bigl[ (1) \left[ (1) - \yh \right] S^2  \nll
& & +~ (1-\yl) \left[ (1-\yl) + \yh \right] S^2
 - 2 M^{2}(\Qh+2m^2) \Bigr]   \nll
& & +~2m^2 S^2 \Bigl[ \left[ (1)-\yh \right]
    \left[(1-\yl) + \yh \right]
 +   (1)(1-\yl) \Bigr]  \Biggr]
\nll
& &-~2m^{2} \frac{B_{2}}{C_{2}^{3/2}}
     \Bigl[    (1) \left[ (1) - \yh \right] S^2 - M^{2}\Qh \Bigr]
\Biggr\}
- \frac{2M^2}{\sqrt{A_2}}
\nll
& &
-~\Biggl\{ (1) \leftrightarrow - (1-\yl)  \Biggr\},
\\
{\cal S}_{3}({\cal{E, I}})
&=&  \Biggl\{\frac{S}{\sqrt{C_{2}}}
\Biggl[\frac{2  \Qh ( \Qh+2m^2)
\left[ (1) + (1-\yl)  \right] }
{\Qh-\Ql}
   - 2 (1-\yl) \Qh -  \yh(\Qh+\Ql)\Biggr] \nonumber \\
& &-~ 2m^2 S \Qh\frac{B_{2}}{C_{2}^{3/2}}
   \left[ 2 (1) - \yh \right] \Biggr\}
+ ~\Biggl\{ (1) \leftrightarrow - (1-\yl) \Biggr\},
\label{eq316}
\end{eqnarray}
with the following abbreviations:
\begin{eqnarray}
A_2 &=& \yl^2S^2+4M^2\Ql,
\nonumber       \\
B_2 &=&\left\{ 2 M^2 \Ql ( \Ql - \Qh )
      +     (1-\yl) (\yl \Qh - \yh \Ql) S^2
 +S^2 (1) \Ql (\yl - \yh) \right\}
\nll
&\equiv&~
-~B_1
        \left\{ (1) \leftrightarrow - (1-\yl) \right\}, \nonumber  \\
C_2 &=&  \left\{  (1-\yl) \Qh - \Ql \left[ (1)- \yh \right] \Bigr]^2
S^2
     + 4m^2 \Bigl[ (\yl-\yh)(\yl \Qh-\yh \Ql) S^2
- M^2(Q^2_h - Q^2_l )^2  \right\}
\nll
   &\equiv&~
C_1 \left\{  (1) \leftrightarrow - (1-y_l) \right\}.
\nonumber
\end{eqnarray}
The expression for the correction factorizing off the
Born term, ${\cal
  L}^{\mr{IR}}(\yl,\ql,\yh,\Qh)$, reads
(exact in both masses $m$ and $M$):
\ba
{\cal L}^{{IR}}(\yl,\ql,\yh,\Qh)
&=&
\frac{\Ql+2m^2}{\Ql-\Qh} \Bigl(\ \frac{1}{\sqrt{C_1}}
   -\frac{1}{\sqrt{C_2}} \Bigr)\
   -m^{2}\Bigl(\ \frac{B_{1}}{C_{1}^{3/2}}
                +\frac{B_{2}}{C_{2}^{3/2}} \Bigr).
\label{eqn052a}
\ea
%
The boundaries for the two-dimensional numerical integration
in~(\ref{*052}) are derived in appendix B.2.3 of~\cite{MI}.

The above formulae are realized in file {\tt teradl.f}.
%
\subsubsection{Mixed variables
\label{MImixv}
}
%
In {\em mixed} variables, the following functions have to be used
in~(\ref{611}):
\ba
\delta_{{VR}}({\cal E}_m)
&=& \delta_{{inf}}({\cal E}_m)
- \frac{1}{2}
 \mbox{ln}^{2}\left(\frac{1-\yh}{1-\xm}\right)
-~{\mr {Li}}_{2} {\left[ \frac {\xm (1-\yh)}{\xm - 1} \right]}
+\frac{3}{2} \mbox{ln}\frac{\Ql}{m^2} -2 \;,
\label{67}
\ea
where
\ba
\delta_{{inf}}({\cal E}_m)=
\left( \mbox{ln}\frac{\Ql}{m^2} - 1 \right)
 \mbox{ln}{\left[ ( 1 - \yh ) ( 1 - \xm )\right]}.
\label{68}
\ea
The final formula for $ d^{2}\sigma^{brems}/d\yh d\Ql $
can be written in the form:
\ba
\frac {d^2 \sigma^{brems}} {d\yh d\ql}
&=&\frac{2 \alpha^{3} S^2}{\lambda_S}
   \Biggl\{\int_{Q_h^{2\min}}^{Q_l^2}
    d\Qh \sum_{i=1}^3 \Bigl[
   {\cal A}_i(\xh,\Qh)\frac{1}{Q^4_h}
    {\cal S}_{i}^{\mr{I}} (\yl,\Ql,\yh,\Qh)
\nll & &
-~{\cal A}_i(x_m,\ql)\frac{1}{Q^4_l}
    {\cal S}_{i}^{\mr B} (\yh,\Ql)
    {\cal L}_{\mr I}^{\mr {IR}}(\yl,\ql,\qh)\Bigr]
\nll
& &+~\int_{Q_l^2}^{y_h S}
d\Qh\sum_{i=1}^3 \Bigl[\
    {\cal A}_i(\xh,\Qh)\frac{1}{Q^4_h}
    {\cal S}_{i}^{\mr{II}}(\Ql,\yh,\Qh)
\nll & &
-~{\cal A}_i(x_m,\Ql) \frac{1}{Q^4_l}
    {\cal S}_{i}^{\mr B} (\Ql,\yh)
    {\cal L}_{\mr{II}}^{\mr {IR}}(\Ql,\yh,\Qh) \Bigr]\
   \Biggr\}.
\label{eqn515}
\ea
The integration boundary $Q_h^{2\min}$ for the remaining
one-dimensional integration may be found in appendix B.3.3 of \cite{MI}.
The integration region must be split into
two regions~$I$ and~$II$ as long as the condition $\ql < \yh S$ is
fulfilled, which corresponds to $\xm \leq 1$.
If instead $\xm > 1$, there is only the region~$I$.
The factorizing part of the cross section differs in the two
kinematical regions:
\ba
{\cal L}^{{IR}}_{I}(\Ql,\yh,\Qh)
&=&\frac {1}{S(\Ql-\Qh)}
   \left[ -1 -\frac{\Ql}{\Qh}({\mr L}_t+{\mr L}_2)
   +({\mr L}  +{\mr L}_t+{\mr L}_1)\right] ,
\\
{\cal L}^{{IR}}_{{II}}(\Ql,\yh,\Qh)
&=&\frac {1}{S(\Ql-\Qh)}
   \left[\ \frac{\Ql}{\Qh}+{\mr L}_1-\frac{\Ql}{\Qh}
   ({\mr L} +{\mr L}_2)\right]\ .
\label{eqn513}
\ea
The following abbreviations are used:
\ba
\nll
{\mr L}    =\mbox{ ln}\frac{\Ql}{m^2}, \hspace{.7cm}
{\mr L}_{t}=\mbox{ ln}\frac{\Ql}{\Qh},\hspace{.7cm}
{\mr L}_{T}=\mbox{ ln}\frac{1 }{\yh},
\label{eqnLOG}
\ea
\ba
\nll
{\mr L}_{1}=\mbox{ln}\frac{\Qh-\yh \ql}{ \mid \Ql-\Qh \mid}
,\hspace{.7cm}
{\mr L}_{2}=\mbox{ln}\frac{(1- \yh)\qh}{\mid \Ql-\Qh \mid} .
\label{eqn514}
\ea
The non-factorizing hard bremsstrahlung functions in the two
integration regions are:
\ba
{\cal S}_{1}^{I}(\Ql,\yh,\Qh)
&=&\frac{1}{S}\Biggl[\frac{1}{2}\frac {\Ql}{\Ql-\Qh}
\left( 1+\frac{ Q_h^4}{ Q_l^4} \right)
       \left( {\mr L}  +{\mr L}_1-{\mr L}_2-1 \right)
\nll
 & &-~\frac{1}{2}\frac{\Ql}{\Qh} \left( 1+ \frac{ Q_h^4}{ Q_l^4}
\right)
       \left( {\mr L}_t+{\mr L}_2 \right) +{\mr L}_{T}-{\mr L}_t
     +\frac{1}{2} \left( 1-\frac{\Qh}{\Ql} \right) \Biggr] ,
\label{m57}
\\
{\cal S}_{2}^{I}(\Ql,\yh,\Qh)
&=&
S \Biggl\{
\frac{1}{\Ql}\frac{\Qh-\yh\Ql}{\Ql-\Qh}
     \left( 1+\frac{ Q_h^4}{ Q_l^4} \right)
    \left( {\mr L}  +{\mr L}_1-{\mr L}_2-1 \right)
\nll
& &+~\frac{1}{\Qh}
\left( 1+\frac{ Q_h^4}{ Q_l^4} \right) \left[\yh(1+\frac{\Ql}{\Qh})
   -
\left( 1+\frac{\Ql}{\Qh} + \frac{\Qh}{\Ql} \right) \right]
     \left( {\mr L}_t+{\mr L}_2 \right)
\nll
& &+~\frac {1}{\Qh}
\left(1+\frac{3}{2}\frac{\Qh}{\Ql}+3\frac{Q_h^4}{Q_l^4}
   +  2\frac{Q_h^6}{Q_l^6} \right)
\nll
& &-~ \frac{\yh}{\Qh}\left(
   \frac{\Ql}{\Qh }+2+4\frac{\Qh}{\Ql}+3\frac{Q_h^4}{Q_l^4}
     \right)
+~\frac{1}{2} \frac{ y_h^2}{\Qh} \left(
   2+\frac{\Ql}{\Qh}+2\frac{\Qh}{\Ql} \right)
\Biggr\},
\\
{\cal S}_{3}^{I}(\Ql,\yh,\Qh)
&=& \frac {\Ql}{\Ql-\Qh}\left( 1+\frac{ Q_h^4}{ Q_l^4}\right)
     \left( 2\frac{\Qh}{\Ql}-\yh \right)
   \left( {\mr L}   +{\mr L}_1 - {\mr L}_2-1 \right)
\nll
& &-~ \frac{\Ql}{\Qh} \left( 1+\frac{ Q_h^4}{ Q_l^4} \right)
\left( 2\frac{\Qh}{\Ql} -\yh +2   \right)
     \left( {\mr L}_t+{\mr L}_2 \right)
\nll
& &+~\frac{\ql}{\Qh}{\left(1+\frac{\Qh}{\Ql}\right)}^2 \left(
    2\frac{\Qh}{\Ql}-\yh \right)
   -\yh\left( \frac{\Ql}{\Qh}+1+2\frac{\Qh}{\Ql} \right)  ,
\label{eqn056}
\ea
\ba
{\cal S}_{1}^{{II}}(\Ql,\yh,\Qh)
&=&\frac{1}{S}\Biggl[-\frac{1}{2}\frac {Q_l^4}{\Qh(\Ql-\Qh)}
\left(1+\frac{ Q_h^4}{ Q_l^4}\right)
\left( {\mr L}  -{\mr L}_1+{\mr L}_2-1 \right)
\nll
 & &-~\frac{1}{2}\frac{\Ql}{\Qh}\left(1+ \frac{ Q_h^4}{ Q_l^4}\right)
        {\mr L}_1 +   {\mr L}_{T}
     +\frac{1}{2}\left(1-\frac{\Ql}{\Qh}\right) \Biggr] ,
\\
{\cal S}_{2}^{{II}}(\Ql,\yh,\Qh)
&=&
S\Biggl\{
-\frac{Q_l^4}{Q_h^4} \frac{(1-\yh)}{\Ql-\Qh}
     \left(1+\frac{ Q_h^4}{ Q_l^4}\right)
    \left({\mr L}  -{\mr L}_1+{\mr L}_2-1\right)
\nll
& &+~ \frac{1}{\Qh}\left(1+\frac{ Q_h^4}{ Q_l^4}\right)\left[\yh
\left(1+\frac{\Ql}{\Qh}\right)
   -\left(1+\frac{\Ql}{\Qh} + \frac{\Qh}{\Ql}\right) \right]
       {\mr L}_1
\nll 
& &+~ \frac {1}{\Ql}
\left(\frac{3}{2}+\frac{\Qh}{\Ql}+3\frac{\Ql}{\Qh}
   +  2\frac{Q_l^4}{Q_h^4} \right)
\nll
& &-~  \frac{\yh}{\Qh}\left(
   \frac{3\Ql}{\Qh }+4+2\frac{\Qh}{\Ql}+\frac{Q_h^4}{Q_l^4}
     \right)
\nll
& &+~ \frac{1}{2} \frac{ y_h^2}{\Qh} \left(
   2+2\frac{\Ql}{\Qh}+\frac{\Qh}{\Ql} \right)
\Biggr\},
\\
{\cal S}_{3}^{{II}}(\Ql,\yh,\Qh)
&=& -~\frac {(2-\yh)}{(\Ql-\Qh)}\frac{Q_l^4}{\Qh}\left(1+\frac{
  Q_h^4}{ Q_l^4\
}
\right)
    \left({\mr L}   -{\mr L}_1 + {\mr L}_2-1\right)
\nll
& &+~  \left(1+\frac{ Q_h^4}{ Q_l^4}\right)\left[
     \yh\frac{\Ql}{\Qh} -2\left(1+\frac{\Ql}{\Qh}\right) \right]
       {\mr L}_1
\nll
& & +~2(1-\yh)\frac{\Ql}{\Qh}{\left(1+\frac{\Qh}{\Ql}\right)}^2
     +\yh\left( 1-  \frac{\Ql}{\Qh} \right) .
\label{eqn511}
\ea

The above formulae are realized in file {\tt teradm.f}.
%
\subsubsection{Hadronic variables
\label{MIhadv}
}
%
There exists a closed expression for
the complete leptonic QED
corrections in hadronic variables
to order \oa\ with soft-photon exponentiation:
\ba
\frac{d^2 \sigma^{QED}}{d \yh d \qh}
=
\frac{d^2 \sigma^{B}}{d \yh d \qh }
\exp \left[ \frac{\alpha}{\pi} \delta_{{inf}} (\yh,\qh) \right]
+ \frac{2 \alpha^3}{S} \sum_{i=1}^3 \frac{1}{Q_h^4}
{\cal A}_i(\xh,\qh) {\cal S}_i(\yh,\qh).
\nonumber
\label{d2wh}
\ea
The soft photon corrections are:
\ba
\delta_{ {inf}} (\yh,\qh)
=
\left( \ln\frac{\qh}{m^2}   - 1 \right) \ln(1-\yh).
\label{dinwh}
\ea
The hard bremsstrahlung corrections are taken into account to order
\oa:
\ba
{\cal S}_1(\yh,\qh)
&=&
\qh
\Biggl[
\frac{1}{4}\ln^2\yh -\frac{1}{2}\ln^2(1-\yh) -\ln\yh\ln(1-\yh)
-\frac{3}{2}\litwo(\yh) +\frac{1}{2}\litwo(1)
\nll & &
- \frac{1}{2} \ln \yh \lh + \frac{1}{4} \left( 1 + \frac{2}{\yh}
\right)
\lhone
+\left(1- \frac{1}{\yh} \right) \ln \yh
   - \left(1+\frac{1}{4\yh}\right)
\Biggr],
\nll
\label{w1h}
\nll
{\cal S}_2(\yh,\qh)
&=&
S^2
\Biggl[
-\frac{1}{2}\yh\ln^2\yh -(1-\yh)\ln^2(1-\yh) -2(1-\yh)\ln\yh\ln(1-\yh)
-\yh\litwo(1)
\nll & &
-~(2-3\yh)\litwo(\yh)
+ \yh  \ln \yh \lh
+ \frac{\yh}{2} (1-\yh) \lhone
-\frac{\yh}{2} (2-\yh) \ln \yh
\Biggr],
\nll
\label{w2h}
\nll
{\cal S}_3(\yh,\qh)
&=&
S \qh
\Biggl\{
-(2-\yh)\Biggl[
-\frac{1}{2}\ln^2\yh +\ln^2(1-\yh) +2\ln\yh\ln(1-\yh)
+3\litwo(\yh)
\nll & &
-~\litwo(1) +\ln\yh  +\ln\yh \lh
\Biggr]
+ \frac{3}{2} \yh \lhone
   + \yh - \frac{7}{2}
+ 2~(1-2\yh) \ln \yh
\Biggr\},
\nll
\label{w3h}
\ea
with $\lh = \ln (\qh/m^2)$ and $\lhone = \lh + \ln[\yh/(1-\yh)]$.

Since the structure functions depend on the hadronic variables and
are hardly to be integrated over together with radiative corrections,
such a compact result may be obtained  in
hadronic  variables only.

The above formulae are realized in files {\tt teradh.f} and {\tt
  terhin.f} using two different analytical expressions.

\subsubsection{Jaquet-Blondel variables
\label{MIjbv}
}
The net cross section is:
\ba
\frac {d^2 \sigma^{QED}}{d\yjb d\qjb}
&=&
\frac{d^{2} {\sigma}^{B}}    {d\yjb d\qjb}
\left\{
\exp \left[\frac{\alpha}{\pi}
 \delta_{{inf}}(\yjb,\qjb) \right]
 -1+\frac{\alpha}{\pi}\Bigl[
 \delta_{{VR}} (\yjb,\qjb)
-\delta_{{inf}}(\yjb,\qjb)\Bigr]
\right\}
\nll
& &+~
\frac{d^{2}{\sigma}^{brems}}{d\yjb d\qjb}.
\ea
The net factorizing part is:
\ba
\delta_{{VR}}(\yjb,\qjb)
&=&
\delta_{{inf}}(\qjb,\yjb)
-\mbox{ln}^2 \xjb
+\mbox{ln} \xjb        \left[\mbox{ln}(1-\yjb)-1\right]
\nll  \vspace{0.3cm}
& &-~\frac{1}{2}
 \mbox{ln}^2 \left[ \frac{(1 - \xjb)(1 - \yjb)}{\xjb\yjb} \right]
\nll
& &+~ \frac{3}{2}\ljb
-   {\mr{ Li}}_2  (\xjb       )
- 2 {\mr{ Li}}_2 ( 1-\xjb ) - 1 ,
\label{dvrj}
\ea
where
\bq
\delta_{{inf}}(\yjb,\qjb)=\ln\frac{(1-\xjb)(1-\yjb)}{1-\xjb(1-\yjb)}
\left(\ln\frac{\qjb}{m^2}-1\right).
\label{dinfj}
\eq
The non-factorizing part of the radiative cross section in Jaquet-Blondel
variables is:
\ba
\frac {d^2 \sigma^{brems}}{d\yjb d\qjb}
&=& \frac {2\alpha^3}{S}\int d\tau\sum_{i=1}^{3} \Biggl[ {\cal
  A}_{i}(\xh,\qh\
)
\frac{1}{Q^4_h} {\cal S}_{i}(\yjb,\qjb,\tau)
\nll
& &-~ {\cal A}_{i}(\xjb,\qjb)\frac{1}{ \qjbf  } \;
     {\cal S}_{i}^{B}(\yjb,\qjb)\;
     {\cal F}^{\mr {IR}}(\yjb,\qjb,\tau) \Biggr] ,
\label{fjb}
\ea
with
\ba
{\cal S}_1(\qjb,\yjb,\tau) & = &
    \qjb
    \left[ \frac{1}{z_2} \left(\LQZ -2\right)
 +  \frac{z_2 }{4 \tau^2} \right]
 +  \frac{ 1 - 8 \yjb }{4(1-\yjb)}  
 +  \LQZ \left( \frac{z_2}{2 \Qt }+\frac{\yjb}{1-\yjb}\right),
\label{S1}
\\
{\cal S}_2(\qjb,\yjb,\tau) & = &
      S^2 \Biggl\{  2(1-\yjb)
    \left[ \frac{1}{z_2} \left( \LQZ - 2 \right)
  + \frac{z_2}{4\tau^2} \right]   
  - \frac{1}{\Qt} \left[ 2( \LQZ -2)
  +  \frac{1}{2}(1-\yjbs ) \right] \nll
 & &+~\frac{\qjb} {\mbox{$Q^4_{\mr  \tau} $}} (1-\yjb)
    \left[ 1 -  (1+\yjb)(\LQZ -3) \right]
  - \frac{\qjbf    }
         {\mbox{$Q^6_{\mr \tau}$}} (1-\yjb)^2 (\LQZ-3)
                     \Biggr\},
\nll
\label{S2}
\\
{\cal S}_3(\qjb,\yjb,\tau) & = &
      S \Biggl\{ 2 \qjb (2-\yjb)
      \left[ \frac{1}{z_2} (\LQZ - 2)
    + \frac{z_2}{4\tau^2} \right]             
 +  \frac{\yjb( 1 + \yjbs )} { 1-\yjb } \LQZ +5   \nll
& &-~ \frac{ 7 \yjb }{2(1-\yjb)}-(1-\yjb)(5+2\yjb)  \nll
& &+~ \frac{\qjb}{\Qt} (1-\yjb) \left[(1-\yjb)
                              \left( 3-\frac{2 \qjb }{ \Qt }\right)
  (\LQZ - 2)
  + 12-5 \LQZ \right]
                     \Biggr\}.
\label{S3}
\ea

The integral over
$\tau$ is performed in \he\ numerically since one cannot neglect
the difference between $\qjb$ and $Q_h^2$ --    the structure
functions
are depending on $Q_h^2$ and thus on $\tau$.

The above formulae are realized in file {\tt teradj.f}.
\subsubsection{Cuts
\label{teradlcut}
}
In the {\tt TERAD} approach it is possible to reject the
production of final states with a
$Q_h^2$ and/or an invariant hadronic mass $W_h^2$ below some
bound.
This is realized in \he\ for all the sets of kinematical variables.
See also the description of flag {\tt IHCU} in
section~\ref{smtc}.

\bigskip

The QED corrections in {\em leptonic} variables in the
{\tt TERAD} part of \he\ allow, in addition,  the application of cuts
on the photon kinematics.
Details of the formulation may be found in section 8.3 and appendix
B.1.2 of~\cite{MI}.
In \he\ this is done by the variable {\tt GCUT} which
is described in
section~\ref{smtc}.
\subsection
[{\tt DISEP}:
Complete \oa\ corrections in
the quark parton model approach
]
{{\tt DISEP}:
Complete \oa\ corrections in
the quark parton model approach
\label{secQPM}
}
Complete order \oa\ corrections in the quark-parton approach
are calculated in the {\tt DISEP} part of \he.
They include a large variety of contributions:
\begin{itemize}
\item[(i)]
Electroweak corrections, including the running QED coupling.
\item[(ii)]
Photonic corrections due to the
emission of photons from the leptons
together with the corresponding vertex corrections.
\item[(iii)]
Photonic corrections due to the
emission of photons from the quark lines
together with the corresponding vertex corrections.
Their  treatment deserves some care in view of the unknown
quark masses leading to logarithmic mass singularities in the
on-mass-shell scheme.
We refer for details to the
literature~\cite{RUJU,LLA1,KP,LLAmix,hsllaq}.
In the branches selected by
{\tt IDIS=1,2} these terms do not contribute.
\item[(iv)]
The interference of photon emission from leptons and quarks
together with the corresponding $\gamma \gamma, \gamma Z, \gamma W$
box corrections.
\item[(v)]
Exponentiation of soft photon corrections due to~(ii).
\end{itemize}
In the present version there are no QCD corrections taken into
account in the parton distributions in this part of \he.

The additional bremsstrahlung diagrams for $\cal NC$ scattering are shown in
figure~\ref{fig6}.
The interference of them among themselves is combined with vertex
corrections from the analogue of figure~\ref{fig1}(b), and the
interference of them with the leptonic radiation (figure~\ref{fig4})
has to be combined with the $\gamma Z$ and $\gamma \gamma$ boxes in
the $\cal NC$ case and of $\gamma W$ boxes in the $\cal CC$ case.

The $\cal CC$ process has some specific features.
The leptonic vertex correction comes from a diagram, where the virtual
photon connects the incoming electron with the virtual $W$ boson; the
final state neutrino has no photonic interactions.
The virtual photonic corrections from the quark side arise analogously.
Furthermore, there are contributions from a Feynman diagram with real
photon emission from the virtual $W$ boson.
All these contributions are taken into account in \he.
Some details of gauge invariance and other peculiarities may be found
in~\cite{zfpc44}.

\begin{figure}[t]
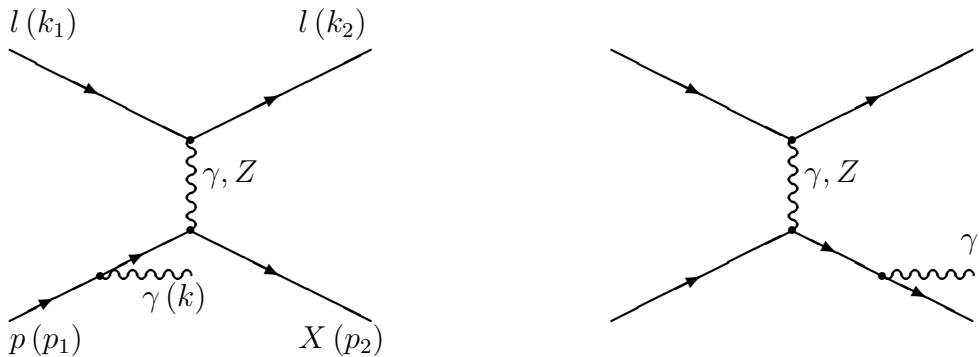

\vspace{0.9cm}
\begin{minipage}[t]{16cm}{
\begin{center}
\begin{Feynman}{160,65}{0,0}{0.8}
\put(00,55){\fermiondrr}
\put(00,58){$l \, (k_1)$}
\put(48,58){$l \, (k_2)$}
\put(32,33){$\gamma, Z $}
\put(22,12,5){$\gamma \, (k)$}
\put(60,55){\fermionurr}
\put(30,40){\circle*{1,5}}
\put(30,25){\circle*{1,5}}
\put(30,25){\fermiondrr}
\put(30,25){\fermionurrhalf}
\put(15,17.5){\fermionurrhalf}
\put(48,5.5){$X \, (p_2)$}
\put(00,5.5){$p \, (p_1)$}
\put(30,25){\photonuphalf}
\put(15,17.5){\photonrighthalf}
\put(15,17.5){\circle*{1,5}}
\put(100,55){\fermiondrr}
\put(132,33){$\gamma, Z$}
\put(160,55){\fermionurr}
\put(130,40){\circle*{1,5}}
\put(130,25){\circle*{1,5}}
\put(130,25){\photonuphalf}
      \put(145,17.5){\photonrighthalf}
\put(158,22){$\gamma$}
\put(145,17.5){\circle*{1,5}}
\put(145,17.5){\fermiondrrhalf}
\put(130,25){\fermiondrrhalf}
\put(130,25){\fermionurr}
\end{Feynman}
\end{center}
}\end{minipage}
\caption{\it
QED Bremsstrahlung off  quarks
\label{fig6}
}
\vspace*{0.5cm}
\end{figure}

The electroweak corrections~(i) are taken into account by generalized
structure functions as is described in section~\ref{impsec}.

If identical structure functions are chosen,
the leptonic QED corrections~(ii) agree with those calculated in the
model independent approach of section~\ref{secMI} although the
underlying analytical expressions look quite differently.
There is no access in the model independent approach to the
corrections~(iii) and~(iv).

The leading logarithmic approximation reduces the complete \oa\
corrections to terms proportional to $\ln (Q^2/m_e^2)$ in~(ii), to
$\ln (Q^2/m_q^2)$ in~(iii). In~(iv) no mass singularities are
contained.

At present, in {\tt HECTOR} are the following branches with QPM based
complete calculations available:
\begin{itemize}
\item[(i)]
${\cal NC}$ and ${\cal CC}$ deep inelastic scattering in {\em leptonic}
variables;
\item[(ii)]
${\cal NC}$ deep inelastic scattering in {\em mixed} variables.
\end{itemize}

%
\subsubsection{$\cal NC$ scattering in leptonic
  variables
\label{QPMlepv}
}
%
The QPM based electroweak corrections to $\cal NC$ scattering in
leptonic variables are described in~\cite{zfpc42}.

The virtual corrections are a part of the generalized structure
functions in the improved Born approximation,
calculated in subroutine {\tt sigbrn.f}.
They are described in
section~\ref{impsec}\footnote{Here we deviate slightly from~\cite{zfpc42}.}.

In addition, the QED corrections have to be determined.
In the {\tt DISEP} part, they have the following structure:
\begin{equation}
\frac{d^2\sigma^{QED}}{d\xl d\yl} =
\frac{2\pi \alpha^2(\ql) S \xl}{Q_l^4}  \sum_{B} \sum_{b} \sum_{Q,{\bar Q}}
c_b {\cal K}(B,p) \left[ {\cal V}(B,p) R_b^V(B) + p {\cal A}(B,p)
R_b^A(B) \right],
\end{equation}
where the sum over $Q,\bar Q$ extends over quarks and anti-quarks and
$B=\{ \gamma, I, Z \} $.
The sum over $b$ covers bremsstrahlung from leptons,
lepton-quark interference, and
quarks ($b= \{ e,i,q \} )$.

The parton distributions are accessed via the subroutine {\tt prodis.f}
directly.
The explicit use of QCD corrections to the parton densities
is not
foreseen here.

The parameter $p$ and the
couplings ${\cal K}(B,p), {\cal V}(B,p), {\cal
  A}(B,p)$ are defined in section~\ref{impsec}.
The coefficients
$c_b$ contain charges:
\ba
c_e &=& Q_e^2,
\\
c_i &=& Q_e Q_q,
\\
c_q &=& Q_q^2.
\label{cb}
\ea
The QED corrections are described by
the functions $R_b^{V,A}(B)$:
\begin{equation}
R_b^{V,A}(B) = R_b(B;1,1-\yl) + s_b^{V,A} R_b(B;\yl-1,-1),
\end{equation}
with
\ba
s_e^V = -s_e^A = s_i^A = -s_i^V = s_q^V = -s_q^A = 1.
\label{sbB}
\ea
The  factors $R_b(B;1,1-y)$ have the following
structure:
\ba
R_c(B;a,b) &=& \frac{\alpha}{\pi}
\Biggl\{ S_c(B;a,b) f_q(\xl,\ql )
+\int_{\xl}^1 d x_h  \Biggl[
\frac{\xl}{x_h} f_q(x_h,\ql) U_c(B;a,b)
\nll &&
+~\frac {r_c x_h f_q(x_h,\ql) - \xl  f_q(\xl,\ql)}
{x_h - \xl}   T_c(B;a,b)
\Biggr] \Biggr\} ,
\label{stu}
\ea
with
\ba
r_c = 1/\xh,\,  1,\, 1 \hspace{1cm} \mathrm{for}\,\, c=e,i,q.
\ea
The $S,T,U$ are results of a two-fold
analytical integrations over photonic angles.
They are the analogues to the kinematical functions ${\cal S}_i$ and
${\cal L}^{\mathrm{IR}}$ shown
in~(\ref{eq314}) to~(\ref{eqn052a}).
Because there are many of them, we refer for explicit expressions to
the original literature.
The six independent functions are:

\bigskip

$R_e(\gamma;a,b), \,\, R_e(I;a,b),\,\,  R_e(Z;a,b),\,\,
R_i(\gamma;a,b),\,\,  R_i(Z;a,b),\,\,  R_q(\gamma;a,b)$.

\bigskip

They are composed in the
subroutines
{\tt SGGiL, SGZiL, SZZiL}, {\tt i}={\tt L,I,Q}
of \he\,
 and the constituents
$S, T, U$ are calculated in functions {\tt FGGiL, FGZiL, FZZiL}, {\tt
  i}={\tt L,I,Q}.
The others are expressed by the following identities:
\ba
 R_i(I;a,b) &=& \frac{1}{2} \left[R_i(\gamma;a,b) + R_i(Z;a,b)\right],
\\
R_q(\gamma;a,b) &=& R_q(I;a,b) = R_q(Z;a,b).
\label{Reiq}
\ea

At the end of this section we should mention that the virtual
corrections, by which the QED corrections are accompanied via the weak
and electromagnetic couplings,
 may also be calculated with the
use of the
actual kinematical variables, i.e. folding them together with the
functions $T$ and $U$.
This may be achieved by chosing {\tt ICONV}=1. The resulting computation
is very time
consuming and  leads to
numerical values which differ only insignificantly from those obtained
without using this option.

In the QPM approach in leptonic
variables the virtuality of the parton densities was choosen by
the outer variable $Q_l^2$.
For hard photon emission, one actually would have
to perform integrations over
weighted structure functions at reduced kinematical variables.
The structure functions depend on the hadron kinematics, i.e. on $x_h$
and $Q_h^2$. In the {\tt DISEP} approach
$Q_h^2$ is to be integrated over analytically,
which leads, however, to the problem that the quark distribution
functions depend on some $Q^2$, but not on the correct (hadronic)
 $Q_h^2$.
The  same procedure is followed in the LLA calculations.
In this sense, the {\tt DISEP}
 approach has a reduced accuracy in leptonic
variables compared to the model independent approach.
For a high precision calculation it may be useful to calculate the
largest corrections~(ii) from {\tt TERAD} and the remaining terms
from {\tt DISEP}.
\subsubsection{$\cal CC$ scattering in leptonic
  variables
\label{QPMlepc}
}
%
The QPM based electroweak corrections in \he\ to $\cal CC$ scattering in
leptonic variables are described in~\cite{zfpc44}.
As in the case of $\cal NC$ scattering, the improved $\cal CC$ Born
cross section is determined in subroutine {\tt sigbrn.f}.

The QED corrections have access to the parton distributions not
through the generalized structure functions of file  {\tt
  gccstf.f} but instead directly through {\tt prodis.f}.
The QED corrections are:
\begin{equation}
\frac{d^2\sigma^{QED}}{d\xl d\yl} =
\frac{G_{\mu}^2 S\xl}{\pi} \left [\frac{M_W^2}{\ql+ M_W^2} \right]^2
\frac{1+\lambda}{2}
\sum_{b}
\sum_{Q,{\bar Q}} \theta(-Q_lQ_q) c_b \rho_c^2(p)
\left[\frac{1+p}{2}  R_b + \frac{1-p}{2}  {\bar R}_b \right] .
\end{equation}
The notations are those being
introduced in sections~\ref{impsec} and~\ref{QPMlepv}.
The generic definition of the
functions $R_b$ is~(\ref{stu}).
They are calculated in subroutines
{\tt SWWL, SWWI, {\rm and} SWWQ}
of \he.
Explicit expressions may be found in appendix~B of~\cite{zfpc44}.

As was  mentioned already in the case of $\cal NC$ scattering,
the QED corrections in leptonic variables in the {\tt DISEP} approach are
calculated with parton distributions having a $Q^2$ dependence on
$\ql$ instead of the correct variable $\qh$.

%
\subsubsection{$\cal NC$ scattering
  in mixed   variables
\label{QPMmixv}
}
%
%
The improved Born cross section is as described in
sections~\ref{borsec} and~\ref{impsec}.

The QPM based QED corrections to $\cal NC$ scattering in mixed
variables have been described in~\cite{LLAmix}.
The structure of the corrections is the following:
\ba
\frac{d^2 \sigma^{QED}}{d\xm d\ym}
&=&
\frac{2\alpha^3 S \xm}{\qm^4}
\, \, \sum_{Q,\bar Q}
\Biggl\{
B_0(\ym,1)  f_q(\xm,\qm)
\sum_{a=e,i,q} c_a S_a(\xm,\ym|m_a^2)
\nll &&+~
 \sum_{a=e,i,q}
\Biggl[
\int_1^{1/\xm}dz \left[{\bar B}_a^V(z;\xm,\ym)+  p {\bar
  B}_a^A(z;\xm,\ym)\right]
\nll &&
+~
\int_{\ym}^1 dz \left[B_a^V(z;\xm,\ym)+  p B_a^A(z;\xm,\ym)\right]
 \Biggr]
\Biggr\}
+
\frac{d^2 \sigma^{box}}{d\xm d\ym},
\label{e4}
\ea
where
\ba
B_0(\ym,z) &=&  V_0 \, Y_{+}\left(\frac{\ym}{z}\right)
                     + p A_0 Y_{+}\left(\frac{\ym}{z}\right).
\label{byz}
\ea
Much of the notations can be found in the previous sections and
section~\ref{impsec}.
In addition, we use for the vector couplings the following modifications:
\ba
V_0&=&V(1,1),
\nll
V(a,b)&=& \sum_{B=\gamma,I,Z} {K}(a,b;B,p)\, {\cal V}(B,p),
\\
K(a,b;B,p)&=&\chi_{B_1}(a) \, \chi_{B_2}(b) \, {\cal K}(B,p),
\nll
\chi_B(a)&=& {G_\mu \over\sqrt{2}}{M_{Z}^{2} \over{8\pi\alpha(Q^2)}}
    \frac{\qm}{a\qm + M_B^2}.
\nonumber
\ea
The axial couplings $A(a,b)$ are related to the couplings ${\cal
  A}(B,p)$
of~(\ref{va}) in the same manner as is explained above for the vector
couplings.

The functions $S_e,S_q$ contain factorizing soft photon corrections and
the corresponding vertex corrections.

The hard bremsstrahlung functions in~(\ref{e4}) are:
\ba
{\stackrel {(-)}{B}}
_a^{V,A}(z;\xm,\ym)
&=&
\left\{ \begin{array}{c}V_a\\A_a\end{array}\right\}
\left\{
\sum_n Reg\left[ {\stackrel{(-)}{F}}_{an}^{V,A}(z,\ym) f_Q(z\xm)\right]
{\stackrel {(-)}{L}}_{an}  +
{\stackrel {(-)}{U}}_a^{V,A}(z,\ym)f_Q(z\xm)
\right\} ,
\nll
\label{e11}
\ea
with the modifications of weak neutral coupling factors; for the
vector couplings:
\ba
V_e&=&c_e \, V(z,z), \hspace{0.7cm} V_i=c_i \, V(1,z), \hspace{0.7cm}
V_q=c_q \, V(1,1).
\label{e13}
\ea
The bremsstrahlung axial couplings are defined analogously.
For $a=e,i,q$,  index $n$ in~(\ref{e11}) runs from 1 to $2,1,2$.
The functions $F_{an}$ are regularized by a
subtraction:
\ba
Reg\left[ {\stackrel{(-)}{F}}_{an}^{V,A}(z,\ym) f_Q(z\xm)\right]
&=&
\frac{\ym}{(1-z)}
\left[
{ \stackrel{(-)}{F}}_{an}^{V,A}(z,\ym) f_Q(z\xm)
-
{\stackrel{(-)}{F}}_{an}^{V,A}(1,\ym) f_Q(\xm)
\right].
\nll
\label{e12}
\ea

Explicit expressions for the functions $S,F,L,U$ may be found
in~\cite{LLAmix}.
In \he\ they are realized in subroutines
{\tt SGGiM, SGZiM, SZZiM}, {\tt i}={\tt L,I,Q}.

\subsection{{\tt TERADLOW}: QED corrections at low $Q^2$}
In \he, the double differential cross section of the photoproduction
process is calculated in leptonic variables
with the aid of file {\tt terlow.f}.
For details of the derivation of formulae and of the kinematics, which
has to be treated exactly in both the electron and proton masses,
 we
refer to~\cite{MI}.

The process kinematics is described by $Q_l^2$ as defined
from the leptonic variables and the invariant mass $W$ of the
system consisting both
of the photon and the hadrons,
\ba
Q^2 = \ql &=& (k_1 - k_2)^2,
\nll
W^2 &=& -(Q_l + p_1)^2.
\label{w2}
\ea
For a fixed value of $W^2 \ll S$, the minimal value of $Q_l^2$
may become extremely small, but remains non-vanishing:
\ba
\ql^{\min}(W^2) \approx m^2 \, \frac{[W^2-M^2]^2}{S^2}
> 0.
\label{q2min}
\ea
This leaves the integrated deep inelastic cross section
finite.
The net cross section is:
\ba
\frac {d^2 \sigma_{\mr R}}{d \yl d \ql}
=
\frac {d^2 \sigma^{\mr {anom}}} {d\yl d\Ql} +
\frac{d^{2}{\sigma}_{\mr R}^{\mr F}}{d\yl d\ql}
+
\frac{d^{2} {\sigma}^{\mr B}}    {d\yl d\ql}
\left\{
{\mr {exp}} \left[\frac{\alpha}{\pi}
 \delta^{\mr {inf}}({\cal E}) \right]
 -1+\frac{\alpha}{\pi}\Bigl[
 \delta^{\mr {VR}} ({\cal E})
-\delta^{\mr {inf}}({\cal E})\Bigr]
\right\} .
\nll
\ea
The contribution from the anomalous magnetic moment
is:
\ba
 \frac {d^2 \sigma^{\mr {anom}}} {d\yl d\Ql} =
        \frac{2  \alpha^{3}S}{{\lambda_S Q_l^4}}
\frac{m^2}{\xl Q_l^2}
{\cal V}^{\mr{anom}} (\beta)
        \left[
        2\beta^2\yl^2\xl F_1(\xl,\ql)
   -(2-\yl)^2
           F_2 (\xl,\ql)   \right],
\label{68-2}
\ea
where
\ba
{\cal V}^{\mr{anom}} (\beta)
&=&
- \frac{{\mr L}_{\beta}}{\beta},
\label{anomm}
\\
   \Lb &=& \mbox{ ln } \frac{\beta +1}{\beta -1},
\\
   \beta&=&\sqrt{1+\frac{4m^2}{\Ql}}.
\label{k120v2}
\ea
The factorizing photonic corrections from bremsstrahlung and the
usual, exact vertex correction are:
\ba
\delta^{\mr {VR}} (\yl,\Ql )
 &=& \delta^{\mr {inf}}(\yl,\Ql)
+~\frac{1}{2 \beta_{1}} \;
  \mbox{ln}\frac{1+\beta_1}{1-\beta_1}
 +\frac{1}{2 \beta_{2}} \;
  \mbox{ln}\frac{1+\beta_2}{1-\beta_2}
+ {\cal S}_{\Phi}
\nll & &
  +~\frac{3}{2} {\beta}\Lb  -2
  - \frac{1+{\beta}^2}{2\beta}\left[\Lb
 \mbox{ln}{\frac{4{\beta}^2}{{\beta}^2-1}}
 + {\mr {Li}}_{2}\left(\frac{1+\beta}{1-\beta}\right)
 - {\mr {Li}}_{2}\left(\frac{1-\beta}{1+\beta}\right)\right] ,
\nll
\label{65-2}
\ea
with
\ba
\delta^{\mr {inf}} (\yl,\Ql)=
     2 \ln { \frac{ W^2-(M+m_{\pi})^2}{m\sqrt{W^2}}}
    \left(\frac{1+{\beta}^2}{2\beta}\Lb-1 \right) ,
\label{66-2}
\ea
and
\ba
\beta_1 &=& \sqrt{1-\frac{4m^2M^2}{(S-\Ql)^2}},
\\
\beta_2&=&\sqrt{1-\frac{4m^2M^2}{[S(1-\yl)+\Ql]^2}},
\label{67-2}
\ea

\ba
{\cal S}_{\Phi}
&=& \frac{1}{2} \left(Q^2+2m^2\right)
 \int_{0}^{1} \frac {d\alpha}{ \beta_{\alpha}(-k_{\alpha}^2)} \;
  \mbox{ln}{ \frac{1-\beta_{\alpha}}{1+\beta_{\alpha}} } .
\label{soft3}
\ea
\ba
k_{\alpha} = k_1 \alpha + k_2 (1- \alpha),
\label{kal7}
\ea

\ba
\beta_{\alpha} &=& \frac {\left|{\vec
    k}_{\alpha}\right|}{k_{\alpha}^0}.
\label{kalp2}
\ea

Finally, the infrared finite part of the real bremsstrahlung
can be written as follows:
\ba
\frac {d^2 \sigma_{\mr R}^{\mr F}}{d \yl d \ql}
&=& \frac {2\alpha^3 S }{\lambda_S}
     \int dM_h^2 d\qh  \sum_{i=1}^{3} \Biggl[ {\cal A}_{i}(\xh,\qh)
     \frac{1}{Q^4_h} {\cal S}_{i}(\yl,\ql,\yh,\qh)
\nll
& &-~{\cal A}_{i}(\xl,\ql)\frac{1}{Q_{l}^4} \;
     {\cal S}_{i}^{B}(\yl,\ql)\;
     {\cal L}^{\mr {IR}}(\yl,\ql,\yh,\Qh) \Biggr] ,
\label{*052p}
\ea
where $ {\cal S}_{i}(\yl,\ql,\yh,\Qh)\;(i=1,2,3)$ are given
by~(\ref{eq313})--(\ref{eq316}), and
$ {\cal L}^{\mr{IR}}(\yl,\ql,\yh,\Qh) $ is defined by~(\ref{eqn513}).
We would like to remind that these expressions are
exact in both masses $m$ and $M$ for the photon exchange contributions.
\subsubsection{Cuts
\label{teradcuts}
}
In {\tt TERADLOW} it is possible to reject the
production of final states with a
$Q_h^2$ and/or an invariant hadronic mass $W_h^2$ below     some
bound,~(cf.
the description of flag {\tt IHCU} in section~\ref{smtc}).
%

\subsection{Other reactions
\label{others}
}
In the LLA approach to leptonic
QED corrections the Born cross section is
folded with splitting functions at rescaled kinematic variables.
This factorization may be used also for exclusive reactions having
the same leptonic tensors as in the case of deep inelastic scattering.
Examples for this class of reactions are e.g.
heavy quark production,
deep inelastic scattering including $Z'$ exchange, and Higgs boson
production. These processes were studied previously in this
context~\cite{LLA, ZPRIM, HIGGS}.

Sometimes one is interested even in the LLA contributions either due to
initial {\it or} final state radiation {\it only}. For this case it is
sufficient that the subsystem (`Born') cross section has a charged
lepton line either in the initial or final state for which the LLA terms
shall be calculated.

In both these cases the respective corrections can be obtained setting
the flag {\tt IBRN}=0. The user has to supply a subroutine
{\tt usrbrn.f} describing the subsystem process.
\clearpage

\clearpage

\section{Structure of the Code
\label{sotc}
}
\ezero
%
%
%
%
The Fortran program \he\ consists of several subdirectories.
They are shown in figure~\ref{hecflow}.
Some of them contain several files, others only one.
Some of the files contain several subroutines and functions.
Exceptions, which are not shown in figure~\ref{hecflow}, are:
the subdirectory {\tt DIZET}, which is used for the
calculation of the electroweak radiative corrections
and the subdirectory {\tt HECAUX} with some auxiliary routines.
Besides the subdirectories, the two files {\tt HECTOR.INP} and {\tt
  HECTOR.OUT} are added in the flowchart.


The subdirectories are used for the following tasks:
\begin{tabbing}
\underline{{\tt CMAIN}}:~~~~\=
{}~~ Organization of the program flow for combined use of the
branches {\tt
  HELIOS} and \\ \> ~~
{\tt TERAD}
                                \\
\underline{{\tt DISEP}}: \>
 ~~ Package for calculation of
complete order \oa\ QED corrections with soft \\ \> ~~
photon exponentiation determined in the quark parton model approach
                                \\
\underline{{\tt DIZET}}: \>
 ~~ Package for the calculation of weak virtual corrections
                                \\
\underline{{\tt GENSTF}}: \>
 ~~ Calculation of the generalized structure functions for {\tt HELIOS},
{\tt TERAD}, and {\tt DISEP}
\\
\underline{{\tt GSFLOW}}: \>
 ~~ Calculation of generalized structure functions
for {\tt TERADLOW}
\\
\underline{{\tt HECAUX}}: \>
 ~~ Some auxiliary functions
\\
 \underline{{\tt HECFFR}}: \>
 ~~ Reads input data cards from file {\tt HECTOR.INP} using {\tt FFREAD}
\\
 \underline{{\tt HECOUT}}: \>
 ~~ Writes output to file {\tt HECTOR.OUT}, partly using {\tt FFREAD}
\\
\underline{{\tt HECSET}}: \>
 ~~ Setting of initialization parameters
\\
\underline{{\tt HECTOR}}: \>
 ~~ The main program
\\
\underline{{\tt HELIOS}}: \>
 ~~ Package for calculation of QED corrections in the leading
 logarithmic                     \\           \> ~~
 approximation, including higher order corrections and soft photon
 exponentiation
\\
 \underline{{\tt HMAIN}}:  \>
 ~~ Organization of the
program flow for branch {\tt HELIOS}
\\
 \underline{{\tt LOWLIB}}: \>
 ~~ Package for the optional modification of structure functions at low
 $Q^2$ values
\\
 \underline{{\tt PRODIS}}: \>
 ~~ Calculation of structure functions for electroweak corrections in all
 approaches   \\                             \> ~~
 and for QED corrections in package {\tt DISEP}
\\
 \underline{{\tt PDFLIB}}:\>
 ~~ Standard package of parton distribution functions
\\
 \underline{{\tt PDFACT}}:\>
 ~~ Default package of parton distribution functions
\\
 \underline{{\tt SIGBRN}}: \>
 ~~ Calculation of the (effective) Born cross section
\\
 \underline{{\tt STRUFC}}: \>
 ~~ Calculation of structure functions from parton distribution functions
\\
 \underline{{\tt TERAD}}:  \>
 ~~ Package for model independent calculation of
complete leptonic order \oa\ QED \\ \> ~~ corrections  with soft photon
exponentiation
\\
 \underline{{\tt TERADLOW}}: \>
 ~~ Package for model independent calculation of
complete leptonic order \oa\ QED \\  \> ~~
corrections  with soft photon
exponentiation intended for the region of extremely \\  \> ~~
low $Q^2$
\\
 \underline{{\tt TMAIN}}:   \>
 ~~ Organization of the
program flow for the
branch {\tt TERAD}, which may
call the QED                     \\    \> ~~
 corrections of the packages {\tt DISEP}, {\tt TERAD}, or
{\tt TERADLOW}
\end{tabbing}

The user supplied subdirectories are installed containing
 trivial versions
only:
\begin{tabbing}
\underline{{\tt USRINI}}:~~~~\=
{}~~ File for input modifications, which go
beyond the change of data cards foreseen
\\
\underline{{\tt USROUT}}: \> ~~ Modified output specifications
 \\
\underline{{\tt USRBRN}}: \> ~~ User supplied Born cross section
 \\
\underline{{\tt USRSTR}}: \> ~~ User supplied structure functions
 \\
\underline{{\tt USRPDF}}: \> ~~ User supplied parton distribution functions
\end{tabbing}

The two files in figure~\ref{hecflow} are:
\begin{tabbing}
\underline{{\tt HECTOR.INP}}: \= ~~ File with the input data card contents
\\
\underline{{\tt HECTOR.OUT}}: \> ~~ Standard output file
\end{tabbing}

\bigskip

%
%
Figure~\ref{genst} is devoted to the interplay of the subdirectories
{\tt GENSTF}, {\tt DIZET},  and {\tt LOWLIB}.

The files may be characterized as follows:
\begin{tabbing}
\underline{{\tt CCOUPL}}: ~~~~\=
 ~~ Calculation of charged current Born couplings or effective couplings
\\
\underline{{\tt DIZET}}: \> ~~ Standard Model Electroweak Library
\\
\underline{{\tt GCCSTF}}: \> ~~ Calculation of generalized structure
functions for charged current reactions
\\
\underline{{\tt GENSTF}}: \> ~~ Calculation of generalized structure
functions for neutral current reactions
\\
\underline{{\tt LOWLIB}}: \> ~~ Files with different structure
function modifications at low $Q^2$
\\
\underline{{\tt MIDSTF}}: \> ~~ Interpolation between the structure
function from {\tt GENSTF} and the low $Q^2$  \\ \> ~~
modified structure function
\\
\underline{{\tt NCOUPL}}: \> ~~ Calculation of neutral current Born
couplings or effective couplings
\\
\underline{{\tt POLYN}}:  \> ~~ Auxiliary function for {\tt MIDSTF}
\\
\underline{{\tt PRODIS}:}\> ~~ Parton distributions of the proton
(used by {\tt DISEP})
\\
\underline{{\tt STRFBS}}: \> ~~ File with the Brasse/Stein modification
of structure functions
\\
\underline{{\tt USRSTR}}: \> ~~ User supplied structure functions
\end{tabbing}


The entry labelled `3' connects the present figure with
figure~\ref{strufc}.


%
%

\bigskip

We show in figure~\ref{strufc} how the actual structure
functions are built from the parton distribution functions.
The subdirectories {\tt STRUFC}, {\tt PDFACT}, {\tt USRPDF}, {\tt
  PDFLIB} are shown as boxes. Their interaction is organized by the flags
{\tt ISTR}, {\tt ISCH}, {\tt ISSE}.

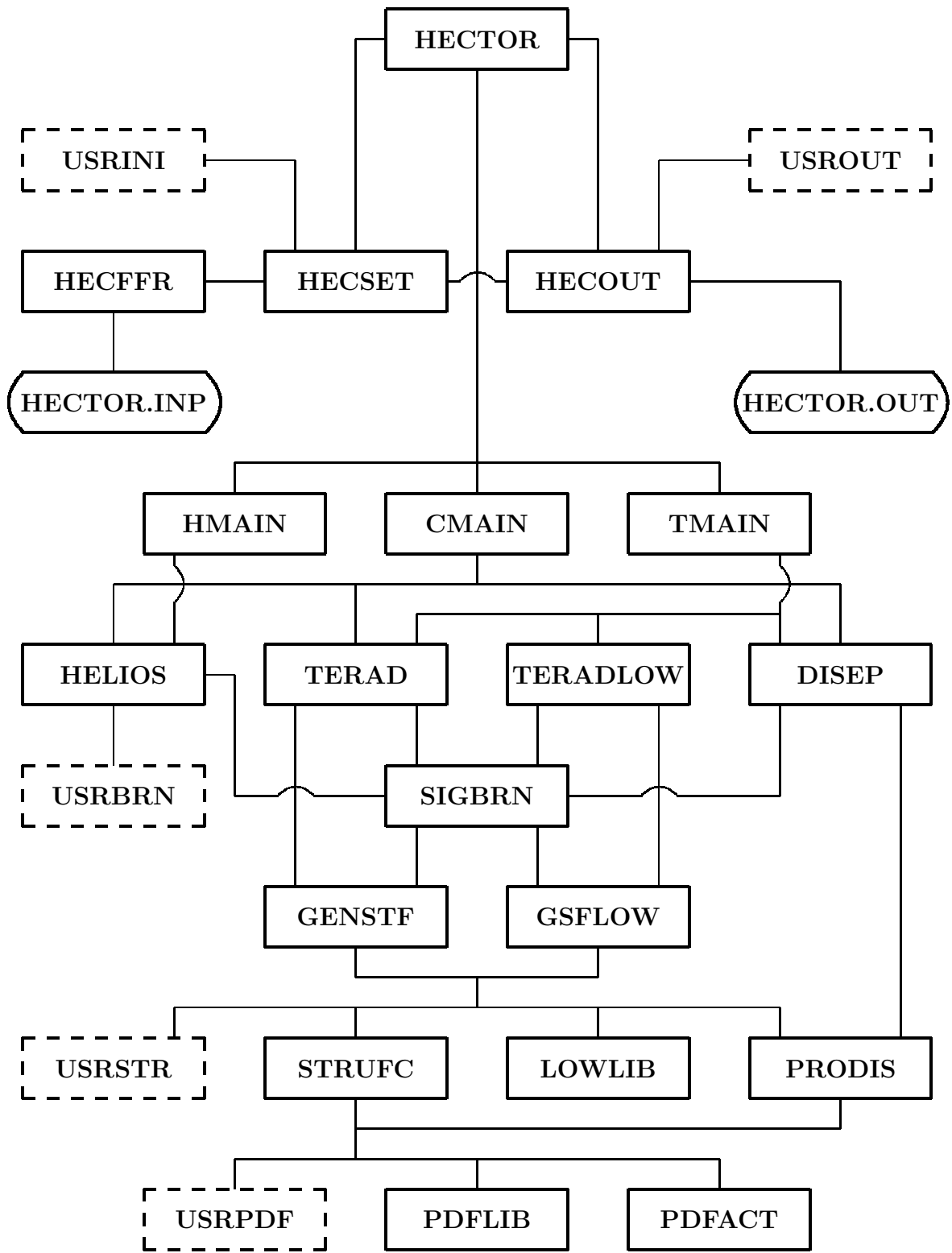
\begin{figure}
\unitlength=1.00mm
\linethickness{1pt}
\begin{picture}(168.00,240.00)(-20,-95)
\put(45.00,160.00){\line(1,0){30.00}}
\put(75.00,160.00){\line(0,-1){10.00}}
\put(75.00,150.00){\line(-1,0){30.00}}
\put(45.00,150.00){\line(0,1){10.00}}
\put(-15.00,120.00){\line(1,0){30.00}}
\put(15.00,120.00){\line(0,-1){10.00}}
\put(15.00,110.00){\line(-1,0){30.00}}
\put(-15.00,110.00){\line(0,1){10.00}}
\put(25.00,120.00){\line(1,0){30.00}}
\put(55.00,120.00){\line(0,-1){10.00}}
\put(55.00,110.00){\line(-1,0){30.00}}
\put(25.00,110.00){\line(0,1){10.00}}
\put(65.00,120.00){\line(1,0){30.00}}
\put(95.00,120.00){\line(0,-1){10.00}}
\put(95.00,110.00){\line(-1,0){30.00}}
\put(65.00,110.00){\line(0,1){10.00}}
\put(-15.00,100.00){\line(1,0){30.00}}
\put(15.00,90.00){\line(-1,0){30.00}}
\put(105.00,100.00){\line(1,0){30.00}}
\put(135.00,90.00){\line(-1,0){30.00}}
\put(5.00,80.00){\line(1,0){30.00}}
\put(35.00,80.00){\line(0,-1){10.00}}
\put(35.00,70.00){\line(-1,0){30.00}}
\put(5.00,70.00){\line(0,1){10.00}}
\put(45.00,80.00){\line(1,0){30.00}}
\put(75.00,80.00){\line(0,-1){10.00}}
\put(75.00,70.00){\line(-1,0){30.00}}
\put(45.00,70.00){\line(0,1){10.00}}
\put(85.00,80.00){\line(1,0){30.00}}
\put(115.00,80.00){\line(0,-1){10.00}}
\put(115.00,70.00){\line(-1,0){30.00}}
\put(85.00,70.00){\line(0,1){10.00}}
\put(5.00,80.00){\line(1,0){30.00}}
\put(35.00,80.00){\line(0,-1){10.00}}
\put(35.00,70.00){\line(-1,0){30.00}}
\put(5.00,70.00){\line(0,1){10.00}}
\put(45.00,80.00){\line(1,0){30.00}}
\put(75.00,80.00){\line(0,-1){10.00}}
\put(75.00,70.00){\line(-1,0){30.00}}
\put(45.00,70.00){\line(0,1){10.00}}
\put(85.00,80.00){\line(1,0){30.00}}
\put(115.00,80.00){\line(0,-1){10.00}}
\put(115.00,70.00){\line(-1,0){30.00}}
\put(85.00,70.00){\line(0,1){10.00}}
\put(-15.00,55.00){\line(1,0){30.00}}
\put(15.00,55.00){\line(0,-1){10.00}}
\put(15.00,45.00){\line(-1,0){30.00}}
\put(-15.00,45.00){\line(0,1){10.00}}
\put(25.00,55.00){\line(1,0){30.00}}
\put(55.00,55.00){\line(0,-1){10.00}}
\put(55.00,45.00){\line(-1,0){30.00}}
\put(25.00,45.00){\line(0,1){10.00}}
\put(65.00,55.00){\line(1,0){30.00}}
\put(95.00,55.00){\line(0,-1){10.00}}
\put(95.00,45.00){\line(-1,0){30.00}}
\put(65.00,45.00){\line(0,1){10.00}}
\put(105.00,55.00){\line(1,0){30.00}}
\put(135.00,55.00){\line(0,-1){10.00}}
\put(135.00,45.00){\line(-1,0){30.00}}
\put(105.00,45.00){\line(0,1){10.00}}
\put(-15.00,55.00){\line(1,0){30.00}}
\put(15.00,55.00){\line(0,-1){10.00}}
\put(15.00,45.00){\line(-1,0){30.00}}
\put(-15.00,45.00){\line(0,1){10.00}}
\put(25.00,55.00){\line(1,0){30.00}}
\put(55.00,55.00){\line(0,-1){10.00}}
\put(55.00,45.00){\line(-1,0){30.00}}
\put(25.00,45.00){\line(0,1){10.00}}
\put(65.00,55.00){\line(1,0){30.00}}
\put(95.00,55.00){\line(0,-1){10.00}}
\put(95.00,45.00){\line(-1,0){30.00}}
\put(65.00,45.00){\line(0,1){10.00}}
\put(105.00,55.00){\line(1,0){30.00}}
\put(135.00,55.00){\line(0,-1){10.00}}
\put(135.00,45.00){\line(-1,0){30.00}}
\put(105.00,45.00){\line(0,1){10.00}}
\put(25.00,15.00){\line(1,0){30.00}}
\put(55.00,15.00){\line(0,-1){10.00}}
\put(55.00,5.00){\line(-1,0){30.00}}
\put(25.00,5.00){\line(0,1){10.00}}
\put(65.00,15.00){\line(1,0){30.00}}
\put(95.00,15.00){\line(0,-1){10.00}}
\put(95.00,5.00){\line(-1,0){30.00}}
\put(65.00,5.00){\line(0,1){10.00}}
\put(25.00,15.00){\line(1,0){30.00}}
\put(55.00,15.00){\line(0,-1){10.00}}
\put(55.00,5.00){\line(-1,0){30.00}}
\put(25.00,5.00){\line(0,1){10.00}}
\put(65.00,15.00){\line(1,0){30.00}}
\put(95.00,15.00){\line(0,-1){10.00}}
\put(95.00,5.00){\line(-1,0){30.00}}
\put(65.00,5.00){\line(0,1){10.00}}
\put(25.00,-10.00){\line(1,0){30.00}}
\put(55.00,-10.00){\line(0,-1){10.00}}
\put(55.00,-20.00){\line(-1,0){30.00}}
\put(25.00,-20.00){\line(0,1){10.00}}
\put(65.00,-10.00){\line(1,0){30.00}}
\put(95.00,-10.00){\line(0,-1){10.00}}
\put(95.00,-20.00){\line(-1,0){30.00}}
\put(65.00,-20.00){\line(0,1){10.00}}
\put(105.00,-10.00){\line(1,0){30.00}}
\put(135.00,-10.00){\line(0,-1){10.00}}
\put(135.00,-20.00){\line(-1,0){30.00}}
\put(105.00,-20.00){\line(0,1){10.00}}
\put(25.00,-10.00){\line(1,0){30.00}}
\put(55.00,-10.00){\line(0,-1){10.00}}
\put(55.00,-20.00){\line(-1,0){30.00}}
\put(25.00,-20.00){\line(0,1){10.00}}
\put(65.00,-10.00){\line(1,0){30.00}}
\put(95.00,-10.00){\line(0,-1){10.00}}
\put(95.00,-20.00){\line(-1,0){30.00}}
\put(65.00,-20.00){\line(0,1){10.00}}
\put(105.00,-10.00){\line(1,0){30.00}}
\put(135.00,-10.00){\line(0,-1){10.00}}
\put(135.00,-20.00){\line(-1,0){30.00}}
\put(105.00,-20.00){\line(0,1){10.00}}
\put(45.00,-35.00){\line(1,0){30.00}}
\put(75.00,-35.00){\line(0,-1){10.00}}
\put(75.00,-45.00){\line(-1,0){30.00}}
\put(45.00,-45.00){\line(0,1){10.00}}
\put(85.00,-35.00){\line(1,0){30.00}}
\put(115.00,-35.00){\line(0,-1){10.00}}
\put(115.00,-45.00){\line(-1,0){30.00}}
\put(85.00,-45.00){\line(0,1){10.00}}
\put(45.00,-35.00){\line(1,0){30.00}}
\put(75.00,-35.00){\line(0,-1){10.00}}
\put(75.00,-45.00){\line(-1,0){30.00}}
\put(45.00,-45.00){\line(0,1){10.00}}
\put(85.00,-35.00){\line(1,0){30.00}}
\put(115.00,-35.00){\line(0,-1){10.00}}
\put(115.00,-45.00){\line(-1,0){30.00}}
\bezier{56}(105.00,90.00)(100.00,95.00)(105.00,100.00)
\bezier{56}(135.00,90.00)(140.00,95.00)(135.00,100.00)
\bezier{56}(15.00,100.00)(20.00,95.00)(15.00,90.00)
\bezier{56}(-15.00,90.00)(-20.00,95.00)(-15.00,100.00)
\put(105.00,136.00){\line(0,-1){2.00}}
\put(105.00,130.00){\line(0,1){2.00}}
\put(105.00,130.00){\line(1,0){2.00}}
\put(109.00,130.00){\line(1,0){2.00}}
\put(113.00,130.00){\line(1,0){2.00}}
\put(117.00,130.00){\line(1,0){2.00}}
\put(121.00,130.00){\line(1,0){2.00}}
\put(125.00,130.00){\line(1,0){2.00}}
\put(129.00,130.00){\line(1,0){2.00}}
\put(133.00,130.00){\line(1,0){2.00}}
\put(135.00,130.00){\line(0,1){2.00}}
\put(135.00,134.00){\line(0,1){2.00}}
\put(135.00,138.00){\line(0,1){2.00}}
\put(135.00,140.00){\line(-1,0){2.00}}
\put(131.00,140.00){\line(-1,0){2.00}}
\put(127.00,140.00){\line(-1,0){2.00}}
\put(123.00,140.00){\line(-1,0){2.00}}
\put(119.00,140.00){\line(-1,0){2.00}}
\put(115.00,140.00){\line(-1,0){2.00}}
\put(111.00,140.00){\line(-1,0){2.00}}
\put(107.00,140.00){\line(-1,0){2.00}}
\put(105.00,140.00){\line(0,-1){2.00}}
\put(-15.00,136.00){\line(0,-1){2.00}}
\put(-15.00,130.00){\line(0,1){2.00}}
\put(-15.00,130.00){\line(1,0){2.00}}
\put(-11.00,130.00){\line(1,0){2.00}}
\put(-7.00,130.00){\line(1,0){2.00}}
\put(-3.00,130.00){\line(1,0){2.00}}
\put(1.00,130.00){\line(1,0){2.00}}
\put(5.00,130.00){\line(1,0){2.00}}
\put(9.00,130.00){\line(1,0){2.00}}
\put(13.00,130.00){\line(1,0){2.00}}
\put(15.00,130.00){\line(0,1){2.00}}
\put(15.00,134.00){\line(0,1){2.00}}
\put(15.00,138.00){\line(0,1){2.00}}
\put(15.00,140.00){\line(-1,0){2.00}}
\put(11.00,140.00){\line(-1,0){2.00}}
\put(7.00,140.00){\line(-1,0){2.00}}
\put(3.00,140.00){\line(-1,0){2.00}}
\put(-1.00,140.00){\line(-1,0){2.00}}
\put(-5.00,140.00){\line(-1,0){2.00}}
\put(-9.00,140.00){\line(-1,0){2.00}}
\put(-13.00,140.00){\line(-1,0){2.00}}
\put(-15.00,140.00){\line(0,-1){2.00}}
\put(5.00,-39.00){\line(0,-1){2.00}}
\put(5.00,-45.00){\line(0,1){2.00}}
\put(5.00,-45.00){\line(1,0){2.00}}
\put(9.00,-45.00){\line(1,0){2.00}}
\put(13.00,-45.00){\line(1,0){2.00}}
\put(17.00,-45.00){\line(1,0){2.00}}
\put(21.00,-45.00){\line(1,0){2.00}}
\put(25.00,-45.00){\line(1,0){2.00}}
\put(29.00,-45.00){\line(1,0){2.00}}
\put(33.00,-45.00){\line(1,0){2.00}}
\put(35.00,-45.00){\line(0,1){2.00}}
\put(35.00,-41.00){\line(0,1){2.00}}
\put(35.00,-37.00){\line(0,1){2.00}}
\put(35.00,-35.00){\line(-1,0){2.00}}
\put(31.00,-35.00){\line(-1,0){2.00}}
\put(27.00,-35.00){\line(-1,0){2.00}}
\put(23.00,-35.00){\line(-1,0){2.00}}
\put(19.00,-35.00){\line(-1,0){2.00}}
\put(15.00,-35.00){\line(-1,0){2.00}}
\put(11.00,-35.00){\line(-1,0){2.00}}
\put(7.00,-35.00){\line(-1,0){2.00}}
\put(5.00,-35.00){\line(0,-1){2.00}}
\put(-15.00,31.00){\line(0,-1){2.00}}
\put(-15.00,25.00){\line(0,1){2.00}}
\put(-15.00,25.00){\line(1,0){2.00}}
\put(-11.00,25.00){\line(1,0){2.00}}
\put(-7.00,25.00){\line(1,0){2.00}}
\put(-3.00,25.00){\line(1,0){2.00}}
\put(1.00,25.00){\line(1,0){2.00}}
\put(5.00,25.00){\line(1,0){2.00}}
\put(9.00,25.00){\line(1,0){2.00}}
\put(13.00,25.00){\line(1,0){2.00}}
\put(15.00,25.00){\line(0,1){2.00}}
\put(15.00,29.00){\line(0,1){2.00}}
\put(15.00,33.00){\line(0,1){2.00}}
\put(15.00,35.00){\line(-1,0){2.00}}
\put(11.00,35.00){\line(-1,0){2.00}}
\put(7.00,35.00){\line(-1,0){2.00}}
\put(3.00,35.00){\line(-1,0){2.00}}
\put(-1.00,35.00){\line(-1,0){2.00}}
\put(-5.00,35.00){\line(-1,0){2.00}}
\put(-9.00,35.00){\line(-1,0){2.00}}
\put(-13.00,35.00){\line(-1,0){2.00}}
\put(-15.00,35.00){\line(0,-1){2.00}}
\put(-15.00,-14.00){\line(0,-1){2.00}}
\put(-15.00,-20.00){\line(0,1){2.00}}
\put(-15.00,-20.00){\line(1,0){2.00}}
\put(-11.00,-20.00){\line(1,0){2.00}}
\put(-7.00,-20.00){\line(1,0){2.00}}
\put(-3.00,-20.00){\line(1,0){2.00}}
\put(1.00,-20.00){\line(1,0){2.00}}
\put(5.00,-20.00){\line(1,0){2.00}}
\put(9.00,-20.00){\line(1,0){2.00}}
\put(13.00,-20.00){\line(1,0){2.00}}
\put(15.00,-20.00){\line(0,1){2.00}}
\put(15.00,-16.00){\line(0,1){2.00}}
\put(15.00,-12.00){\line(0,1){2.00}}
\put(15.00,-10.00){\line(-1,0){2.00}}
\put(11.00,-10.00){\line(-1,0){2.00}}
\put(7.00,-10.00){\line(-1,0){2.00}}
\put(3.00,-10.00){\line(-1,0){2.00}}
\put(-1.00,-10.00){\line(-1,0){2.00}}
\put(-5.00,-10.00){\line(-1,0){2.00}}
\put(-9.00,-10.00){\line(-1,0){2.00}}
\put(-13.00,-10.00){\line(-1,0){2.00}}
\put(-15.00,-10.00){\line(0,-1){2.00}}
\put(0.00,30.00){\makebox(0,0)[cc] {\bf USRBRN}}
\put(0.00,-15.00){\makebox(0,0)[cc]{{\bf USRSTR}}}
\put(0.00,135.00){\makebox(0,0)[cc]{{\bf USRINI}}}
\put(120.00,135.00){\makebox(0,0)[cc]{{\bf USROUT}}}
\put(0.00,115.00){\makebox(0,0)[cc]{{\bf HECFFR}}}
\put(40.00,115.00){\makebox(0,0)[cc]{{\bf HECSET}}}
\put(80.00,115.00){\makebox(0,0)[cc]{{\bf HECOUT}}}
\put(0.00,95.00){\makebox(0,0)[cc]{{\bf HECTOR.INP}}}
\put(120.00,95.00){\makebox(0,0)[cc]{{\bf HECTOR.OUT}}}
\put(20.00,75.00){\makebox(0,0)[cc]{{\bf HMAIN}}}
\put(60.00,75.00){\makebox(0,0)[cc]{{\bf CMAIN}}}
\put(100.00,75.00){\makebox(0,0)[cc]{{\bf TMAIN}}}
\put(0.00,50.00){\makebox(0,0)[cc]{{\bf HELIOS}}}
\put(40.00,50.00){\makebox(0,0)[cc]{{\bf TERAD}}}
\put(80.00,50.00){\makebox(0,0)[cc]{{\bf TERADLOW}}}
\put(120.00,50.00){\makebox(0,0)[cc]{{\bf DISEP}}}
\put(80.00,10.00){\makebox(0,0)[cc]{{\bf GSFLOW}}}
\put(40.00,10.00){\makebox(0,0)[cc]{{\bf GENSTF}}}
\put(40.00,-15.00){\makebox(0,0)[cc]{{\bf STRUFC}}}
\put(80.00,-15.00){\makebox(0,0)[cc]{{\bf LOWLIB}}}
\put(120.00,-15.00){\makebox(0,0)[cc]{{\bf PRODIS}}}
\put(60.00,-40.00){\makebox(0,0)[cc]{{\bf PDFLIB}}}
\put(100.00,-40.00){\makebox(0,0)[cc]{{\bf PDFACT}}}
\put(60.00,155.00){\makebox(0,0)[cc]{{\bf HECTOR}}}
\put(20.00,-40.00){\makebox(0,0)[cc]{{\bf USRPDF}}}
\put(60.00,30.00){\makebox(0,0)[cc]{{\bf SIGBRN}}}
\put(45.00,35.00){\line(1,0){30.00}}
\put(75.00,35.00){\line(0,-1){10.00}}
\put(75.00,25.00){\line(-1,0){30.00}}
\put(45.00,25.00){\line(0,1){10.00}}
\linethickness{.5pt}
\put(85.00,-45.00){\line(0,1){10.00}}
\put(75.00,155.00){\line(1,0){5.00}}
\put(80.00,155.00){\line(0,-1){35.00}}
\put(90.00,120.00){\line(0,1){15.00}}
\put(90.00,135.00){\line(1,0){15.00}}
\put(95.00,115.00){\line(1,0){25.00}}
\put(120.00,115.00){\line(0,-1){15.00}}
\put(60.00,80.00){\line(0,1){70.00}}
\put(99.00,85.00){\line(-1,0){79.00}}
\put(20.00,85.00){\line(0,-1){5.00}}
\put(55.00,115.00){\line(1,0){2.00}}
\put(65.00,115.00){\line(-1,0){2.00}}
\bezier{32}(63.00,115.00)(60.00,118.00)(57.00,115.00)
\put(60.00,70.00){\line(0,-1){5.00}}
\put(120.00,55.00){\line(0,1){10.00}}
\put(120.00,65.00){\line(-1,0){110.00}}
\put(40.00,65.00){\line(0,-1){10.00}}
\put(50.00,55.00){\line(0,1){5.00}}
\put(50.00,60.00){\line(1,0){60.00}}
\put(110.00,60.00){\line(0,-1){5.00}}
\put(80.00,55.00){\line(0,1){5.00}}
\put(10.00,70.00){\line(0,-1){2.00}}
\bezier{32}(10.00,68.00)(13.00,65.00)(10.00,62.00)
\put(10.00,62.00){\line(0,-1){7.00}}
\put(0.78,65.00){\line(1,0){14.44}}
\put(0.00,65.00){\line(0,-1){10.00}}
\put(0.00,65.00){\line(1,0){1.00}}
\put(110.00,60.00){\line(0,1){2.00}}
\put(110.00,62.00){\line(0,-1){2.00}}
\put(110.00,62.00){\line(0,-1){2.00}}
\put(110.00,68.00){\line(0,1){2.00}}
\put(40.00,5.00){\line(0,-1){5.00}}
\put(40.00,0.00){\line(1,0){40.00}}
\put(80.00,0.00){\line(0,1){5.00}}
\put(60.00,0.00){\line(0,-1){5.00}}
\put(40.00,-10.00){\line(0,1){5.00}}
\put(80.00,-5.00){\line(0,-1){5.00}}
\put(110.00,-10.00){\line(0,1){5.00}}
\put(110.00,-5.00){\line(-1,0){100.00}}
\put(10.00,-5.00){\line(0,-1){5.00}}
\put(20.00,-35.00){\line(0,1){5.00}}
\put(20.00,-30.00){\line(1,0){80.00}}
\put(100.00,-30.00){\line(0,-1){5.00}}
\put(40.00,-30.00){\line(0,1){10.00}}
\put(40.00,-25.00){\line(1,0){80.00}}
\put(120.00,-25.00){\line(0,1){5.00}}
\put(15.00,135.00){\line(1,0){15.00}}
\put(30.00,135.00){\line(0,-1){15.00}}
\put(40.00,120.00){\line(0,1){35.00}}
\put(40.00,155.00){\line(1,0){5.00}}
\put(15.00,115.00){\line(1,0){10.00}}
\put(0.00,110.00){\line(0,-1){10.00}}
\put(60.00,-30.00){\line(0,-1){5.00}}
\put(100.00,80.00){\line(0,1){5.00}}
\put(100.00,85.00){\line(0,0){0.00}}
\put(98.00,85.00){\line(0,0){0.00}}
\put(96.00,85.00){\line(1,0){4.00}}
\put(30.00,45.00){\line(0,-1){30.00}}
\put(50.00,15.00){\line(0,1){10.00}}
\put(50.00,45.00){\line(0,-1){10.00}}
\put(70.00,35.00){\line(0,1){10.00}}
\put(70.00,25.00){\line(0,-1){10.00}}
\put(90.00,15.00){\line(0,1){30.00}}
\put(20.00,50.00){\line(0,-1){20.00}}
\put(20.00,50.00){\line(-1,0){5.00}}
\put(0.00,45.00){\line(0,-1){10.00}}
\put(20.00,30.00){\line(1,0){7.00}}
\put(33.00,30.00){\line(1,0){12.00}}
\put(75.00,30.00){\line(1,0){12.00}}
\bezier{32}(33.00,30.00)(30.00,33.00)(27.00,30.00)
\bezier{32}(93.00,30.00)(90.00,33.00)(87.00,30.00)
\bezier{32}(10.00,68.00)(13.00,65.00)(10.00,62.00)
\bezier{32}(110.00,62.00)(113.00,65.00)(110.00,68.00)
\put(130.00,45.00){\line(0,-1){55.00}}
\put(110.00,45.00){\line(0,-1){15.00}}
\put(93.00,30.00){\line(1,0){17.00}}
\end{picture}

\vspace{-4.5cm}
\caption
[\it Basic {\tt HECTOR} flowchart]
{\it
{Basic {\tt HECTOR} flowchart. Shown are subdirectories.}
\label{hecflow}
}
\end{figure}


\bigskip

\he\ consists of four branches as may be seen in figure~\ref{hecflow}:
{\tt HELIOS},
{\tt TERAD},
{\tt DISEP},
{\tt TERADLOW}.
The first three branches of \he\, which contain the different packages for
the calculation of the QED corrections are shown in
figure~\ref{branches}.
The subdirectories {\tt HELIOS}, {\tt TERAD}, {\tt DISEP} are
symbolized by the big boxes.
The settings of flags {\tt IMEA} and {\tt IDSP} define the choice of
the type of corrections.
{\tt HELIOS} consists of one file, while
inside {\tt TERAD} and {\tt DISEP}, the file structure is shown.
The index i in the figure is a short--hand notation.
For i=1, it is {\tt IORD}=2 and {\tt IEPC}=0, while i=2 corresponds to
{\tt IORD}=2 and {\tt IEPC}=1.

The branch {\tt TERADLOW} is attained by {\tt ILOW}=1 together with
{\tt IMEA}=1 and {\tt IOPT}=2.


\bigskip

Figure~\ref{maip} illustrates the influence of the flags {\tt IUSR},  {\tt
ITV1},  {\tt ITV2},  {\tt IBIN},  {\tt IOPT}, {\tt IORD} on the
initialization of \he.
Shown are the subdirectories {\tt HECSET} and {\tt HECTOR}, together
with some selected files (in boxes).
We should comment shortly on the following of them:

\begin{tabbing}
\underline{{\tt FILBIN}}: ~~~\=
{}~~The selected lattice of kinematical points is
filled
\\
\underline{\tt HECCAL}: \>
{}~~Calls the branch of \he\ chosen and initializes
the actual calculation
\\
\underline{\tt HECRUN}: \>
{}~~Runs the job
\\
\underline{\tt HECTER}: \>
{}~~The common blocks of {\tt TERAD} are filled
\\
\underline{\tt USRBIN}: \>
{}~~Optional filling of a user supplied
kinematical lattice
\\
\underline{\tt USRINT}: \>
{}~~A user supplied integration over bins is performed

\end{tabbing}

The entries 1 and 2 are links to figure~\ref{heou}, where the output
organization is illustrated.

\begin{figure}
\unitlength=1.00mm
\linethickness{1pt}
\begin{picture}(168.00,240.00)(-40,-50)
\put(102.00,110.00){\line(1,0){20.00}}
\put(122.00,110.00){\line(0,-1){10.00}}
\put(122.00,100.00){\line(-1,0){20.00}}
\put(102.00,100.00){\line(0,1){10.00}}
\put(37.00,100.00){\line(1,0){20.00}}
\put(57.00,100.00){\line(0,-1){10.00}}
\put(57.00,90.00){\line(-1,0){20.00}}
\put(37.00,90.00){\line(0,1){10.00}}
\put(37.00,75.00){\line(1,0){20.00}}
\put(57.00,75.00){\line(0,-1){10.00}}
\put(57.00,65.00){\line(-1,0){20.00}}
\put(37.00,65.00){\line(0,1){10.00}}
\put(102.00,-10.00){\line(0,1){2.00}}
\put(102.00,-6.00){\line(0,1){2.00}}
\put(102.00,-2.00){\line(0,1){2.00}}
\put(104.00,0.00){\line(1,0){2.00}}
\put(108.00,0.00){\line(1,0){2.00}}
\put(112.00,0.00){\line(1,0){2.00}}
\put(116.00,0.00){\line(1,0){2.00}}
\put(120.00,0.00){\line(1,0){2.00}}
\put(122.00,-2.00){\line(0,-1){2.00}}
\put(122.00,-6.00){\line(0,-1){2.00}}
\put(122.00,-10.00){\line(-1,0){2.00}}
\put(118.00,-10.00){\line(-1,0){2.00}}
\put(114.00,-10.00){\line(-1,0){2.00}}
\put(110.00,-10.00){\line(-1,0){2.00}}
\put(106.00,-10.00){\line(-1,0){2.00}}
\put(-2.00,170.00){\line(1,0){20.00}}
\put(18.00,170.00){\line(0,-1){10.00}}
\put(18.00,160.00){\line(-1,0){20.00}}
\put(-2.00,160.00){\line(0,1){10.00}}
\put(-2.00,40.00){\line(1,0){20.00}}
\put(18.00,40.00){\line(0,-1){10.00}}
\put(18.00,30.00){\line(-1,0){20.00}}
\put(-2.00,30.00){\line(0,1){10.00}}
\put(-27.00,40.00){\line(1,0){20.00}}
\put(-7.00,40.00){\line(0,-1){10.00}}
\put(-7.00,30.00){\line(-1,0){20.00}}
\put(-27.00,30.00){\line(0,1){10.00}}
\put(-14.00,0.00){\line(1,0){20.00}}
\put(6.00,0.00){\line(0,-1){10.00}}
\put(6.00,-10.00){\line(-1,0){20.00}}
\put(-14.00,-10.00){\line(0,1){10.00}}
\put(-27.00,170.00){\line(1,0){20.00}}
\put(-7.00,170.00){\line(0,-1){10.00}}
\put(-27.00,160.00){\line(0,1){10.00}}
\put(-27.00,160.00){\line(1,0){20.00}}
\put(37.00,20.00){\line(1,0){20.00}}
\put(57.00,20.00){\line(0,-1){10.00}}
\put(57.00,10.00){\line(-1,0){20.00}}
\put(37.00,10.00){\line(0,1){10.00}}
\put(8.00,35.00)   {\makebox(0,0)[cc]{NCOUPL}}
\put(47.00,15.00)  {\makebox(0,0)[cc]{DIZET  }}
\put(-17.00,35.00) {\makebox(0,0)[cc]{ CCOUPL}}
\put(47.00,70.00)  {\makebox(0,0)[cc]{POLYN  }}
\put(47.00,95.00)  {\makebox(0,0)[cc]{MIDSTF }}
\put(112.00,105.00){\makebox(0,0)[cc]{STRFBS }}
\put(8.00,165.00)  {\makebox(0,0)[cc]{GENSTF }}
\put(112.00,-5.00) {\makebox(0,0)[cc]{USRSTR }}
\put(-4.00,-5.00)  {\makebox(0,0)[cc]{PRODIS }}
\put(-17.00,165.00){\makebox(0,0)[cc]{GCCSTF }}
\put(56.00,42.00){\makebox(0,0)  [cc] {\bf DIZET}}
\put(112.00,118.00){\makebox(0,0)[cc] {\bf LOWLIB}}
\put(-30.00,192.00){\makebox(0,0)[lc] {\bf GENSTF}}
\put(72.00,107.00){\makebox(0,0)[lc]{\tt IVAR=2   }}
\put(72.00,182.00){\makebox(0,0)[lc]{\tt IVAR=$-$1}}
\put(21.00,107.00){\makebox(0,0)[cc]{\tt IVAR     }}
\put(133.00,40.00){\makebox(0,0)[rc]{\tt IMOD=0   }}
\put(113.00,40.00){\makebox(0,0)[rc]{\tt IMOD=1   }}
\put(72.00,132.00){\makebox(0,0)[lc]{\tt IVAR=1   }}
\put(72.00,157.00){\makebox(0,0)[lc]{\tt IVAR=0   }}
\put(112.00,50.00){\makebox(0,0)[cc]{\tt IWEA=0   }}
\put(-4.00,57.00) {\makebox(0,0)[cc]{\tt IWEA=1   }}
\put(71.00,147.00){\makebox(0,0)[lc]{\tiny DAMPING        }}
\put(16.00,102.00){\makebox(0,0)[lc]{\tiny CHOICE OF      }}
\put(16.00,99.00){\makebox(0,0)[lc]{\tiny DAMPING         }}
\put(16.00,96.00){\makebox(0,0)[lc]{\tiny SCHEME          }}
\put(16.00,93.00){\makebox(0,0)[lc]{\tiny FOR $Q^2< Q_0^2$}}
\put(71.00,153.00){\makebox(0,0)[lc]{\tiny VOLKONSKY/     }}
\put(71.00,150.00){\makebox(0,0)[lc]{\tiny PROKHOROV      }}
\put(71.00,128.00){\makebox(0,0)[lc]{\tiny STEIN DAMPING  }}
\put(71.00,103.00){\makebox(0,0)[lc]{\tiny BRASSE/STEIN   }}
\put(71.00,100.00){\makebox(0,0)[lc]{\tiny PARAMETERIZATION}}
\put(71.00,178.00){\makebox(0,0)[lc]{\tiny QUARK          }}
\put(71.00,175.00){\makebox(0,0)[lc]{\tiny DISTRIBUTIONS  }}
\put(71.00,172.00){\makebox(0,0)[lc]{\tiny ARE FROZEN     }}
\put(71.00,169.00){\makebox(0,0)[lc]{\tiny AT $Q^2_0=Q^2_{_{min}}$ }}
\put(71.00,165.00){\makebox(0,0)[lc]{\tiny FOR $Q^2<Q^2_0$}}
\put(130.00,36.00){\makebox(0,0)[rc]{\tiny PDF'S ARE      }}
\put(130.00,33.00){\makebox(0,0)[rc]{\tiny USED FOR       }}
\put(130.00,30.00){\makebox(0,0)[rc]{\tiny THE TARGET     }}
\put(130.00,27.00){\makebox(0,0)[rc]{\tiny STRUCTURE      }}
\put(130.00,24.00){\makebox(0,0)[rc]{\tiny FUNCTIONS      }}
\put(110.00,36.00){\makebox(0,0)[rc]{\tiny USER           }}
\put(110.00,33.00){\makebox(0,0)[rc]{\tiny SUPPLIED       }}
\put(110.00,30.00){\makebox(0,0)[rc]{\tiny STRUCTURE      }}
\put(110.00,27.00){\makebox(0,0)[rc]{\tiny FUNCTIONS      }}
\put(132.00,-5.00){\circle{6.00}}
\put(132.00,-5.00){\makebox(0,0)[cc]{{$3$}}}
\linethickness{0.5pt}
\put(34.00,105.00){\line(0,-1){10.00}}
\put(34.00,95.00){\line(1,0){3.00}}
\put(57.00,95.00){\line(1,0){3.00}}
\put(60.00,95.00){\line(0,1){10.00}}
\put(15.00,105.00){\line(1,0){12.00}}
\put(-24.00,160.00){\line(0,-1){110.00}}
\put(-24.00,50.00){\line(1,0){11.00}}
\put(4.00,50.00){\line(1,0){17.00}}
\put(15.00,105.00){\line(0,1){55.00}}
\put(112.00,60.00){\line(0,-1){7.00}}
\put(47.00,75.00){\line(0,1){15.00}}
\put(102.00,105.00){\line(-1,0){75.00}}
\put(102.00,50.00){\line(-1,0){83.00}}
\put(122.00,50.00){\line(1,0){10.00}}
\put(132.00,50.00){\line(0,-1){8.00}}
\put(-7.00,50.00){\line(1,0){5.00}}
\put(112.00,42.00){\line(0,1){5.00}}
 \put(-10.00,61.00){\line(0,1){99.00}}
 \put(1.00,61.00){\line(0,1){99.00}}
\put(112.00,0.00){\line(0,1){37.00}}
\put(132.00,37.00){\line(0,-1){39.00}}
\put(-10.00,0.00){\line(0,1){30.00}}
\put(1.00,30.00){\line(0,-1){30.00}}
\put(15.00,30.00){\line(0,-1){13.00}}
\put(15.00,17.00){\line(1,0){22.00}}
\put(4.00,13.00){\line(1,0){33.00}}
\put(-7.00,13.00){\line(1,0){5.00}}
\put(-13.00,13.00){\line(-1,0){11.00}}
\put(-24.00,13.00){\line(0,1){17.00}}
\put(30.00,180.00){\line(1,0){65.00}}
\put(95.00,155.00){\line(-1,0){65.00}}
\put(95.00,130.00){\line(-1,0){65.00}}
\put(30.00,105.00){\line(0,1){75.00}}
\put(-10.00,61.00){\line(0,-1){2.00}}
\put(1.00,59.00){\line(0,1){2.00}}
\put(1.00,54.00){\line(0,-1){14.00}}
\put(-10.00,40.00){\line(0,1){14.00}}
\bezier{32}(-13.00,50.00)(-10.00,53.00)(-7.00,50.00)
\bezier{32}(-2.00,50.00)(1.00,53.00)(4.00,50.00)
\bezier{32}(-2.00,13.00)(1.00,16.00)(4.00,13.00)
\bezier{32}(-13.00,13.00)(-10.00,16.00)(-7.00,13.00)
\linethickness{0.1pt}
\put(127.00,123.00){\line(-1,0){30.00}}
\put(127.00,60.00){\line(0,1){63.00}}
\put(97.00,123.00){\line(0,-1){63.00}}
\put(97.00,60.00){\line(1,0){30.00}}
\put(-31.00,195.00){\line(0,-1){132.00}}
\put(64.00,63.00){\line(0,1){132.00}}
\put(64.00,195.00){\line(-1,0){95.00}}
\put(-31.00,63.00){\line(0,-1){2.00}}
\put(-31.00,61.00){\line(1,0){95.00}}
\put(-31.00,45.00){\line(1,0){95.00}}
\put(-31.00,5.00){\line(1,0){95.00}}
\put(-31.00,5.00){\line(0,1){40.00}}
\put(64.00,5.00){\line(0,1){40.00}}
\put(64.00,61.00){\line(0,1){2.00}}
\end{picture}

\vspace{-3cm}
\caption
{\it
Logical structure of {\tt GENSTF}
\label{genst}
}
\end{figure}
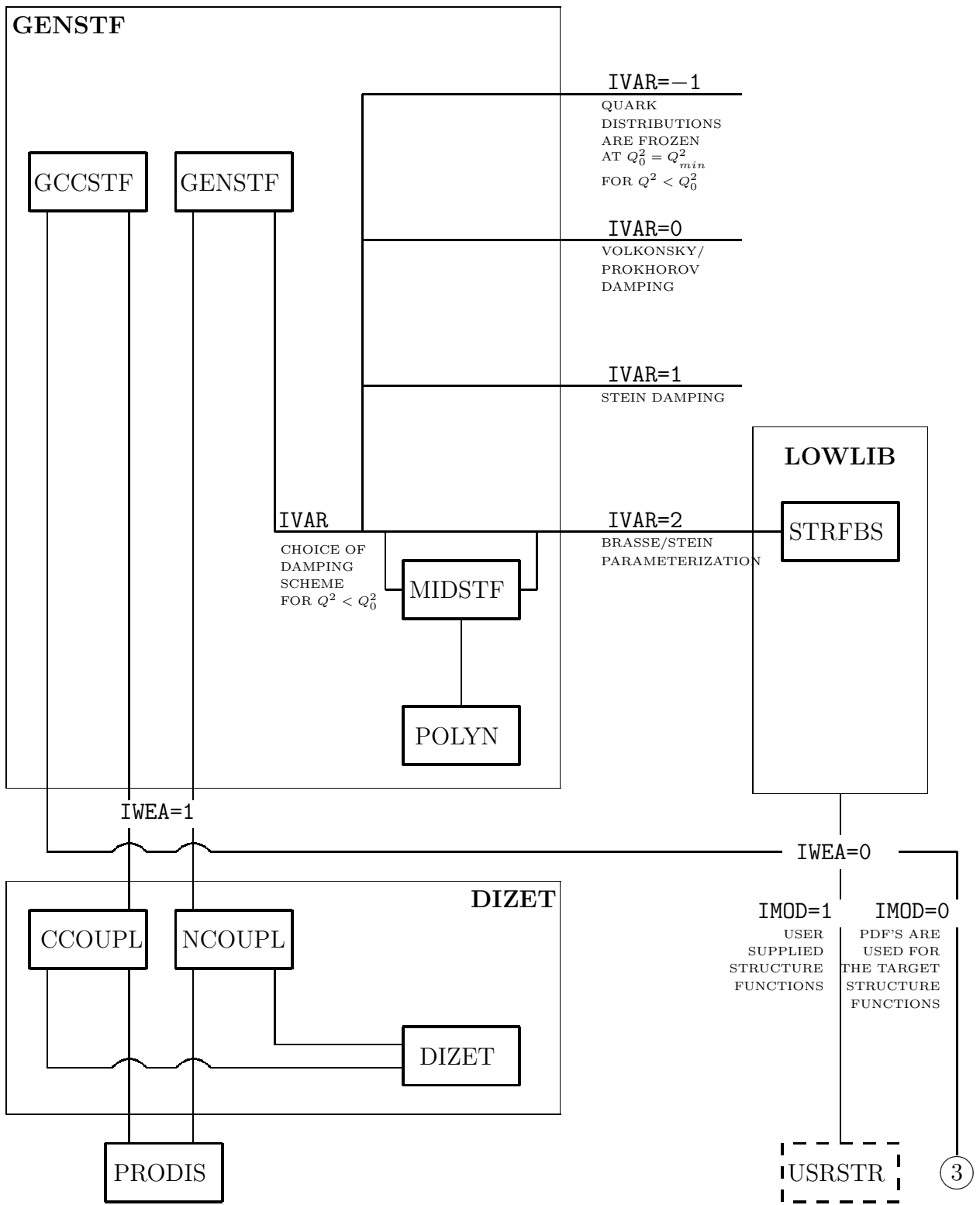
\clearpage

\begin{figure}
\unitlength=1.00mm
\linethickness{1pt}
\begin{picture}(168.00,240.00)(20,0)
\put(39.33,57.00)  {\makebox(0,0)[cc]{ STRUFC     }}
\put(105.33,190.00){\makebox(0,0)[cc]{ PDFACT     }}
\put(111.00,92.00) {\makebox(0,0)[cc]{ USRPDF     }}
\put(104.33,50.00) {\makebox(0,0)[cc]{ STRUCTM    }}
\put(77.33,143.00) {\makebox(0,0)[cc]{ PDFNUL     }}
\put(105.33,202.00){\makebox(0,0)[cc]{ ALPQCD     }}
\put(175.33,165.00){\makebox(0,0)[cc]{ MRS        }}
\put(175.33,195.00){\makebox(0,0)[cc]{CTEQ        }}
\put(175.33,180.00){\makebox(0,0)[cc]{GRV         }}
\put(175.33,240.00){\makebox(0,0)[cc]{CTEQ        }}
\put(175.33,225.00){\makebox(0,0)[cc]{GRV         }}
\put(175.33,120.00){\makebox(0,0)[cc]{MRS         }}
\put(175.33,150.00){\makebox(0,0)[cc]{CTEQ        }}
\put(175.33,135.00){\makebox(0,0)[cc]{GRV         }}
\put(100.50,103.00){\makebox(0,0)[cc]{\bf USRPDF  }}
\put(37.41,194.25) {\makebox(0,0)[cc]{\bf STRUFC  }}
\put(100.50,73.00) {\makebox(0,0)[cc]{\bf PDFLIB  }}
\put(100.50,246.00){\makebox(0,0)[cc]{\bf PDFACT  }}
\put(124.33,57.00){\makebox(0,0)[lc]{\tt ISET  }}
\put(66.33,193.00){\makebox(0,0)[lc]{\tt ISTR=0}}
\put(66.33,95.00){\makebox(0,0)[lc]{\tt ISTR=1 }}
\put(66.33,53.00){\makebox(0,0)[lc]{\tt ISTR=2 }}
\put(122.33,138.00){\makebox(0,0)[lc]{\tt ISCH=2 }}
\put(124.33,33.00){\makebox(0,0)[lc]{\tt ISCH  }}
\put(152.33,42.00){\makebox(0,0)[lc]{\tt ISCH=0 }}
\put(152.33,33.00){\makebox(0,0)[lc]{\tt ISCH=1 }}
\put(152.33,22.00){\makebox(0,0)[lc]{\tt ISCH=2 }}
\put(122.33,228.00){\makebox(0,0)[lc]{\tt ISCH=0 }}
\put(122.33,183.00){\makebox(0,0)[lc]{\tt ISCH=1 }}
\put(145.50,243.00){\makebox(0,0)[lc]{\tt ISSE=1 }}
\put(145.50,228.00){\makebox(0,0)[lc]{\tt ISSE=2 }}
\put(145.50,198.00){\makebox(0,0)[lc]{\tt ISSE=1 }}
\put(145.50,183.00){\makebox(0,0)[lc]{\tt ISSE=2 }}
\put(145.50,168.00){\makebox(0,0)[lc]{\tt ISSE=3 }}
\put(145.50,153.00){\makebox(0,0)[lc]{\tt ISSE=1 }}
\put(145.50,138.00){\makebox(0,0)[lc]{\tt ISSE=2 }}
\put(145.50,123.00){\makebox(0,0)[lc]{\tt ISSE=3-5}}
\put(124.33,73.00) {\makebox(0,0)[lc]{\tt IGPR     }}
\put(56.33,60.00)  {\makebox(0,0)[cc]{\tt ISTR       }}
\put(66.33,90.00)  {\makebox(0,0)[lc]{\tiny USER     }}
\put(66.33,87.00)  {\makebox(0,0)[lc]{\tiny SUPPLIED }}
\put(66.33,84.00)  {\makebox(0,0)[lc]{\tiny PARTON   }}
\put(66.33,81.00)  {\makebox(0,0)[lc]{\tiny DENSITY  }}
\put(66.33,78.00)  {\makebox(0,0)[lc]{\tiny FUCNTIONS}}
\put(66.33,48.00)  {\makebox(0,0)[lc]{\tiny PDFLIB IS}}
\put(66.33,45.00)  {\makebox(0,0)[lc]{\tiny USED FOR }}
\put(66.33,42.00)  {\makebox(0,0)[lc]{\tiny PARTON   }}
\put(66.33,39.00)  {\makebox(0,0)[lc]{\tiny DENSITIES}}
\put(124.33,68.00) {\makebox(0,0)[lc]{\tiny PDFLIB PARAMETER,          }}
\put(124.33,65.00) {\makebox(0,0)[lc]{\tiny IT DEFINES AN AUTHOR       }}
\put(124.33,52.00) {\makebox(0,0)[lc]{\tiny IT DEFINES A SPECIFIC      }}
\put(124.33,49.00) {\makebox(0,0)[lc]{\tiny PARAMETERIZATION SET WITHIN }}
\put(124.33,46.00) {\makebox(0,0)[lc]{\tiny THE GROUP DEFINED BY IGPR  }}
\put(151.33,37.00) {\makebox(0,0)[lc]{\tiny LEADING ORDER QCD          }}
\put(151.33,28.00) {\makebox(0,0)[lc]{\tiny DIS FACTORIZATION          }}
\put(151.33,26.00) {\makebox(0,0)[lc]{\tiny SCHEME                     }}
\put(151.33,17.00) {\makebox(0,0)[lc]{\tiny MS FACTORIZATION           }}
\put(151.33,15.00) {\makebox(0,0)[lc]{\tiny SCHEME                     }}
\put(121.33,223.00){\makebox(0,0)[lc]{\tiny LEADING ORDER              }}
\put(121.33,221.00){\makebox(0,0)[lc]{\tiny QCD                        }}
\put(121.33,178.00){\makebox(0,0)[lc]{\tiny DIS                        }}
\put(121.33,176.00){\makebox(0,0)[lc]{\tiny FACTORIZATION              }}
\put(121.33,174.00){\makebox(0,0)[lc]{\tiny SCHEME                     }}
\put(121.33,172.00){\makebox(0,0)[lc]{\tiny (NLO)                     }}
\put(66.33,188.00) {\makebox(0,0)[lc]{\tiny USE OF                     }}
\put(66.33,185.00) {\makebox(0,0)[lc]{\tiny RECENT                     }}
\put(66.33,182.00) {\makebox(0,0)[lc]{\tiny PARAMETRI-                 }}
\put(66.33,179.00) {\makebox(0,0)[lc]{\tiny ZATION  FOR                }}
\put(66.33,176.00) {\makebox(0,0)[lc]{\tiny PARTON                     }}
\put(66.33,173.00) {\makebox(0,0)[lc]{\tiny DISTRIBU-                  }}
\put(66.33,170.00) {\makebox(0,0)[lc]{\tiny TIONS                      }}
\put(66.33,167.00) {\makebox(0,0)[lc]{\tiny SELECTED BY                }}
\put(66.33,164.00) {\makebox(0,0)[lc]{\tiny FLAG ISSE                  }}
\put(121.33,133.00){\makebox(0,0)[lc]{\tiny MS                         }}
\put(121.33,131.00){\makebox(0,0)[lc]{\tiny FACTORIZATION              }}
\put(121.33,129.00){\makebox(0,0)[lc]{\tiny SCHEME                     }}
\put(121.33,127.00){\makebox(0,0)[lc]{\tiny (NLO)                     }}
\put(124.33,62.00) {\makebox(0,0)[lc]{\tiny GROUP IN THE LIBRARY       }}
\put(123.33,28.00) {\makebox(0,0)[lc]{\tiny QCD                        }}
\put(123.33,25.00) {\makebox(0,0)[lc]{\tiny FACTORIZATION              }}
\put(123.33,22.00) {\makebox(0,0)[lc]{\tiny SCHEME                     }}
\put(77.00,150.00) {\makebox(0,0)[cb]{\tiny LINK WITH                  }}
\linethickness{1pt}
\put(29.00,52.00){\line(0,1){10.00}}
\put(29.00,62.00){\line(1,0){20.00}}
\put(49.00,62.00){\line(0,-1){10.00}}
\put(49.00,52.00){\line(-1,0){20.00}}
\put(92.00,45.00){\line(0,1){10.00}}
\put(92.00,55.00){\line(1,0){25.00}}
\put(117.00,55.00){\line(0,-1){10.00}}
\put(117.00,45.00){\line(-1,0){25.00}}
\put(67.33,138.00){\line(1,0){20.00}}
\put(87.33,148.00){\line(-1,0){20.00}}
\put(67.33,138.00){\line(0,1){10.00}}
\put(87.33,148.00){\line(0,-1){10.00}}
\put(95.00,185.00){\line(1,0){20.00}}
\put(115.00,185.00){\line(0,1){10.00}}
\put(115.00,195.00){\line(-1,0){20.00}}
\put(95.00,195.00){\line(0,-1){10.00}}
\put(95.00,197.00){\line(1,0){20.00}}
\put(115.00,197.00){\line(0,1){10.00}}
\put(115.00,207.00){\line(-1,0){20.00}}
\put(95.00,207.00){\line(0,-1){10.00}}
\linethickness{0.5pt}
\put(29.00,57.00){\line(-1,0){6.00}}
\put(94.00,184.00){\line(1,0){22.00}}
\put(116.00,184.00){\line(0,1){24.00}}
\put(116.00,208.00){\line(-1,0){22.00}}
\put(94.00,208.00){\line(0,-1){24.00}}
\linethickness{0.1pt}
\put(90.00,82.00){\line(1,0){45.00}}
\put(135.00,82.00){\line(0,1){25.00}}
\put(135.00,107.00){\line(-1,0){45.00}}
\put(90.00,107.00){\line(0,-1){25.00}}
\put(90.00,110.00){\line(1,0){100.00}}
\put(190.00,110.00){\line(0,1){140.00}}
\put(190.00,250.00){\line(-1,0){100.00}}
\put(90.00,250.00){\line(0,-1){140.00}}
\put(27.00,32.00){\line(1,0){62.00}}
\put(90.00,77.00){\line(1,0){100.00}}
\put(190.00,77.00){\line(0,-1){65.00}}
\put(190.00,12.00){\line(-1,0){100.00}}
\put(90.00,12.00){\line(0,1){65.00}}
\put(89.08,32.00){\line(0,1){165.83}}
\put(89.08,197.83){\line(-1,0){62.08}}
\put(27.00,197.83){\line(0,-1){165.83}}
\put(93.00,54.00){\line(1,0){23.00}}
\put(116.00,54.00){\line(0,-1){8.00}}
\put(116.00,46.00){\line(-1,0){23.00}}
\put(93.00,46.00){\line(0,1){8.00}}
\linethickness{1pt}
\put(102.00,97.00){\line(0,-1){2.00}}
\put(102.00,93.00){\line(0,-1){2.00}}
\put(102.00,89.00){\line(0,-1){2.00}}
\put(102.00,97.00){\line(1,0){2.00}}
\put(106.00,97.00){\line(1,0){2.00}}
\put(110.00,97.00){\line(1,0){2.00}}
\put(114.00,97.00){\line(1,0){2.00}}
\put(118.00,97.00){\line(1,0){2.00}}
\put(120.00,97.00){\line(0,-1){2.00}}
\put(120.00,93.00){\line(0,-1){2.00}}
\put(120.00,89.00){\line(0,-1){2.00}}
\put(116.00,87.00){\line(-1,0){2.00}}
\put(112.00,87.00){\line(-1,0){2.00}}
\put(108.00,87.00){\line(-1,0){2.00}}
\put(104.00,87.00){\line(-1,0){2.00}}
\put(118.00,87.00){\line(1,0){2.00}}
\linethickness{.5pt}
\put(165.00,240.00){\line(-1,0){20.00}}
\put(145.00,240.00){\line(0,-1){15.00}}
\put(120.00,225.00){\line(1,0){45.00}}
\put(165.00,195.00){\line(-1,0){20.00}}
\put(145.00,195.00){\line(0,-1){30.00}}
\put(145.00,165.00){\line(1,0){20.00}}
\put(120.00,180.00){\line(1,0){45.00}}
\put(145.00,150.00){\line(1,0){20.00}}
\put(145.00,120.00){\line(1,0){20.00}}
\put(120.00,135.00){\line(1,0){45.00}}
\put(145.00,150.00){\line(0,-1){30.00}}
\put(120.00,225.00){\line(0,-1){90.00}}
\put(116.00,190.00){\line(1,0){4.00}}
\linethickness{1pt}
\put(165.00,245.00){\line(1,0){20.00}}
\put(185.00,245.00){\line(0,-1){10.00}}
\put(185.00,235.00){\line(-1,0){20.00}}
\put(165.00,235.00){\line(0,1){10.00}}
\put(165.00,230.00){\line(1,0){20.00}}
\put(185.00,230.00){\line(0,-1){10.00}}
\put(185.00,220.00){\line(-1,0){20.00}}
\put(165.00,220.00){\line(0,1){10.00}}
\put(165.00,200.00){\line(1,0){20.00}}
\put(185.00,200.00){\line(0,-1){10.00}}
\put(185.00,190.00){\line(-1,0){20.00}}
\put(165.00,190.00){\line(0,1){10.00}}
\put(165.00,185.00){\line(1,0){20.00}}
\put(185.00,185.00){\line(0,-1){10.00}}
\put(185.00,175.00){\line(-1,0){20.00}}
\put(165.00,175.00){\line(0,1){10.00}}
\put(165.00,170.00){\line(1,0){20.00}}
\put(185.00,170.00){\line(0,-1){10.00}}
\put(185.00,160.00){\line(-1,0){20.00}}
\put(165.00,160.00){\line(0,1){10.00}}
\put(165.00,155.00){\line(1,0){20.00}}
\put(185.00,155.00){\line(0,-1){10.00}}
\put(185.00,145.00){\line(-1,0){20.00}}
\put(165.00,145.00){\line(0,1){10.00}}
\put(165.00,140.00){\line(1,0){20.00}}
\put(185.00,140.00){\line(0,-1){10.00}}
\put(185.00,130.00){\line(-1,0){20.00}}
\put(165.00,130.00){\line(0,1){10.00}}
\put(165.00,125.00){\line(1,0){20.00}}
\put(185.00,125.00){\line(0,-1){10.00}}
\put(185.00,115.00){\line(-1,0){20.00}}
\put(165.00,115.00){\line(0,1){10.00}}
\linethickness{0.5pt}
\put(65.00,190.00){\line(1,0){29.00}}
\put(65.00,190.00){\line(0,-1){140.00}}
\put(65.00,50.00){\line(1,0){27.00}}
\put(102.00,92.00){\line(-1,0){37.00}}
\put(49.00,57.00){\line(1,0){16.00}}
\linethickness{0.25pt}
\put(85.00,148.00){\line(0,1){42.00}}
\put(85.00,138.00){\line(0,-1){46.00}}
\put(122.00,70.00){\line(1,0){63.00}}
\put(122.00,54.00){\line(1,0){63.00}}
\put(122.00,30.00){\line(1,0){63.00}}
\put(149.00,39.00){\line(1,0){36.00}}
\put(149.00,19.00){\line(1,0){36.00}}
\put(149.00,39.00){\line(0,-1){20.00}}
\put(122.00,30.00){\line(0,1){40.00}}
\put(117.00,50.00){\line(1,0){5.00}}
\put(20.00,57.00){\circle{6.00}}
\put(20.00,57.00){\makebox(0,0)[cc]{3}}
\end{picture}
\vspace{-1cm}
\caption
{\it
Building structure functions in \he
\label{strufc}
}
\end{figure}
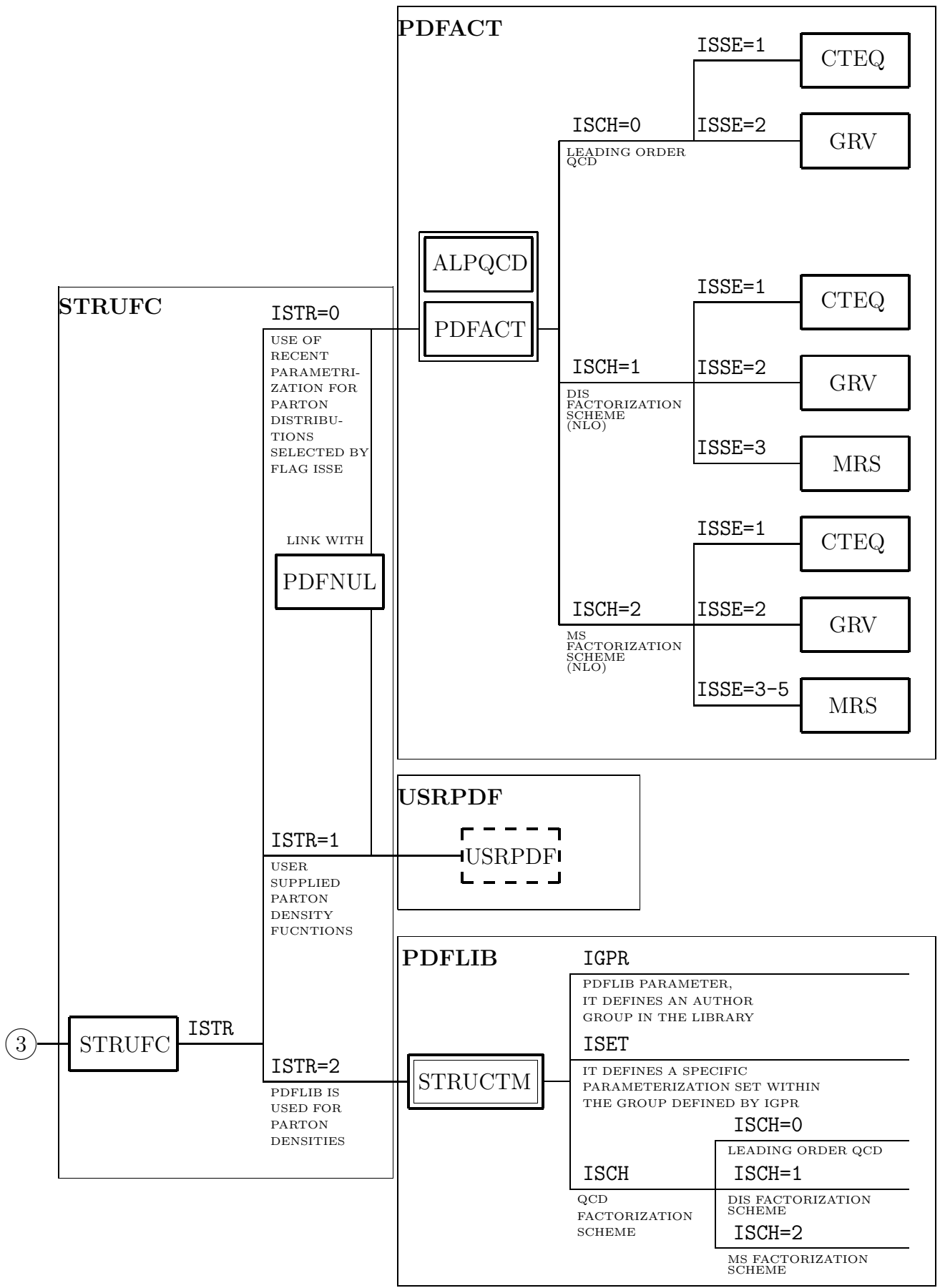

\begin{figure}
\unitlength=1.00mm
\linethickness{1pt}
\begin{picture}(168.00,240.00)(-20,-5)
\put(-7.00,55.00){\line(1,0){20.00}}
\put(13.00,55.00){\line(0,-1){10.00}}
\put(13.00,45.00){\line(-1,0){20.00}}
\put(-7.00,45.00){\line(0,1){10.00}}
\put(28.00,55.00){\line(1,0){20.00}}
\put(48.00,55.00){\line(0,-1){10.00}}
\put(48.00,45.00){\line(-1,0){20.00}}
\put(28.00,45.00){\line(0,1){10.00}}
\put(-7.00,35.00){\line(1,0){20.00}}
\put(13.00,35.00){\line(0,-1){10.00}}
\put(13.00,25.00){\line(-1,0){20.00}}
\put(-7.00,25.00){\line(0,1){10.00}}
\put(28.00,35.00){\line(1,0){20.00}}
\put(48.00,35.00){\line(0,-1){10.00}}
\put(48.00,25.00){\line(-1,0){20.00}}
\put(28.00,25.00){\line(0,1){10.00}}
\put(-7.00,95.00){\line(1,0){20.00}}
\put(13.00,95.00){\line(0,-1){10.00}}
\put(13.00,85.00){\line(-1,0){20.00}}
\put(-7.00,85.00){\line(0,1){10.00}}
\put(28.00,95.00){\line(1,0){20.00}}
\put(48.00,95.00){\line(0,-1){10.00}}
\put(48.00,85.00){\line(-1,0){20.00}}
\put(28.00,85.00){\line(0,1){10.00}}
\put(-7.00,75.00){\line(1,0){20.00}}
\put(13.00,75.00){\line(0,-1){10.00}}
\put(13.00,65.00){\line(-1,0){20.00}}
\put(-7.00,65.00){\line(0,1){10.00}}
\put(28.00,75.00){\line(1,0){20.00}}
\put(48.00,75.00){\line(0,-1){10.00}}
\put(48.00,65.00){\line(-1,0){20.00}}
\put(28.00,65.00){\line(0,1){10.00}}
\put(-7.00,195.00){\line(1,0){20.00}}
\put(13.00,195.00){\line(0,-1){10.00}}
\put(13.00,185.00){\line(-1,0){20.00}}
\put(-7.00,185.00){\line(0,1){10.00}}
\put(28.00,195.00){\line(1,0){20.00}}
\put(48.00,195.00){\line(0,-1){10.00}}
\put(48.00,185.00){\line(-1,0){20.00}}
\put(28.00,185.00){\line(0,1){10.00}}
\put(88.00,195.00){\line(1,0){20.00}}
\put(108.00,195.00){\line(0,-1){10.00}}
\put(108.00,185.00){\line(-1,0){20.00}}
\put(88.00,185.00){\line(0,1){10.00}}
\put(-7.00,215.00){\line(1,0){20.00}}
\put(13.00,215.00){\line(0,-1){10.00}}
\put(13.00,205.00){\line(-1,0){20.00}}
\put(-7.00,205.00){\line(0,1){10.00}}
\put(28.00,215.00){\line(1,0){20.00}}
\put(48.00,215.00){\line(0,-1){10.00}}
\put(48.00,205.00){\line(-1,0){20.00}}
\put(28.00,205.00){\line(0,1){10.00}}
\put(88.00,215.00){\line(1,0){20.00}}
\put(108.00,215.00){\line(0,-1){10.00}}
\put(108.00,205.00){\line(-1,0){20.00}}
\put(88.00,205.00){\line(0,1){10.00}}
\put(123.00,215.00){\line(1,0){20.00}}
\put(143.00,215.00){\line(0,-1){10.00}}
\put(143.00,205.00){\line(-1,0){20.00}}
\put(123.00,205.00){\line(0,1){10.00}}
\put(-7.00,175.00){\line(1,0){20.00}}
\put(13.00,175.00){\line(0,-1){10.00}}
\put(13.00,165.00){\line(-1,0){20.00}}
\put(-7.00,165.00){\line(0,1){10.00}}
\put(28.00,175.00){\line(1,0){20.00}}
\put(48.00,175.00){\line(0,-1){10.00}}
\put(48.00,165.00){\line(-1,0){20.00}}
\put(28.00,165.00){\line(0,1){10.00}}
\put(-7.00,155.00){\line(1,0){20.00}}
\put(13.00,155.00){\line(0,-1){10.00}}
\put(13.00,145.00){\line(-1,0){20.00}}
\put(-7.00,145.00){\line(0,1){10.00}}
\put(28.00,155.00){\line(1,0){20.00}}
\put(48.00,155.00){\line(0,-1){10.00}}
\put(48.00,145.00){\line(-1,0){20.00}}
\put(28.00,145.00){\line(0,1){10.00}}
\put(88.00,155.00){\line(1,0){20.00}}
\put(108.00,155.00){\line(0,-1){10.00}}
\put(108.00,145.00){\line(-1,0){20.00}}
\put(88.00,145.00){\line(0,1){10.00}}
\put(123.00,155.00){\line(1,0){20.00}}
\put(143.00,155.00){\line(0,-1){10.00}}
\put(143.00,145.00){\line(-1,0){20.00}}
\put(123.00,145.00){\line(0,1){10.00}}
\put(-7.00,135.00){\line(1,0){20.00}}
\put(13.00,135.00){\line(0,-1){10.00}}
\put(13.00,125.00){\line(-1,0){20.00}}
\put(-7.00,125.00){\line(0,1){10.00}}
\put(28.00,135.00){\line(1,0){20.00}}
\put(48.00,135.00){\line(0,-1){10.00}}
\put(48.00,125.00){\line(-1,0){20.00}}
\put(28.00,125.00){\line(0,1){10.00}}
\put(-7.00,115.00){\line(1,0){20.00}}
\put(13.00,115.00){\line(0,-1){10.00}}
\put(13.00,105.00){\line(-1,0){20.00}}
\put(-7.00,105.00){\line(0,1){10.00}}
\put(28.00,115.00){\line(1,0){20.00}}
\put(48.00,115.00){\line(0,-1){10.00}}
\put(48.00,105.00){\line(-1,0){20.00}}
\put(28.00,105.00){\line(0,1){10.00}}
\put(88.00,102.00){\line(1,0){20.00}}
\put(108.00,102.00){\line(0,-1){10.00}}
\put(108.00,92.00){\line(-1,0){20.00}}
\put(88.00,92.00){\line(0,1){10.00}}
\put(88.00,88.00){\line(1,0){20.00}}
\put(108.00,88.00){\line(0,-1){10.00}}
\put(108.00,78.00){\line(-1,0){20.00}}
\put(88.00,78.00){\line(0,1){10.00}}
\put(123.00,75.00){\line(1,0){20.00}}
\put(143.00,75.00){\line(0,-1){10.00}}
\put(143.00,65.00){\line(-1,0){20.00}}
\put(123.00,65.00){\line(0,1){10.00}}
\put(3.00,210.00){\makebox(0,0)[cc]{LEPNCi}}
\put(38.00,210.00){\makebox(0,0)[cc]{LEPNC}}
\put(3.00,190.00){\makebox(0,0)[cc]{JETNCi}}
\put(38.00,190.00){\makebox(0,0)[cc]{JETNC}}
\put(3.00,170.00){\makebox(0,0)[cc]{JETCCi}}
\put(38.00,170.00){\makebox(0,0)[cc]{JETCC}}
\put(3.00,150.00){\makebox(0,0)[cc]{MIXEDi}}
\put(38.00,150.00){\makebox(0,0)[cc]{MIXNC}}
\put(3.00,130.00){\makebox(0,0)[cc]{DOUBAi}}
\put(38.00,130.00){\makebox(0,0)[cc]{DOUBAN}}
\put(3.00,110.00){\makebox(0,0)[cc]{ANYJBi}}
\put(38.00,110.00){\makebox(0,0)[cc]{ANYJB}}
\put(133.00,70.00){\makebox(0,0)[cc]{DISEPW}}
\put(98.00,83.00){\makebox(0,0)[cc]{TERHIN}}
\put(98.00,97.00){\makebox(0,0)[cc]{TERADH}}
\put(98.00,150.00){\makebox(0,0)[cc]{TERADM}}
\put(133.00,150.00){\makebox(0,0)[cc]{DISEPM}}
\put(98.00,190.00){\makebox(0,0)[cc]{TERADJ}}
\put(98.00,210.00){\makebox(0,0)[cc]{TERADL}}
\put(133.00,210.00){\makebox(0,0)[cc]{DISEPL}}
\put(3.00,50.00){\makebox(0,0)[cc]{SIGMAi}}
\put(38.00,50.00){\makebox(0,0)[cc]{SIGMA}}
\put(3.00,30.00){\makebox(0,0)[cc]{XSIGMi}}
\put(38.00,30.00){\makebox(0,0)[cc]{XSIGMA}}
\put(3.00,70.00){\makebox(0,0)[cc]{LEPCCi}}
\put(38.00,70.00){\makebox(0,0)[cc]{LEPCC}}
\put(3.00,90.00){\makebox(0,0)[cc]{HADNCi}}
\put(38.00,90.00){\makebox(0,0)[cc]{HADNC}}
\put(20.00,228.00){\makebox(0,0)[cc]{\bf HELIOS}}
\put(98.00,228.00){\makebox(0,0)[cc]{\bf TERAD}}
\put(133.00,228.00){\makebox(0,0)[cc]{\bf DISEP}}
\put(68.00,212.00){\makebox(0,0)[cc]{\tt IMEA=1}}
\put(68.00,192.00){\makebox(0,0)[cc]{\tt IMEA=2}}
\put(68.00,172.00){\makebox(0,0)[cc]{\tt IMEA=3}}
\put(68.00,152.00){\makebox(0,0)[cc]{\tt IMEA=4}}
\put(68.00,132.00){\makebox(0,0)[cc]{\tt IMEA=5}}
\put(68.00,112.00){\makebox(0,0)[cc]{\tt IMEA=6}}
\put(68.00,92.00){\makebox(0,0)[cc]{\tt IMEA=7}}
\put(68.00,72.00){\makebox(0,0)[cc]{\tt IMEA=8}}
\put(98.00,72.00){\makebox(0,0)[cc]{\tt IDSP=2}}
\put(98.00,142.00){\makebox(0,0)[cc]{\tt IDSP=2}}
\put(98.00,202.00){\makebox(0,0)[cc]{\tt IDSP=2}}
\put(68.00,52.00){\makebox(0,0)[cc]{\tt IMEA=9}}
\put(68.00,32.00){\makebox(0,0)[cc]{\tt IMEA=10}}
\put(68.00,88.00){\makebox(0,0)[cc]{\tiny HADRONIC, NC}}
\put(68.00,208.00){\makebox(0,0)[cc]{\tiny LEPTONIC, NC}}
\put(68.00,188.00){\makebox(0,0)[cc]{\tiny JB, NC}}
\put(68.00,168.00){\makebox(0,0)[cc]{\tiny JB, CC}}
\put(68.00,148.00){\makebox(0,0)[cc]{\tiny MIXED, NC}}
\put(68.00,128.00){\makebox(0,0)[cc]{\tiny DOUBLE ANGLE}}
\put(68.00,125.00){\makebox(0,0)[cc]{\tiny METHOD, NC}}
\put(68.00,108.00){\makebox(0,0)[cc]{\tiny LEPTONIC ANGLE}}
\put(68.00,105.00){\makebox(0,0)[cc]{\tiny AND {\small
        $y_{_{JB}}$}, NC}}
\put(68.00,68.00){\makebox(0,0)[cc]{\tiny LEPTONIC, CC}}
\put(-7.00,16.00){\makebox(0,0)[lc]
{{{\small i=1:}{$\cal O$}\small$(\alpha^2 L^2)$ {\small QED} {\tiny
      CORRECTIONS}}}}
\put(68.00,48.00){\makebox(0,0)[cc]{\tiny $\Sigma$ - METHOD }}
\put(68.00,28.00){\makebox(0,0)[cc]{\tiny $e\Sigma$ - METHOD}}
\put(-7.00,12.00){\makebox(0,0)[lc]
{{{\small i=2:}{$\cal O$}\small$(\alpha^2 L^2)\,e^-e^+ $ - {\tiny
      CONVERSION}}}}
\linethickness{0.5pt}
\put(3.00,205.00){\line(0,-1){5.00}}
\put(53.00,200.00){\line(0,1){10.00}}
\put(3.00,185.00){\line(0,-1){5.00}}
\put(53.00,180.00){\line(0,1){10.00}}
\put(3.00,165.00){\line(0,-1){5.00}}
\put(3.00,145.00){\line(0,-1){5.00}}
\put(3.00,125.00){\line(0,-1){5.00}}
\put(3.00,105.00){\line(0,-1){5.00}}
\put(53.00,160.00){\line(0,1){10.00}}
\put(53.00,140.00){\line(0,1){10.00}}
\put(53.00,120.00){\line(0,1){10.00}}
\put(53.00,100.00){\line(0,1){10.00}}
\put(3.00,200.00){\line(1,0){50.00}}
\put(53.00,180.00){\line(-1,0){50.00}}
\put(3.00,160.00){\line(1,0){50.00}}
\put(3.00,140.00){\line(1,0){50.00}}
\put(3.00,120.00){\line(1,0){50.00}}
\put(3.00,100.00){\line(1,0){50.00}}
\put(48.00,210.00){\line(1,0){40.00}}
\put(48.00,190.00){\line(1,0){40.00}}
\put(48.00,170.00){\line(1,0){35.00}}
\put(48.00,150.00){\line(1,0){40.00}}
\put(48.00,130.00){\line(1,0){35.00}}
\put(48.00,110.00){\line(1,0){35.00}}
\put(83.00,90.00){\line(-1,0){35.00}}
\put(123.00,70.00){\line(-1,0){75.00}}
\put(133.00,145.00){\line(0,-1){5.00}}
\put(133.00,140.00){\line(-1,0){50.00}}
\put(83.00,140.00){\line(0,1){10.00}}
\put(133.00,205.00){\line(0,-1){5.00}}
\put(133.00,200.00){\line(-1,0){50.00}}
\put(83.00,200.00){\line(0,1){10.00}}
\put(3.00,65.00){\line(0,-1){5.00}}
\put(53.00,60.00){\line(0,1){10.00}}
\put(3.00,60.00){\line(1,0){50.00}}
\put(3.00,85.00){\line(0,-1){5.00}}
\put(53.00,80.00){\line(0,1){10.00}}
\put(3.00,80.00){\line(1,0){50.00}}
\put(88.00,83.00){\line(-1,0){5.00}}
\put(83.00,83.00){\line(0,1){14.00}}
\put(83.00,97.00){\line(1,0){5.00}}
\put(3.00,45.00){\line(0,-1){5.00}}
\put(53.00,40.00){\line(0,1){10.00}}
\put(3.00,40.00){\line(1,0){50.00}}
\put(48.00,50.00){\line(1,0){35.00}}
\put(3.00,25.00){\line(0,-1){5.00}}
\put(53.00,20.00){\line(0,1){10.00}}
\put(3.00,20.00){\line(1,0){50.00}}
\put(48.00,30.00){\line(1,0){35.00}}
\linethickness{0.1pt}
\put(51.00,235.00){\line(0,-1){180.10}}
\put(-9.86,235.00){\line(1,0){60.95}}
\put(145.86,235.00){\line(0,-1){180.00}}
\put(84.90,55.00){\line(0,1){180.00}}
\put(111.00,235.00){\line(-1,0){26.10}}
\put(111.00,235.00){\line(0,-1){180.00}}
\put(111.10,55.00){\line(-1,0){26.19}}
\put(120.00,55.00){\line(1,0){25.86}}
\put(146.00,235.00){\line(-1,0){25.86}}
\put(120.14,235.00){\line(0,-1){180.00}}
\put(-10.00,10.00){\line(1,0){61.00}}
\put(51.00,10.00){\line(0,1){46.00}}
\put(-10.00,235.00){\line(0,-1){0.11}}
\put(-10.00,10.00){\line(0,1){0.11}}
\put(-10.00,235.00){\line(0,-1){224.83}}
\end{picture}

\vspace{-1cm}
\caption
[\it
The branches of \he
]
{
\it
The branches of \he.
Different treatments of QED corrections are
chosen by settings of flags {\tt IMEA} and {\tt IDSP}.
\label{branches}
}
\end{figure}
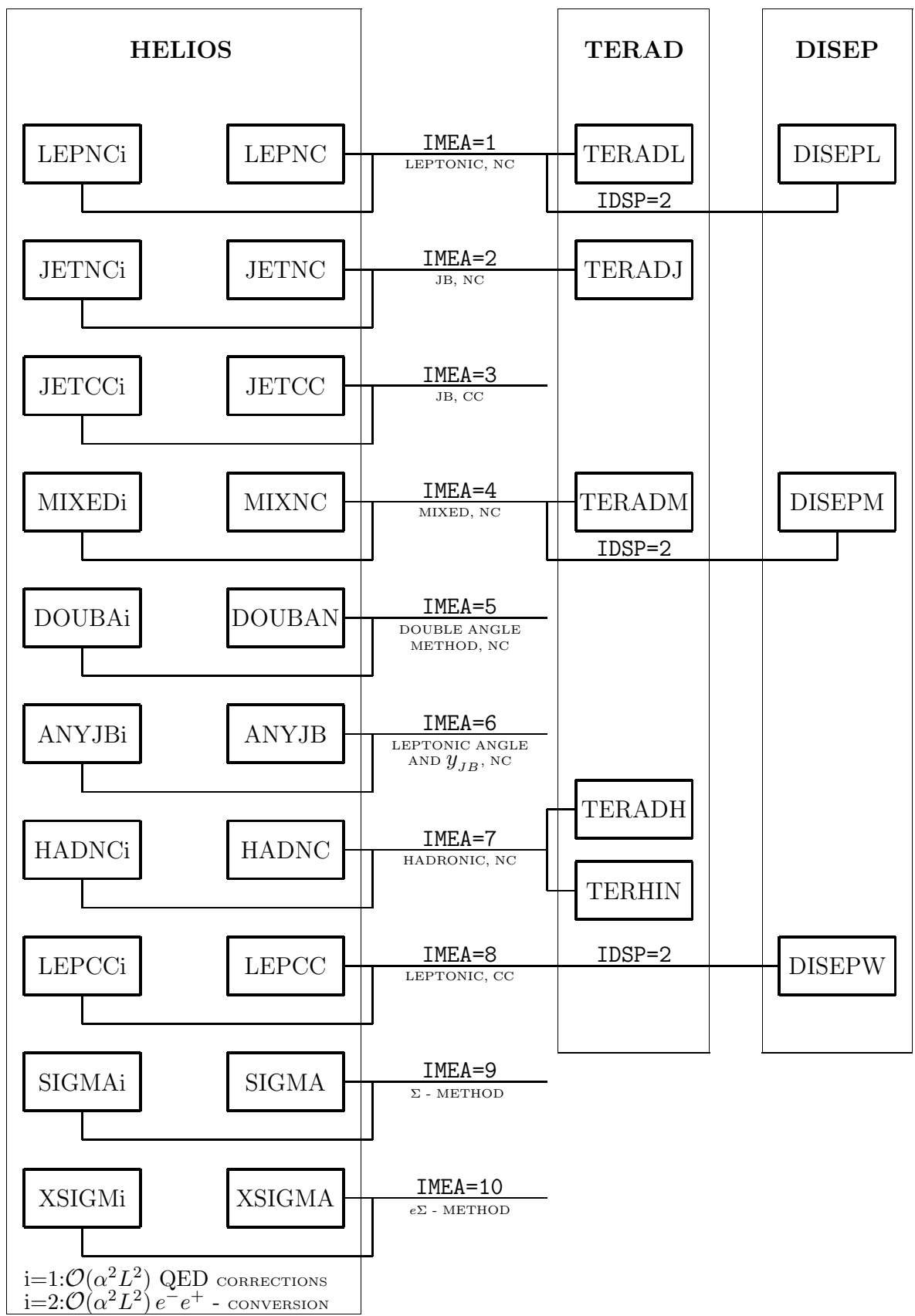
\clearpage

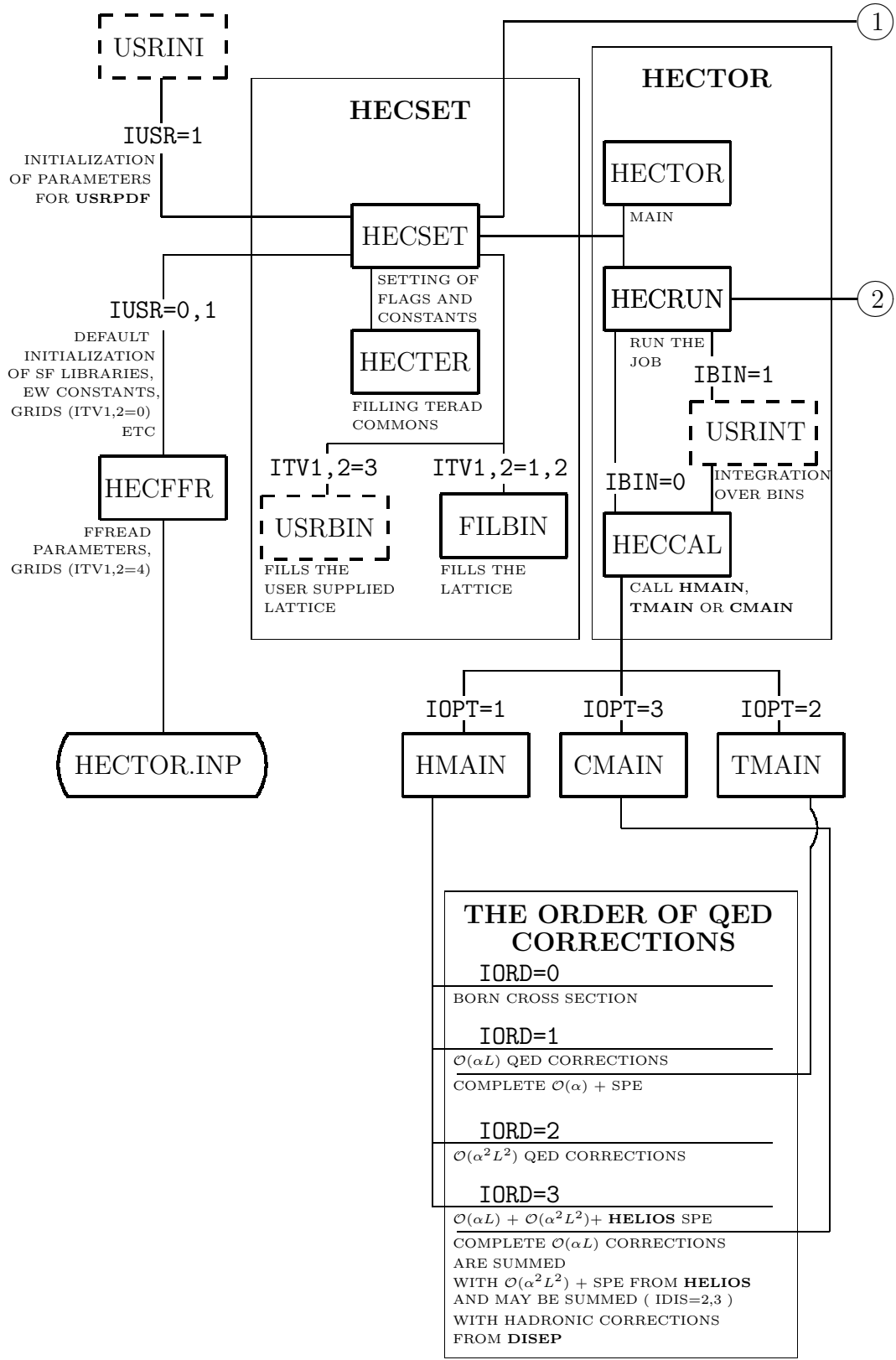
\begin{figure}
\unitlength=1.00mm
\linethickness{1pt}
\begin{picture}(168.00,240.00)(-15,-5)
\put(26.33,254.00){\line(1,0){2.00}}
\put(30.33,254.00){\line(1,0){2.00}}
\put(34.33,254.00){\line(1,0){2.00}}
\put(38.33,254.00){\line(1,0){2.00}}
\put(42.33,254.00){\line(1,0){2.00}}
\put(46.33,254.00){\line(0,-1){2.00}}
\put(46.33,250.00){\line(0,-1){2.00}}
\put(46.33,246.00){\line(0,-1){2.00}}
\put(46.33,244.00){\line(-1,0){2.00}}
\put(42.33,244.00){\line(-1,0){2.00}}
\put(38.33,244.00){\line(-1,0){2.00}}
\put(34.33,244.00){\line(-1,0){2.00}}
\put(30.33,244.00){\line(-1,0){2.00}}
\put(26.33,254.00){\line(0,-1){2.00}}
\put(26.33,250.00){\line(0,-1){2.00}}
\put(26.33,246.00){\line(0,-1){2.00}}
\put(106.33,214.00){\line(1,0){20.00}}
\put(126.33,214.00){\line(0,-1){10.00}}
\put(126.33,204.00){\line(-1,0){20.00}}
\put(106.33,204.00){\line(0,1){10.00}}
\put(106.33,175.00){\line(1,0){20.00}}
\put(126.33,175.00){\line(0,-1){10.00}}
\put(126.33,165.00){\line(-1,0){20.00}}
\put(106.33,165.00){\line(0,1){10.00}}
\put(106.33,234.00){\line(1,0){20.00}}
\put(126.33,234.00){\line(0,-1){10.00}}
\put(126.33,224.00){\line(-1,0){20.00}}
\put(106.33,224.00){\line(0,1){10.00}}
\put(66.33,224.00){\line(1,0){20.00}}
\put(86.33,224.00){\line(0,-1){10.00}}
\put(86.33,214.00){\line(-1,0){20.00}}
\put(66.33,214.00){\line(0,1){10.00}}
\put(66.33,204.00){\line(1,0){20.00}}
\put(86.33,204.00){\line(0,-1){10.00}}
\put(86.33,194.00){\line(-1,0){20.00}}
\put(66.33,194.00){\line(0,1){10.00}}
\put(26.33,184.00){\line(1,0){20.00}}
\put(46.33,184.00){\line(0,-1){10.00}}
\put(46.33,174.00){\line(-1,0){20.00}}
\put(26.33,174.00){\line(0,1){10.00}}
\put(21.33,140.00){\line(1,0){20.00}}
\put(41.33,130.00){\line(-1,0){20.00}}
\put(41.33,140.00){\line(1,0){10.00}}
\put(41.33,130.00){\line(1,0){10.00}}
\put(106.33,214.00){\line(1,0){20.00}}
\put(126.33,214.00){\line(0,-1){10.00}}
\put(126.33,204.00){\line(-1,0){20.00}}
\put(106.33,204.00){\line(0,1){10.00}}
\put(124.33,140.00){\line(1,0){20.00}}
\put(144.33,140.00){\line(0,-1){10.00}}
\put(144.33,130.00){\line(-1,0){20.00}}
\put(124.33,130.00){\line(0,1){10.00}}
\put(124.33,140.00){\line(1,0){20.00}}
\put(144.33,140.00){\line(0,-1){10.00}}
\put(144.33,130.00){\line(-1,0){20.00}}
\put(124.33,130.00){\line(0,1){10.00}}
\put(99.33,140.00){\line(1,0){20.00}}
\put(99.33,130.00){\line(0,1){10.00}}
\put(99.33,130.00){\line(0,1){10.00}}
\put(99.33,140.00){\line(1,0){20.00}}
\put(119.33,140.00){\line(0,-1){10.00}}
\put(119.33,130.00){\line(-1,0){20.00}}
\put(119.33,140.00){\line(0,-1){10.00}}
\put(119.33,130.00){\line(-1,0){20.00}}
\put(74.33,140.00){\line(1,0){20.00}}
\put(94.33,140.00){\line(0,-1){10.00}}
\put(94.33,130.00){\line(-1,0){20.00}}
\put(74.33,130.00){\line(0,1){10.00}}
\put(74.33,140.00){\line(1,0){20.00}}
\put(94.33,140.00){\line(0,-1){10.00}}
\put(94.33,130.00){\line(-1,0){20.00}}
\put(74.33,130.00){\line(0,1){10.00}}
\put(80.33,178.00){\line(1,0){20.00}}
\put(100.33,178.00){\line(0,-1){10.00}}
\put(100.33,168.00){\line(-1,0){20.00}}
\put(80.33,168.00){\line(0,1){10.00}}
\bezier{48}(51.00,140.00)(54.00,135.00)(51.00,130.00)
\bezier{48}(21.00,140.00)(18.00,135.00)(21.00,130.00)
\put(76.33,239.00){\makebox(0,0) [cc]{\bf HECSET     }}
\put(123.33,244.00){\makebox(0,0)[cc]{\bf HECTOR     }}
\put(109.33,111.00){\makebox(0,0)[cc]{\bf THE ORDER OF QED }}
\put(109.33,107.00){\makebox(0,0)[cc]{\bf CORRECTIONS}}
\put(76.33,199.00){\makebox(0,0) [cc]{HECTER}}
\put(61.98,172.71){\makebox(0,0) [cc]{USRBIN}}
\put(76.33,219.00){\makebox(0,0) [cc]{HECSET}}
\put(116.33,229.00){\makebox(0,0)[cc]{HECTOR}}
\put(116.33,209.00){\makebox(0,0)[cc]{HECRUN}}
\put(116.33,170.00){\makebox(0,0)[cc]{HECCAL}}
\put(130.33,188.00){\makebox(0,0)[cc]{USRINT}}
\put(36.33,249.00){\makebox(0,0)[cc]{USRINI }}
\put(36.33,179.00){\makebox(0,0)[cc]{HECFFR }}
\put(36.33,135.00){\makebox(0,0)[cc]{HECTOR.INP }}
\put(116.33,209.00){\makebox(0,0)[cc]{HECRUN}}
\put(134.33,135.00){\makebox(0,0)[cc]{TMAIN }}
\put(109.33,135.00){\makebox(0,0)[cc]{CMAIN }}
\put(84.33,135.00){\makebox(0,0) [cc]{HMAIN }}
\put(90.33,173.00){\makebox(0,0) [cc]{FILBIN}}
\put(134.33,144.00){\makebox(0,0)[cc]{\tt IOPT=2}}
\put(109.33,144.00){\makebox(0,0)[cc]{\tt IOPT=3}}
\put(84.33,144.00){\makebox(0,0) [cc]{\tt IOPT=1}}
\put(86.33,102.00){\makebox(0,0) [lc]{\tt IORD=0}}
\put(86.33,92.00){\makebox(0,0)  [lc]{\tt IORD=1}}
\put(86.33,77.00){\makebox(0,0)  [lc]{\tt IORD=2}}
\put(86.33,67.00){\makebox(0,0)  [lc]{\tt IORD=3}}
\put(106.33,180.00){\makebox(0,0)[lc]{\tt IBIN=0}}
\put(120.33,197.00){\makebox(0,0)[lc]{\tt IBIN=1}}
\put(61.77,182.00) {\makebox(0,0)[cc]{\tt ITV1,2=3}}
\put(89.77,182.00) {\makebox(0,0)[cc]{\tt ITV1,2=1,2}}
\put(36.33,207.00){\makebox(0,0) [cc]{\tt IUSR=0,1}}
\put(110.33,202.00){\makebox(0,0)[lc]{\tiny RUN THE }}
\put(110.33,199.00){\makebox(0,0)[lc]{\tiny JOB     }}
\put(110.33,163.00){\makebox(0,0)[lc]{\tiny CALL {\bf HMAIN},       }}
\put(110.33,160.00){\makebox(0,0)[lc]{\tiny {\bf TMAIN} OR {\bf CMAIN}}}
\put(123.77,181.00){\makebox(0,0)[lc]{\tiny  INTEGRATION            }}
\put(123.77,178.00){\makebox(0,0)[lc]{\tiny OVER BINS               }}
\put(70.33,212.00) {\makebox(0,0)[lc]{\tiny SETTING OF              }}
\put(70.33,209.00) {\makebox(0,0)[lc]{\tiny FLAGS AND               }}
\put(70.33,206.00) {\makebox(0,0)[lc]{\tiny CONSTANTS               }}
\put(110.33,222.00){\makebox(0,0)[lc]{\tiny MAIN                    }}
\put(36.00,231.00){\makebox(0,0)[rc]{\tiny INITIALIZATION           }}
\put(35.33,228.00){\makebox(0,0)[rc]{\tiny OF PARAMETERS           }}
\put(35.33,225.00){\makebox(0,0)[rc]{\tiny FOR {\bf USRPDF        }}}
\put(35.00,203.00){\makebox(0,0)[rc]{\tiny DEFAULT                 }}
\put(36.00,200.00){\makebox(0,0)[rc]{\tiny INITIALIZATION          }}
\put(35.73,197.00){\makebox(0,0)[rc]{\tiny OF SF LIBRARIES,        }}
\put(36.73,194.00){\makebox(0,0)[rc]{\tiny EW CONSTANTS,           }}
\put(35.33,191.00){\makebox(0,0)[rc]{\tiny GRIDS (ITV1,2=0)        }}
\put(36.00,188.00){\makebox(0,0)[rc]{\tiny ETC                     }}
\put(35.33,172.00){\makebox(0,0)[rc]{\tiny FFREAD                  }}
\put(35.33,169.00){\makebox(0,0)[rc]{\tiny PARAMETERS,             }}
\put(35.33,166.00){\makebox(0,0)[rc]{\tiny GRIDS (ITV1,2=4)        }}
\put(82.33,98.00){\makebox(0,0)[lc]{\tiny BORN CROSS SECTION}}
\put(82.33,88.00){\makebox(0,0)[lc]
                        {\tiny ${\cal O}(\alpha L)$ QED CORRECTIONS }}
\put(82.33,84.00){\makebox(0,0)[lc]{\tiny COMPLETE ${\cal O}(\alpha)$
         +  SPE      }}
\put(82.33,73.00){\makebox(0,0)[lc]
                    {\tiny ${\cal O}(\alpha^2 L^2)$ QED CORRECTIONS }}
\put(82.33,63.00){\makebox(0,0)[lc]{\tiny  ${\cal O}(\alpha L )$ +
                         ${\cal O}(\alpha^2 L^2)$+ {\bf HELIOS}
SPE                        }}
\put(82.33,59.00){\makebox(0,0)[lc]{\tiny COMPLETE ${\cal O}(\alpha
                                                  L)$ CORRECTIONS   }}
\put(82.33,53.00){\makebox(0,0)[lc]{\tiny WITH ${\cal O}(\alpha^2
                                        L^2)$ + SPE
          FROM {\bf HELIOS}}}
\put(82.33,50.00){\makebox(0,0)[lc]{\tiny AND MAY BE SUMMED (
                                                        IDIS=2,3 ) }}
\put(80.33,166.00){\makebox(0,0)[lc]{\tiny FILLS THE}}
\put(80.33,163.00){\makebox(0,0)[lc]{\tiny LATTICE}}
\put(52.33,166.00){\makebox(0,0)[lc]{\tiny FILLS THE}}
\put(52.33,163.00){\makebox(0,0)[lc]{\tiny USER SUPPLIED}}
\put(52.33,160.00){\makebox(0,0)[lc]{\tiny LATTICE}}
\put(66.33,192.00){\makebox(0,0)[lc]{\tiny FILLING TERAD}}
\put(66.33,189.00){\makebox(0,0)[lc]{\tiny COMMONS}}
\put(82.33,47.00){\makebox(0,0)[lc] {\tiny WITH HADRONIC CORRECTIONS }}
\put(82.33,56.00){\makebox(0,0)[lc] {\tiny ARE SUMMED}}
\put(82.33,44.00){\makebox(0,0)[lc] {\tiny  FROM {\bf DISEP} }}
\linethickness{.5pt}
\bezier{32}(139.00,122.00)(141.00,125.00)(139.00,128.00)
\put(109.33,224.00){\line(0,-1){10.00}}
\put(86.33,219.00){\line(1,0){23.00}}
\put(36.33,216.00){\line(1,0){30.00}}
\put(36.33,184.00){\line(0,1){20.00}}
\put(36.33,209.00){\line(0,1){7.00}}
\put(36.33,174.00){\line(0,-1){34.00}}
\put(69.33,214.00){\line(0,-1){10.00}}
\put(126.33,209.00){\line(1,0){20.00}}
\put(149.33,209.00){\circle{6.00}}
\put(90.33,216.00){\line(-1,0){4.00}}
\put(90.33,216.00){\line(0,-1){30.00}}
\put(90.33,186.00){\line(0,-1){1.50}}
\put(90.33,186.00){\line(-1,0){28.00}}
\put(62.33,186.00){\line(0,-1){1.50}}
\put(90.33,178.00){\line(0,1){1.50}}
\put(62.33,178.00){\line(0,1){1.50}}
\put(123.33,200.00){\line(0,1){4.00}}
\put(123.33,195.00){\line(0,-1){2.00}}
\put(123.33,182.70){\line(0,-1){7.70}}
\put(86.33,222.00){\line(1,0){4.00}}
\put(90.33,222.00){\line(0,1){31.00}}
\put(90.33,253.00){\line(1,0){56.00}}
\put(149.33,253.00){\circle{6.00}}
\linethickness{.1pt}
\put(104.33,155.00){\line(1,0){38.00}}
\put(142.33,155.00){\line(0,1){94.00}}
\put(142.33,249.00){\line(-1,0){38.00}}
\put(104.33,155.00){\line(0,1){94.00}}
\put(102.33,155.00){\line(0,1){89.00}}
\put(102.33,244.00){\line(-1,0){52.00}}
\put(50.33,244.00){\line(0,-1){89.00}}
\put(50.33,155.00){\line(1,0){52.00}}
\linethickness{.5pt}
\put(108.00,175.00){\line(0,1){3.00}}
\put(108.00,183.00){\line(0,1){21.00}}
\put(149.00,209.00){\makebox(0,0)[cc]{{ 2}}}
\put(149.00,253.00){\makebox(0,0)[cc]{{ 1}}}
\linethickness{1pt}
\put(51.98,177.71){\line(1,0){2.00}}
\put(55.98,177.71){\line(1,0){2.00}}
\put(59.98,177.71){\line(1,0){2.00}}
\put(63.98,177.71){\line(1,0){2.00}}
\put(67.98,177.71){\line(1,0){2.00}}
\put(71.98,177.71){\line(0,-1){2.00}}
\put(71.98,173.71){\line(0,-1){2.00}}
\put(71.98,169.71){\line(0,-1){2.00}}
\put(71.98,167.71){\line(-1,0){2.00}}
\put(67.98,167.71){\line(-1,0){2.00}}
\put(63.98,167.71){\line(-1,0){2.00}}
\put(59.98,167.71){\line(-1,0){2.00}}
\put(55.98,167.71){\line(-1,0){2.00}}
\put(51.98,177.71){\line(0,-1){2.00}}
\put(51.98,173.71){\line(0,-1){2.00}}
\put(51.98,169.71){\line(0,-1){2.00}}
\put(119.98,192.71){\line(1,0){2.00}}
\put(123.98,192.71){\line(1,0){2.00}}
\put(127.98,192.71){\line(1,0){2.00}}
\put(131.98,192.71){\line(1,0){2.00}}
\put(135.98,192.71){\line(1,0){2.00}}
\put(139.98,192.71){\line(0,-1){2.00}}
\put(139.98,188.71){\line(0,-1){2.00}}
\put(139.98,184.71){\line(0,-1){2.00}}
\put(139.98,182.71){\line(-1,0){2.00}}
\put(135.98,182.71){\line(-1,0){2.00}}
\put(131.98,182.71){\line(-1,0){2.00}}
\put(127.98,182.71){\line(-1,0){2.00}}
\put(123.98,182.71){\line(-1,0){2.00}}
\put(119.98,192.71){\line(0,-1){2.00}}
\put(119.98,188.71){\line(0,-1){2.00}}
\put(119.98,184.71){\line(0,-1){2.00}}
\linethickness{.5pt}
\put(109.00,165.00){\line(0,-1){15.00}}
\put(84.00,150.00){\line(1,0){50.00}}
\put(134.00,150.00){\line(0,-1){4.00}}
\put(134.00,140.00){\line(0,1){2.00}}
\put(109.00,142.00){\line(0,-1){2.00}}
\put(84.00,140.00){\line(0,1){2.00}}
\put(84.00,146.00){\line(0,1){4.00}}
\put(109.00,150.00){\line(0,-1){4.00}}
\put(66.00,222.00){\line(-1,0){30.00}}
\put(36.00,222.00){\line(0,1){10.00}}
\put(36.00,244.00){\line(0,-1){7.00}}
\linethickness{.25pt}
\put(83.00,86.00){\line(1,0){56.00}}
\put(109.00,130.00){\line(0,-1){5.00}}
\put(79.00,65.00){\line(1,0){54.00}}
\put(79.00,75.00){\line(1,0){54.00}}
\put(133.00,90.00){\line(-1,0){54.00}}
\put(79.00,100.00){\line(1,0){54.00}}
\put(79.00,130.00){\line(0,-1){65.00}}
\put(83.00,61.00){\line(1,0){59.00}}
\put(142.00,61.00){\line(0,1){64.00}}
\put(142.00,125.00){\line(-1,0){33.00}}
\put(139.00,130.00){\line(0,-1){2.00}}
\put(139.00,122.00){\line(0,-1){36.00}}
\linethickness{.1pt}
\put(137.00,115.00){\line(-1,0){56.00}}
\put(81.00,115.00){\line(0,-1){74.00}}
\put(81.00,41.00){\line(1,0){56.00}}
\put(137.00,41.00){\line(0,1){74.00}}
\put(36.33,235.00){\makebox(0,0)[cc]{\tt IUSR=1}}
\end{picture}
\vspace{-4cm}
\caption
{\it
Input organization and choice of QED treatments
\label{maip}
}
\end{figure}

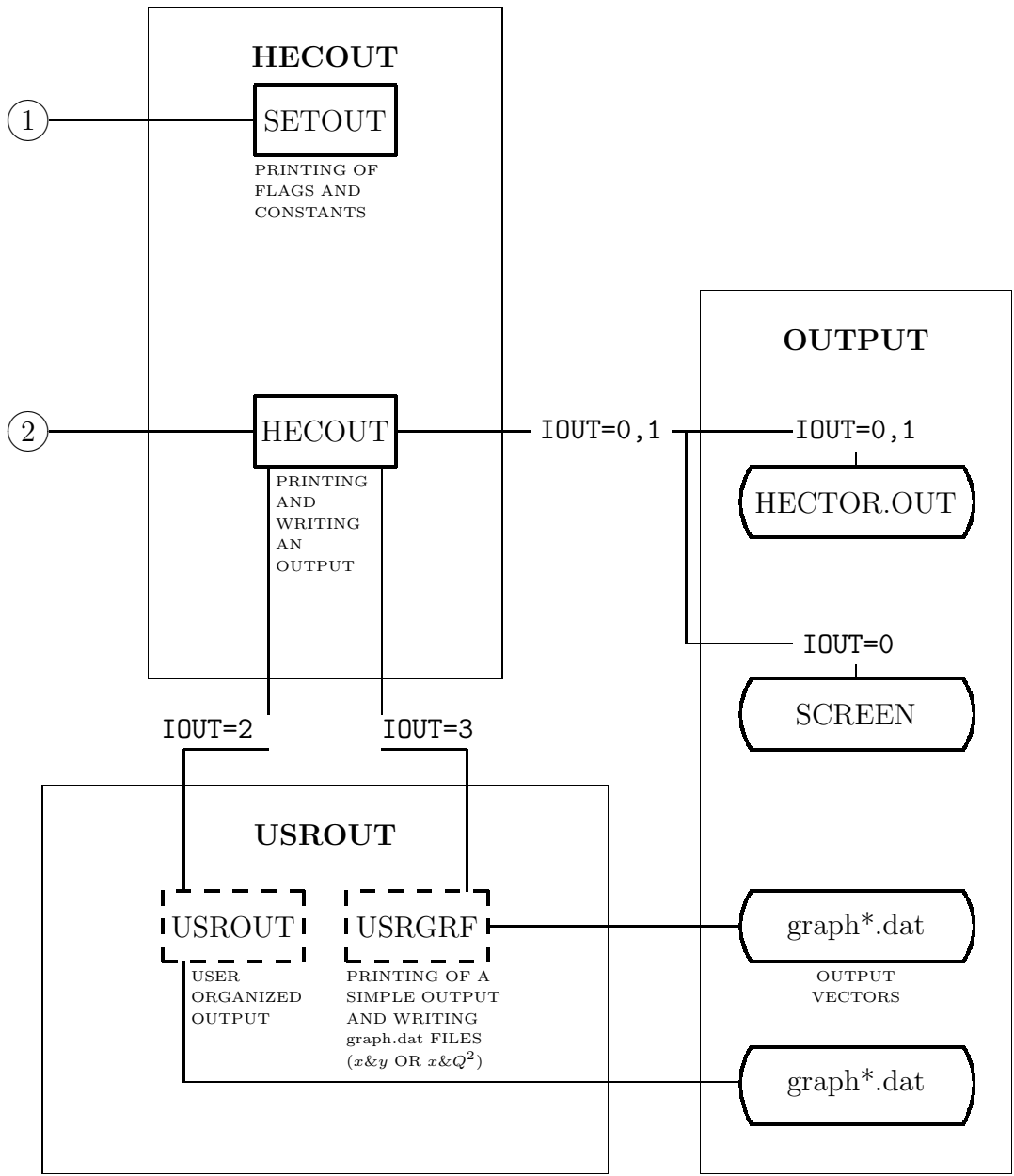
\begin{figure}
\unitlength=1.00mm
\linethickness{1pt}
\begin{picture}(168.00,240.00)(-10,-85)
\put(40.00,144.00){\line(1,0){20.00}}
\put(60.00,144.00){\line(0,1){10.00}}
\put(60.00,154.00){\line(-1,0){20.00}}
\put(40.00,154.00){\line(0,-1){10.00}}
\put(40.00,100.00){\line(0,1){10.00}}
\put(40.00,110.00){\line(1,0){20.00}}
\put(60.00,110.00){\line(0,-1){10.00}}
\put(60.00,100.00){\line(-1,0){20.00}}
\put(110.00,100.00){\line(1,0){30.00}}
\put(140.00,90.00){\line(-1,0){30.00}}
\put(110.00,40.00){\line(1,0){30.00}}
\put(140.00,30.00){\line(-1,0){30.00}}
\put(110.00,70.00){\line(1,0){30.00}}
\put(140.00,60.00){\line(-1,0){30.00}}
\put(110.00,18.00){\line(1,0){30.00}}
\put(140.00,7.00){\line(-1,0){30.00}}
\put(53.00,30.00){\line(1,0){2.00}}
\put(57.00,30.00){\line(1,0){2.00}}
\put(61.00,30.00){\line(1,0){2.00}}
\put(65.00,30.00){\line(1,0){2.00}}
\put(69.00,30.00){\line(1,0){2.00}}
\put(73.00,30.00){\line(0,1){2.00}}
\put(73.00,34.00){\line(0,1){2.00}}
\put(73.00,38.00){\line(0,1){2.00}}
\put(73.00,40.00){\line(-1,0){2.00}}
\put(69.00,40.00){\line(-1,0){2.00}}
\put(65.00,40.00){\line(-1,0){2.00}}
\put(61.00,40.00){\line(-1,0){2.00}}
\put(57.00,40.00){\line(-1,0){2.00}}
\put(53.00,40.00){\line(0,-1){2.00}}
\put(53.00,36.00){\line(0,-1){2.00}}
\put(53.00,32.00){\line(0,-1){2.00}}
\put(27.00,30.00){\line(1,0){2.00}}
\put(31.00,30.00){\line(1,0){2.00}}
\put(35.00,30.00){\line(1,0){2.00}}
\put(39.00,30.00){\line(1,0){2.00}}
\put(43.00,30.00){\line(1,0){2.00}}
\put(47.00,30.00){\line(0,1){2.00}}
\put(47.00,34.00){\line(0,1){2.00}}
\put(47.00,38.00){\line(0,1){2.00}}
\put(47.00,40.00){\line(-1,0){2.00}}
\put(43.00,40.00){\line(-1,0){2.00}}
\put(39.00,40.00){\line(-1,0){2.00}}
\put(35.00,40.00){\line(-1,0){2.00}}
\put(31.00,40.00){\line(-1,0){2.00}}
\put(27.00,40.00){\line(0,-1){2.00}}
\put(27.00,36.00){\line(0,-1){2.00}}
\put(27.00,32.00){\line(0,-1){2.00}}
\bezier{48}(140.00,100.00)(143.00,95.00)(140.00,90.00)
\bezier{48}(140.00,70.00)(143.00,65.00)(140.00,60.00)
\bezier{48}(140.00,40.00)(143.00,35.00)(140.00,30.00)
\bezier{48}(140.00,18.00)(143.00,13.00)(140.00,7.00)
\bezier{48}(110.00,100.00)(107.00,95.00)(110.00,90.00)
\bezier{48}(110.00,70.00)(107.00,65.00)(110.00,60.00)
\bezier{48}(110.00,40.00)(107.00,35.00)(110.00,30.00)
\bezier{48}(110.00,18.00)(107.00,13.00)(110.00,7.00)
\put(125.00,95.00){\makebox(0,0)[cc]{HECTOR.OUT}}
\put(125.00,65.00){\makebox(0,0)[cc]{SCREEN}}
\put(125.00,35.00){\makebox(0,0)[cc]{graph*.dat}}
\put(125.00,13.00){\makebox(0,0)[cc]{graph*.dat}}
\put(50.00,149.00){\makebox(0,0)[cc]{SETOUT}}
\put(50.00,105.00){\makebox(0,0)[cc]{HECOUT}}
\put(63.00,35.00) {\makebox(0,0)[cc]{USRGRF}}
\put(37.00,35.00) {\makebox(0,0)[cc]{USROUT}}
\put(125.00,118.00){\makebox(0,0)[cc]{\bf OUTPUT}}
\put(50.00,158.00) {\makebox(0,0)[cc]{\bf HECOUT}}
\put(50.00,48.00)  {\makebox(0,0)[cc]{\bf USROUT}}
\put(40.00,142.00){\makebox(0,0)[lc]{\tiny PRINTING OF}}
\put(40.00,139.00){\makebox(0,0)[lc]{\tiny FLAGS AND }}
\put(40.00,136.00){\makebox(0,0)[lc]{\tiny CONSTANTS}}
\put(43.00,98.00) {\makebox(0,0)[lc]{\tiny PRINTING }}
\put(43.00,95.00) {\makebox(0,0)[lc]{\tiny AND  }}
\put(43.00,92.00) {\makebox(0,0)[lc]{\tiny WRITING}}
\put(53.00,28.00) {\makebox(0,0)[lc]{\tiny PRINTING OF A}}
\put(53.00,25.00) {\makebox(0,0)[lc]{\tiny SIMPLE OUTPUT}}
\put(53.00,16.00) {\makebox(0,0)[lc]{\tiny ($x\&y$ OR $x\&Q^2$)}}
\put(43.00,89.00) {\makebox(0,0)[lc]{\tiny AN}}
\put(53.00,22.00) {\makebox(0,0)[lc]{\tiny AND WRITING}}
\put(53.00,19.00) {\makebox(0,0)[lc]{\tiny graph.dat FILES}}
\put(43.00,86.00){\makebox(0,0)[lc]{\tiny OUTPUT   }}
\put(31.00,28.00){\makebox(0,0)[lc]{\tiny USER     }}
\put(31.00,25.00){\makebox(0,0)[lc]{\tiny ORGANIZED}}
\put(31.00,22.00){\makebox(0,0)[lc]{\tiny OUTPUT   }}
\put(125.00,28.00){\makebox(0,0)[cc]{\tiny OUTPUT}}
\put(125.00,25.00){\makebox(0,0)[cc]{\tiny VECTORS}}
\put(58.00,63.00)  {\makebox(0,0)[lc]{\tt IOUT=3  }}
\put(42.00,63.00)  {\makebox(0,0)[rc]{\tt IOUT=2  }}
\put(89.00,105.00) {\makebox(0,0)[cc]{\tt IOUT=0,1}}
\put(125.00,75.00) {\makebox(0,0)[cc]{\tt IOUT=0  }}
\put(125.00,105.00){\makebox(0,0)[cc]{\tt IOUT=0,1}}
\linethickness{0.5pt}
\put(8.00,149.00){\circle{6.00}}
\put(8.00,105.00){\circle{6.00}}
\linethickness{0.1pt}
\put(103.00,10.00){\line(0,1){115.00}}
\put(103.00,125.00){\line(1,0){44.00}}
\put(147.00,125.00){\line(0,-1){115.00}}
\put(25.00,70.00){\line(1,0){50.00}}
\put(90.00,55.00){\line(0,-1){45.00}}
\put(10.00,10.00){\line(0,1){45.00}}
\linethickness{.5pt}
\put(42.00,100.00){\line(0,-1){35.00}}
\put(58.00,65.00){\line(0,1){0.00}}
\put(58.00,65.00){\line(0,0){0.00}}
\put(58.00,65.00){\line(0,1){0.00}}
\put(58.00,65.00){\line(0,1){35.00}}
\put(70.00,60.00){\line(0,-1){20.00}}
\put(30.00,60.00){\line(0,-1){20.00}}
\put(60.00,105.00){\line(1,0){18.00}}
\put(73.00,35.00){\line(1,0){35.50}}
\put(11.00,105.00){\line(1,0){29.00}}
\put(40.00,149.00){\line(-1,0){29.00}}
\put(30.00,60.00){\line(1,0){12.00}}
\put(58.00,60.00){\line(1,0){12.00}}
\put(8.00,149.00){\makebox(0,0)[cc]{1}}
\put(8.00,105.00){\makebox(0,0)[cc]{2}}
\put(30.00,30.00){\line(0,-1){17.00}}
\put(30.00,13.00){\line(1,0){78.50}}
\put(100.00,105.00){\line(1,0){16.00}}
\put(99.00,105.00){\line(1,0){1.00}}
\put(78.00,105.00){\line(1,0){1.00}}
\put(125.00,72.00){\line(0,-1){2.00}}
\put(101.00,105.00){\line(0,-1){30.00}}
\put(101.00,75.00){\line(1,0){15.00}}
\put(125.00,102.00){\line(0,-1){2.00}}
\linethickness{0.1pt}
\put(25.00,70.00){\line(0,1){95.00}}
\put(25.00,165.00){\line(1,0){50.00}}
\put(75.00,165.00){\line(0,-1){95.00}}
\put(103.00,0.00){\line(1,0){44.00}}
\put(147.00,0.00){\line(0,1){10.00}}
\put(103.00,0.00){\line(0,1){10.00}}

\put(90.00,0.00){\line(-1,0){80.00}}
\put(90.00,0.00){\line(0,1){10.00}}
\put(10.00,0.00){\line(0,1){10.00}}
\put(10.00,55.00){\line(1,0){80.00}}

\end{picture}
\vspace{-7cm}
\caption
{\it
The output organization of {\tt HECTOR}
\label{heou}
}
\end{figure}


\clearpage

\section{User Options
\label{uo}
}
\ezero
The following set of flags and parameters may be defined
by the user in the file {\tt HECTOR.INP}.
The default values of the flags, their type, names of the
corresponding variables entering the program with the names of
host {\tt COMMON} blocks  one may find in appendix~B,
table~\ref{usropt}.
\subsection{Selection of the scattering process
\label{sotsp}}

\begin{tabbing}
\= {\tt IBEA}=$+$1:~\= anti-lepton beam (positron or $\mu ^+$) \kill
\>  \underline{{\tt IBEA}} \>
                       \\
\= {\tt IBEA}=$+$1:~\= anti-lepton beam (positron or $\mu ^+$)
                       \\
\> {\tt IBEA}=$-$1:\> lepton beam (electron or $\mu ^-$)
                       \\[.2cm]
\> \underline{{\tt ILEP}} \>
                       \\
\> {\tt ILEP}=1: \>
 electron (or positron) scattering
                       \\
\> {\tt ILEP}=2: \> muon scattering
                       \\[.2cm]
\> \underline{{\tt ITAR}} \> \\
\> {\tt ITAR}=1:\> proton target        \\
\> {\tt ITAR}=2:\> deuteron target
\end{tabbing}
\underline{{\tt POLB}} \\
  {\tt POLB}: ~~~degree of lepton beam polarization
$(-1 \leq {\tt POLB} \leq 1)$
                       \\[.2cm]
\underline{{\tt EELE}} \\
  {\tt EELE}: ~~~lepton beam energy in GeV
                       \\[.2cm]
\underline{{\tt ETAR}} \\
  {\tt ETAR}: ~~~energy of target beam in GeV
\begin{tabbing}
\= {\tt IBRN}=1:~\= llllll \kill
\= \underline{{\tt IBRN}} \> \\
\> {\tt IBRN}:\> to be used for {\tt IOPT=1} only \\
\> {\tt IBRN}=0:\> Born cross section of a lepton scattering
                   is defined by the user    \\
\> \> in the subroutine {\tt usrbrn.f}
                      \\
\> {\tt IBRN}=1:\> default Born cross section
                      \\[.2cm]
\end{tabbing}
\subsection{Selection of measurement type and cuts
\label{smtc}
}
\begin{tabbing}
\= {\tt IMEA}=10: \= llllll \kill
\> \underline{{\tt IMEA}}
                       \> \\
\> {\tt IMEA}=1: \>
 lepton measurement (leptonic variables), neutral current DIS
                       \\
\> {\tt IMEA}=2: \>
 jet measurement (Jaquet-Blondel variables), neutral current DIS
                       \\
\> {\tt IMEA}=3: \>
 jet measurement (Jaquet-Blondel variables), charged current DIS
 (only for {\tt IOPT=1})                       \\
\> {\tt IMEA}=4: \>
 mixed variables, neutral current DIS
                       \\
\> {\tt IMEA}=5:
 \> double angle method, neutral current DIS (only for {\tt IOPT=1})
                       \\
\> {\tt IMEA}=6: \>
lepton angle and $\yjb$ measurement, neutral current DIS (only for
{\tt IOPT=1})
                       \\
\> {\tt IMEA}=7: \>
 hadron measurement, neutral current DIS
                       \\
\> {\tt IMEA}=8: \>
 leptonic variables, charged current DIS
                       \\
\> {\tt IMEA}=9: \>
 $\Sigma$ method, neutral current DIS (only for {\tt IOPT=1})
                       \\
\> {\tt IMEA}=10:\>
 $\: e\Sigma$ method, neutral current DIS (only for {\tt IOPT=1})
                       \\ [.8cm]
\> \underline{{\tt ZCUT}}
                       \> \\
\> {\tt ZCUT}:\>
 cut on the reconstructed electron beam energy
 in GeV (only for {\tt IOPT}=1);
                       \\ \> \>
  see section~\ref{zcut}
                       \\[.2cm]
\> \underline{{\tt IHCU}}
                       \> \\
\> {\tt IHCU}=0:\>
 no cut on the hadronic final state
                       \\
\> {\tt IHCU}=1:\>
 rejects hadronic final states with $Q_h^2 \leq$ {\tt Q2CU} [$\GeV^2$]
 and invariant hadronic mass
                       \\  \> \>
 (photon not included) $W_h^2 \leq$ {\tt HM2C} [$\GeV^2$];
 the two desired cut values
                       \\ \> \>
 {\tt Q2CU} and {\tt HM2C} have to be set
                       \\
\pagebreak
\> \underline{{\tt Q2CU}}
                       \> \\
\> {\tt Q2CU}: \>
 cut on the transferred momentum squared, $Q_h^2$, [$\GeV^2$];
see {\tt
   IHCU} above
                       \\
\> \underline{{\tt HM2C}} \> \\
\> {\tt HM2C}: \>
cut on the hadronic invariant mass square, [$\GeV^2$]; see {\tt IHCU}
above
                       \\ [.2cm]
\> \underline{{\tt IGCU}}
                       \> \\
\> {\tt IGCU}=0: \>
 no cut on photon variables
                        \\
\> {\tt IGCU}=1: \>
 rejects all final states with a photon energy $E_\gamma \geq E_{cut}$ [$\GeV$]
 and a photon angle
                        \\  \> \>
 in the interval $(\theta_{cut2}, \theta_{cut1})$
(angles in radians inside
 $(0, \pi )$ with respect to the
                        \\  \> \>
  electron beam); to be used only for {\tt IOPT}=2, {\tt IMEA}=1
                        \\
\> \underline{{\tt GCUT}}
                        \> \\
\> {\tt GCUT}: \>
cut on the photon energy in GeV; see {\tt IGCU} above
                        \\
\> \underline{{\tt THC1}}
                        \> \\
\> {\tt THC1}: \>
 minimal photon angle $\theta_{cut2}$ in rad; see {\tt IGCU} above
                        \\
\> \underline{{\tt THC2}}
                        \> \\
\> {\tt THC2}: \> maximal photon angle $\theta_{cut1}$ in rad;
            see {\tt IGCU} above
\end{tabbing}
\subsection{Selection of the main program paths and related flags
\label{sotmpp}}
\begin{tabbing}
\= {\tt IMEA=1}:~ \= llllll \kill
\> \underline{{\tt IOPT}}
                        \> \\
\> {\tt IOPT}=1:\>
  {\tt HELIOS} branch: QED LLA RC calculated in first and higher
  orders                                    \\
\> {\tt IOPT}=2:\>
  {\tt TERAD} branch: complete calculation of EW and QED RC in
  $\order(\alpha)$
  with
                                  \\  \> \>
  soft photon exponentiation;
the model independent {\tt TERAD} results may be \\ \> \>
  supplemented by hadronic corrections from {\tt DISEP}
  (for {\tt IDIS}=2,3);
                                  \\  \> \>
see description of {\tt IDSP} and {\tt ILOW} below
                                  \\
\> {\tt IOPT}=3:\>
  combined calculation using both branches; here the {\tt TERAD} results
  are
                                  \\ \> \>
  supplemented by {\tt HELIOS} higher order terms beginning with
  $\order(\alpha^2)$; hadronic
                                  \\ \> \>
  corrections from {\tt DISEP} may also be added (for {\tt
    IDIS}=2,3); the soft photon
                                  \\ \>\>
  exponentiation of {\tt TERAD} is switched off
                                  \\[.2cm]
\> \underline{{\tt IORD}}
                                  \>  \\
\> {\tt IORD}=0:\>
  Born cross section (for {\tt IOPT=1})
                                  \\
\> {\tt IORD}=1:\>
  QED corrections to $\order(\alpha L)$ (for {\tt IOPT}=1)
  or to $\order(\alpha)$ with soft photon
                                  \\ \> \>
  exponentiation (for {\tt IOPT}=2)
                                  \\
\> {\tt IORD}=2:\>
  QED corrections to $\order((\alpha L)^2)$ only (for {\tt IOPT}=1)
                                  \\
\> {\tt IORD}=3:\>
 QED corrections to $\order(\alpha L)$ and $\order((\alpha L)^2)$
 are summed and the {\tt HELIOS}
                                  \\ \> \>
 soft photon exponentiation is included (for {\tt IOPT}=1) {\em or} complete
 $\order(\alpha)$ corrections
                                  \\ \> \>
 are summed
 with the $\order((\alpha L)^2)$ and the soft photon exponentiation
 terms from
                                  \\ \> \>
{\tt HELIOS} (for {\tt IOPT}=3)
                                  \\[.2cm]
\> \underline{{\tt IBOS}}
                                  \>  \\
\> {\tt IBOS}:\>
 calculation of RC to different terms in the $\cal NC$ cross section; may
 be used for all
                                  \\ \> \>
 values of {\tt IMEA} but 3,8
                                  \\
\> {\tt IBOS}=1:\>
 only photon $\cal NC$ exchange contribution ($|\gamma ^2|$ term)
                                  \\
\> {\tt IBOS}=2:\>
 only the photon - Z boson interference contribution to the $\cal NC$ exchange
 ($|\gamma Z|$ term)             \\
\> {\tt IBOS}=3:\>
 only the Z boson $\cal NC$ exchange contribution ($|Z^2|$ term)
                                  \\
\> {\tt IBOS}=4:\>
sum of
 all $\cal NC$ exchange contributions
                                  \\ [.2cm]
\> \underline{{\tt ICOR}}\>
                                  \\
\> {\tt ICOR}:\>
 to be used for {\tt IOPT}=1 only
                                  \\
\> {\tt ICOR}=1:\>
 initial state lepton radiation
                                  \\
\> {\tt ICOR}=2:\>
 final state lepton radiation
                                  \\
\> {\tt ICOR}=3:\>
 Compton peak contribution (available for {\tt IMEA}=1 only)
                                  \\
\> {\tt ICOR}=4:\>
 initial plus final state radiation
                                  \\
\> {\tt ICOR}=5:\>
 initial and final state radiation plus Compton peak contribution
                                  \\[.2cm]
\> \underline{{\tt ICMP}}\>
                                  \\
\> {\tt ICMP}:\>
 to be used for {\tt IOPT}=1 only
                                  \\
\> {\tt ICMP}=0:\>
 no Compton term
                                  \\
\> {\tt ICMP}=1:\>
 onefold Compton integral~(\ref{compt1})
                                  \\
\> {\tt ICMP}=2:\>
 twofold Compton integral~(\ref{compt2})
                                  \\[.2cm]
\> \underline{{\tt IEPC}}\>
                                  \\
\> {\tt IEPC}:\>
 to be used for {\tt IOPT}=1,3 only
                                  \\
\> {\tt IEPC}=0:\>
 $e^+e^-$  conversion in $\order ((\alpha L)^2)$ is switched off
                                  \\
\> {\tt IEPC}=1:\>  $e^+e^-$  conversion is included
                                  \\[.2cm]
\> \underline{{\tt IWEA}}\>
                                  \\
\> {\tt IWEA}=0:\>
 EW RC are applied via $\sin\theta_W^{\mr{eff}}$, the effective EW
 mixing angle
                                  \\
\> {\tt IWEA}=1:\>
 the full set of EW form factors is calculated;
                                 \\ \> \>
 {\bf Caution!} -- using this  flag may enlarge
 the CPU time considerably
                                  \\[.2cm]
\> \underline{{\tt IVPL}}\>
                                  \\
\> {\tt IVPL}=0:\>
 $\alpha =1/137\dots$ ~is not running
                                  \\
\> {\tt IVPL}=1:\>
 running $\alpha $ due to the complete fermionic vacuum polarization
                                  \\[.2cm]
\> \underline{{\tt IEXP}}\>
                                  \\
\> {\tt IEXP}=0:\>
 no soft photon exponentiation
                                  \\
\> {\tt IEXP}=1:\>
 soft photon exponentiation is applied; see description of {\tt IOPT}
                                  \\[.2cm]
\> \underline{{\tt IDSP}} \>
                                  \\
\> {\tt IDSP}=0: \>
 the {\tt DISEP} branch in {\tt TERAD} is not used
                                  \\
\> {\tt IDSP}=1: \>
 together with {\tt TERAD}, the {\tt DISEP} branch is used (possible only for
 {\tt IMEA}=1,4,8); see
                                  \\ \> \>
 description of flag {\tt IDIS}
                                  \\
\> {\tt IDSP}=2: \>
 only the {\tt DISEP} branch is used
                                  \\[.2cm]
\> \underline{{\tt IDIS}} \>
                                  \\
\> {\tt IDIS}=1: \>
 {\tt DISEP} leptonic corrections
                                  \\
\> {\tt IDIS}=2: \>
 {\tt DISEP} leptonic and lepton quark interference corrections
                                  \\
\> {\tt IDIS}=3: \>
 {\tt DISEP} leptonic, interference, and quarkonic corrections
                                  \\[.2cm]
\> \underline{{\tt IDQM}} \>
                                  \\
\> {\tt IDQM}: \>
 to be used for {\tt IMEA}=4, {\tt IDIS}=2 only, see~\cite{LLAmix}
                                  \\
\> {\tt IDQM}=1: \>
 quark masses {\tt AMQ}: {\tt AMQ} = $x M$, $M$ the proton mass
                                  \\
\> {\tt IDQM}=2: \>
  {\tt AMQ} = const, no ISR LLA corrections from quarks
                                  \\
\> {\tt IDQM}=3: \>
  {\tt AMQ} = const, no ISR or FSR LLA correction from quarks
                                  \\[.2cm]
\> \underline{{\tt ILOW}} \>
                                  \\
\> {\tt ILOW}: \>
 to be used for {\tt IMEA}=1, {\tt IOPT}=2 only
                                  \\
\> {\tt ILOW}=0: \>
 no use of {\tt TERADLOW}
                                  \\
\> {\tt ILOW}=1: \>
  use of {\tt TERADLOW}
\end{tabbing}
\subsection{Structure function parameterizations
\label{sfp}
}
\begin{tabbing}
\= {\tt IMEA=10}:~ \= llllll \kill
\> \underline{{\tt IMOD}}\>
                                  \\
\> {\tt IMOD}=0:\>
 use of quark distributions for the target structure functions
                                  \\
\> {\tt IMOD}=1:\>
 the parameterization of structure functions is user supplied in the
 file {\tt usrstr.f}
                                  \\ \> \>
 (e.g. for a model
 independent analysis)
                                  \\[.2cm]
\> \underline{{\tt ISTR}}\>
                                  \\
\> {\tt ISTR}:\>
 to be used for {\tt IMOD}=0 only      \\
\> {\tt ISTR}=0:\>
 use of recent parameterizations for parton distributions selected by flags
{\tt ISCH}
                                  \\ \> \>
 and {\tt ISSE}
                                  \\
\> {\tt ISTR}=1:\>
 parton density functions are user supplied via file
 {\tt usrpdf.f}
                                  \\
\> {\tt ISTR}=2:\>
 {\tt PDFLIB}~\cite{pdflib} is used for parton densities
                                  \\[.2cm]
\> \underline{{\tt ISSE}}\>
                                  \\
\> {\tt ISSE}:\>
 to be used for {\tt ISTR}=0
 \\
\> {\tt ISSE}=1:\>  parton distributions {\tt CTEQ3} \cite{cteq3}
                                  \\
\> {\tt ISSE}=2:\>  parton distributions {\tt GRV} \cite{GRV95,GRV92}
                    (cf. the comments in sect.~(\ref{msb}))
                                  \\
\> {\tt ISSE}=3:\>  parton distributions {\tt MRS(A)} \cite{MRS}
                                  \\
\> {\tt ISSE}=4:\>  parton distributions {\tt MRS(A')} \cite{MRS}
                                  \\
\> {\tt ISSE}=5:\>  parton distributions {\tt MRS(G)} \cite{MRS}
                                  \\[.2cm]
\> \underline{{\tt ITER}}\>
                                  \\
\> {\tt ITER}=0:\>
quark distributions and $\alpha_{_{QED}}$ (if running, see flag
{\tt  IVPL}) in leptonic bremsstrahlung
                                  \\   \> \>
are artifically chosen to depend on $Q^2$ as
defined from lepton momenta
                                  \\
\> {\tt ITER}=1:\>
 they depend on $Q^2$ as defined from the hadronic momenta
($Q^2$ is the integration
                                  \\ \> \>
variable)
                                  \\[.2cm]
\> \underline{{\tt IVAR}}\>
                                  \\
\> {\tt IVAR}$<$3:\>
 use of a damping factor for $Q^2 < Q^2_0$: $\lim_{Q^2\to
   0}q(x,Q^2)=0$;
                                  \\ \> \>
the value of $Q^2_0$ depends on the method
                                  \\
\> {\tt IVAR}=$-$1:\>
 the chosen quark distributions are not modified
                                  \\
\> {\tt IVAR}=0:\>
 Volkonsky/Prokhorov damping~\cite{Prokh}
                                  \\
\> {\tt IVAR}=1:\>
 Stein damping ~\cite{stein}
                                  \\
\> {\tt IVAR}=2:\>
 Brasse/Stein parameterization~\cite{brasse,stein}
                                  \\[.2cm]
\> \underline{{\tt IGRP}}\>
                                  \\
\> {\tt IGRP}:\>
 to be used for {\tt ISTR}=2
                                  \\
\> {\tt IGRP}:\>
  a {\tt PDFLIB} parameter; it defines an author group in the library (see
 {\tt PDFLIB}                                 \\ \> \>
 write-up~\cite{pdflib})
                                  \\[.2cm]
\> \underline{{\tt ISET}}\>
                                  \\
\> {\tt ISET}:\>
 to be used for {\tt ISTR}=2
                                  \\
\> {\tt ISET}:\>
 a {\tt PDFLIB} parameter which defines the specific parameterization
 set within  the group
                                  \\ \> \>
 defined by {\tt IGRP}; see~\cite{pdflib})
                                  \\[.2cm]
\> \underline{{\tt ISCH}}\>
                                  \\
\> {\tt ISCH}:\>
 to be used for {\tt ISTR}=0,2
                                  \\
\> {\tt ISCH}:\>
select the
 QCD factorization scheme; the flag has to be chosen in accordance with the
                                  \\ \> \>
 selected parameterization in {\tt PDFACT} or {\tt PDFLIB}
                                  \\
\> {\tt ISCH}=0:\>
 leading order QCD
                                  \\
\> {\tt ISCH}=1:\>
 DIS   factorization scheme
                                  \\
\> {\tt ISCH}=2:\>
 $\overline{\rm MS}$ factorization scheme
\end{tabbing}

\subsection{Bins and grid
\label{bag}
}

\begin{tabbing}
\= {\tt IMEA}=1: \= llllll \kill
\> \underline{{\tt IBIN}}\>
                                  \\
\> {\tt IBIN}=0:\>
 no integration over bins
                                  \\
\> {\tt IBIN}=1:\>
user
  supplied
integration over bins with subroutine {\tt usrint.f}
                                  \\[.2cm]
\> \underline{{\tt IBCO}}\>
                                  \\
\> {\tt IBCO}=1:\>
 bins are defined in the variables {\tt VAR1}=$x$ and {\tt VAR2}=$y$
                                  \\
\> {\tt IBCO}=2:\>
 bins are defined in the variables {\tt VAR1}=$x$ and {\tt VAR2}=$Q^2$
                                  \\
\> {\tt IBCO}=3:\>  bins are defined in the variables {\tt VAR1}=$y$
              and {\tt VAR2}=$Q^2$
                                  \\[.2cm]
\> \underline{{\tt ITV1}}\>
                                  \\
\> {\tt ITV1}=0:\>
 default bins in the first variable ({\tt VAR1}); see~(\ref{defx}) in
appendix~\ref{tuo}
                                  \\
\> {\tt ITV1}=1:\>
 logarithmical grid in the first variable
                                  \\
\> {\tt ITV1}=2:\>
 linear grid in the first variable
                                  \\
\> {\tt ITV1}=3:\>
 the bins in the first variable are defined by the user as a set of numbers
 in {\tt USRBIN};
                                  \\ \> \>
 see {\tt VAR1} below
                                  \\
\> {\tt ITV1}=4:\>
 user supplied bins in variable 1 via {\tt FFREAD}
                                  \\[.2cm]
\> \underline{{\tt ITV2}}\>
                                  \\
\> {\tt ITV2}=0:\>
  default bins in the second variable ({\tt VAR2})
                                  \\
\> {\tt ITV2}=1:\>
 logarithmical grid in the second variable
                                  \\
\> {\tt ITV2}=2:\>
  linear grid in the second variable
                                  \\
\> {\tt ITV2}=3:\>
 bins in the second variable are defined by the user
 as a set of numbers in {\tt USRBIN};
                                  \\ \> \>
 see {\tt VAR2} below
                                  \\
\> {\tt ITV2}=4:\>
 user supplied bins in variable 2 via {\tt FFREAD}
                                  \\[.2cm]
                                  \\[.2cm]
\> \underline{{\tt NVA1}}\>
                                  \\
\> {\tt NVA1}:\>
number of bins $(\leq 100)$ in the first variable,
to be used for {\tt ITV1}=1,2
                                  \\[.2cm]
\> \underline{{\tt NVA2}}
                                  \\
\> {\tt NVA2}:\>
number of bins $(\leq 100)$ in the second variable,
to be used for {\tt ITV2}=1,2
                                  \\[.2cm]
\> \underline{{\tt V1MN}}\>
                                  \\
\> {\tt V1MN}:\>
minimal value of the first variable ({\tt V1MN} $ > 0.0 $),
 to be used for {\tt ITV1}=1,2
                                  \\[.2cm]
\> \underline{{\tt V1MX}}\>
                                  \\
\> {\tt V1MX}:\>
maximal value of the first variable ({\tt V1MX} $< 1.0$),
to be used for {\tt ITV1}=1,2
                                  \\[.2cm]
\> \underline{{\tt V2MN}}\>
                                  \\
\> {\tt V2MN}:\>
minimal value of the second variable ({\tt V2MN} $ > 0.0 $),
 to be used for {\tt ITV2}=1,2
                                  \\[.2cm]
\> \underline{{\tt V2MX}}\>
                                  \\
\> {\tt V2MX}:\>
 maximal value of the second  variable ({\tt V2MX}: $ y < 1.0 $ or
 $ Q^2 < s $), to be used
                                  \\ \> \>
 for {\tt ITV2}=1,2
                                  \\[.2cm]
\> \underline{{\tt VAR1}}\>
                                  \\
\> {\tt VAR1}:\>
user supplied bins in the first  variable,  a set ($N \leq 100 $) of
 {\tt REAL} numbers from
                                  \\ \> \>
 the interval $0.0 < x < 1.0$ for {\tt IBCO}=1,2,
 or $0.0 < y < 1.0$ for {\tt IBCO}=3;
 to be
                                  \\ \> \>
 used for {\tt ITV1}=4
                                  \\[.2cm]
\> \underline{{\tt VAR2}}\>
                                  \\
\> {\tt VAR2}:\>
user supplied bins in the second  variable, a set ($N \leq 100 $) of
 {\tt REAL} numbers
                                  \\ \> \>
 from the interval $0.0 < y < 1.0$ for {\tt IBCO}=1,
 or $Q^2_{min} \le Q^2 < s$ for {\tt IBCO}=2,3,
                                  \\ \> \>
 ($Q^2_{min}$ depends on the structure functions used,
usually $Q^2_{min}$=4 or 5 \GeV $^2$);
                                  \\ \> \>
 to be used
 for {\tt ITV2}=4
\end{tabbing}
\subsection{User initialization\label{ui}}
\begin{tabbing}
\= {\tt IMEA}=1: \= llllll \kill
\> \underline{{\tt IUSR}}
                             \> \\
\> {\tt IUSR}=0: \>
default initialization of libraries, electroweak constants, etc.
                             \\
\> {\tt IUSR}=1: \>
some additional initialization may be done using the external subroutine
 {\tt usrini.f}
                             \\ \> \>
where the user may initialize her/his set of wanted quantities
\end{tabbing}
\subsection{Choice of an output}
\underline{{\tt IOUT}}
                             \\
{\tt IOUT}=0,1: output is written into file {\tt HECTOR.OUT}
              and on the screen
                             \\
{\tt IOUT}=2: output is created by user subroutine {\tt usrout.f}
                             \\
{\tt IOUT}=3: a simple output for creation of figures
                             \\[.4cm]

\bigskip

\noindent
After reading the data cards  supplied by the user the program checks
their compatibility.
The following flags have highest priority:
{\tt IOPT}, {\tt IMOD}, {\tt ITV1}, and {\tt ITV2}.
If there are flags being in conflict with them
they will be ignored.
In other cases of incompatibility
the program will stop and write a message into the
file {\tt HECTOR.OUT}.

\clearpage


\subsection{User subroutines}
The user may organize her/his own interface for
using the code with help of several subroutines, which
can be rewritten by the user. Simple examples
can be found inside the correspondent files.
\\[.15cm]

\noindent
\underline{{\tt usrini.f}}: is called if {\tt IUSR}=1.
\begin{itemize}
\item[]{\tt CALL USRINI}
\end{itemize}
In this subroutine the user may initialize parameters
for later use.
They can be transferred via {\tt COMMON /HECUSR/}. \\[.5cm]

\noindent
\underline{{\tt usrdpf.f}}: is called if {\tt ISTR}=1.
\begin{itemize}
\item[]{\tt CALL USRPDF(X,Q,UV,DV,US,DS,SS,CS,BS,TS,GLU)}
\end{itemize}
In this subroutine the user may put a set of parton density
functions.
\begin{itemize}
\item[]{\tt INPUT: X,Q}, with {\tt Q}=$\sqrt{Q^2}$.
\item[]{\tt OUTPUT: UV,DV,US,DS,SS,CS,BS,TS,GLU}
\end{itemize}
Note that all density functions should be
multiplied by $x$, i.e. {\tt UV} = $x u_v(x,Q^2)$.
If the chosen density functions are
calculated up to the next to leading order one has also to implement
QCD radiative
corrections, in dependence on the factorization scheme used. \\[.5cm]

\noindent
\underline{{\tt usrstf.f}}: is called if {\tt IMOD}=1.
\begin{itemize}
\item[]{\tt CALL USRSTR(X,Q,FL,GL,HL,F2,G2,H2,XG3,XH3,WL,W2,XW3)}
\end{itemize}
In this subroutine the user may put a set of target structure
functions.
\begin{itemize}
\item[]{\tt INPUT: X,Q}, with {\tt Q}=$\sqrt{Q^2}$.
\item[]{\tt OUTPUT: FL,GL,HL,F2,G2,H2,XG3,XH3,WL,W2,XW3}
\end{itemize}
Note that in this case one may not necessarily refer to the parton picture
but use a model independent parameterization of the structure
functions instead. \\[.5cm]

\noindent
\underline{{\tt usrbin.f}}: is called if {\tt ITV1}=3 and/or
{\tt ITV2}=3.
\begin{itemize}
\item[]{\tt CALL USRBIN}
\end{itemize}
In this subroutine the user may define a
set of points in two variables, where the cross sections and
radiative correction are calculated.
Note that the internal definitions of {\tt usrbin.f} ~have
higher priority than the definitions in {\tt HECTOR.INP}.
An example is given.\\[.5cm]

\newpage

\noindent
\underline{{\tt usrint.f}}: is called if {\tt IBIN}=1.
\begin{itemize}
\item[]{\tt CALL USRINT}
\end{itemize}
In this subroutine the user may organize an integration over bins.
A simple example is illustrated by the default routine.
\\[.5cm]

\noindent
\underline{{\tt usrbrn.f}}: is called if {\tt IBRN}=0.
\begin{itemize}
\item[]{\tt CALL USRBRN(X,Y,S,SIG0)}
\end{itemize}
This subroutine can only be used in the {\tt HELIOS} branch of the code
({\tt IOPT}=1).
\begin{itemize}
\item[]{\tt INPUT: X,Y,S}
\item[]{\tt OUTPUT: SIG0}, the Born cross section $d^2\sigma/dxdy$
in [nb].
\end{itemize}
The radiative corrections to the process defined are
then calculated according to the flags defined in {\tt HECTOR.INP}.
\\[.5cm]

\noindent
\underline{{\tt usrout.f}}: is called if {\tt IOUT}=2.
\begin{itemize}
\item[]{\tt CALL USROUT}
\end{itemize}
In this subroutine the user may organize her/his own output.
A simple example, which creates {\tt OUTPUT VECTORS}
to be used e.g. in graphics programs, is given.
\\[.5cm]

\noindent
\underline{{\tt usrgrf.f}}: is called if {\tt IOUT}=3.
\begin{itemize}
\item[]{\tt CALL USRGRF}
\end{itemize}
This is another example of the user supplied {\tt OUTPUT} routine
creating {\tt OUTPUT VECTORS} of a different format.
\\[.5cm]

\clearpage
\section{Examples of numerical output of \he}
In the initialization phase, the user has to decide whether user
supplied input shall be used in addition.
If not, {\tt IUSR}=0 has to be chosen
(see section~\ref{ui}).
The setting of all the electroweak parameters is done in the file
{\tt hecset.f}.
The unexperienced user can refer to the
default flags.
The beam parameters (particle type, charge, polarization, energy)
are selected by   the parameters given in section~\ref{sotsp}.
Furthermore, one has to decide in which variable set the QED
corrections shall   be calculated (flag {\tt IMEA}) and whether
(possible)
cut parameters
shall be set.
Then, the type of corrections to be applied has to be defined,
including a choice of branch(es) ({\tt HELIOS, TERAD, DISEP}, or
a combination
of them) to be used.
This is done by the flags  {\tt IOPT}, {\tt  IORD} etc. being
described in
section~\ref{sotmpp}.
The structure function parameterizations are chosen by the flags of
section~\ref{sfp}.
For {\tt IBIN}=0, the output is calculated for the pre--defined grid
of kinematical points and no integration over bins is
performed.
The flags and parameters of section~\ref{bag} have to be set.

\bigskip

Two example files for {\tt HECTOR.INP}, {\tt HECTOR.INP.SRT} and
{\tt HECTOR.INP.STD} are available. The sample outputs
(appendices~C, D) correspond to these input files (renamed as
{\tt HECTOR.INP}).
In addition, we show in tables~3--6 cross section values
for varying input parameters and in figures~\ref{fcc1}--\ref{fcc3}
radiative corrections in percent for three selected types of variables.
These examples are foreseen to act as  benchmarks for typical
variations in the use of \he.

\sloppy
\begin{table}[hbt]
\centering
\begin{tabular}{||r|r||r |r||}
\hline \hline
\multicolumn{1}{||c|}{{\tt ISSE}}
  & \multicolumn{1}{ c||}{{\tt ISCH}}
  & \multicolumn{1}{c| }{$\sigma_{B}$~[nb]}
  & \multicolumn{1}{c||}{$\delta_{\mathrm{ini}}^{(1)}$~[\%]}  \\
\hline \hline
1  & 0 &0.2246E+06  & 13.01     \\
2  & 0 &0.2441E+06  & 12.94     \\
\hline
1  & 1 &0.2225E+06  & 12.65     \\
3  & 1 &0.2679E+06  & 11.76     \\
\hline
1  & 2 &0.2345E+06  & 12.82     \\
2  & 2 &0.2702E+06  & 12.13     \\
3  & 2 &0.2695E+06  & 11.84     \\
4  & 2 &0.2410E+06  & 13.56     \\
5  & 2 &0.2415E+06  & 13.55     \\
\hline  \hline
\end{tabular}
\normalsize
\caption
[\it
Born cross sections for unpolarized $e^+ p$ scattering
with NLO QCD corrections for different sets of parton
parameterizations
]
{\it
Born cross sections for unpolarized $e^+ p$ scattering
with NLO QCD corrections for different sets of parton
parameterizations.
Parameters:
$x = 10^{-3}, y = 0.1$, $E_e = 26.7~{\rm GeV}$, $E_p = 820.0~{\rm GeV}$,
and
{\tt IVPL} $=1$ (i.e. running $\alpha$).
The low $Q^2$ damping corresponds to {\tt IVAR} $=0$.
\label{T2}
}
\end{table}

\sloppy
\begin{table}[bht]
\centering
\begin{tabular}{||c||c|c||}
\hline \hline
\multicolumn{1}{|| c    ||}  { {\tt IVAR}}
  & \multicolumn{1}{  c |} { $\sigma_{B}$~[nb]       }
  & \multicolumn{1}{  c ||} {$\delta_{\mathrm {ini}}^{(1)}$~ [\%]  }     \\
\hline \hline
 --1 &0.4026E+08     & 9.452     \\
   0  &0.3816E+08     & 9.222     \\
  1  &0.3616E+08     & 9.197     \\
  2  &0.7445E+07     & 32.53     \\
 \hline  \hline
 \end{tabular}

\normalsize
\caption
[\it
Born cross sections for $e^+ p$ scattering with different parton
distribution parameterizations at low $Q^2$
]
{\it
Born cross sections for $e^+ p$ scattering with different parton
distribution parameterizations at low $Q^2$.
Parameters:
$x = 10^{-4}, y = 0.1$, $E_e = 26.7~{\rm GeV}$, $E_p = 820.0~{\rm GeV}$,
{\tt ISSE} $=1$, {\tt ISCH} $=0$, {\tt IVPL} $=1$.
 \label{T3}
}
\end{table}

\newcommand\shh{{\mathrm{sh}}}
\begin{table}[bht]
\centering
\begin{tabular}{||r|r|r||r |r||}
\hline \hline
\multicolumn{1}{||c|}{{\tt IOPT}}
  & \multicolumn{1}{ c||}{{\tt IDSP}}
  & \multicolumn{1}{ c||}{{\tt IWEA}}
  & \multicolumn{1}{c| }{$\sigma_{B}$~[nb]}
  & \multicolumn{1}{c||}{$\delta_{\mathrm{all} }^{(1)}$~[\%]}  \\
\hline \hline
3  & 0  & 0 &0.1260E-02  & -3.980     \\
3  & 0  & 1 &0.1278E-02  & -5.130     \\
\hline
2  & 2  & 0 &0.1260E-02  & -5.129     \\
2  & 2  & 1 &0.1278E-02  & -6.268     \\
\hline  \hline
\end{tabular}
\normalsize
\caption
[\it
Born cross sections for $e^+ p$ scattering for different values of
flag {\tt IWEA}
]
{\it
Born cross sections for $e^+ p$ scattering for different values of
flag {\tt IWEA}.
Parameters:
$x = 0.5, y = 0.5$, $E_e = 26.7~{\rm GeV}$,
$E_p = 820.0~{\rm GeV}$,
{\tt ISSE} $=1$, {\tt ISCH} $=0$, {\tt IVAR} $=0$, {\tt IVPL} $=1$.
\label{T4}
}
\end{table}

\sloppy
\begin{table}\centering
\begin{tabular}{||r|r|r|r||r|r|r|r||c||}
\hline \hline
\multicolumn{1}{||c|}{{\tt IMEA}}
  & \multicolumn{1}{|c|}{{\tt ICOR}}
  & \multicolumn{1}{|c|}{{\tt IDSP}}
  & \multicolumn{1}{|c||}{{\tt IOPT}}
  & \multicolumn{1}{c|}{$\sigma_{B}$~[nb]}
  & \multicolumn{1}{c|}{$\delta^{(1)}$~[\%]}
  & \multicolumn{1}{c|}{$\delta^{(2)}$~[\%]}
  & \multicolumn{1}{c||}{$\delta^{\mathrm {all}     }$~[\%]}
  & \multicolumn{1}{c||}{Comments} \\
\hline \hline
 1  & 1 & 0 & 1 & 0.2246E+06 & 13.01   &--.4092 & 12.60   &     \\
2  & 1 & 0 & 1 & 0.2246E+06 & 3.928  &  .3942 & 4.322   &     \\
3  & 1 & 0 & 1 & 0.1066E+01 & --.0837&  .4420 & .3583  &     \\
4  & 1 & 0 & 1 & 0.2246E+06 & 10.22  &  .6968 & 10.92   &     \\
5  & 1 & 0 & 1 & 0.2246E+06 & 5.512  &  .2490 & 5.761   & $E_{cut}=35~\GeV$\\
6  & 1 & 0 & 1 & 0.2246E+06 & 6.182  &  .2334 & 6.416   & $E_{cut}=35~\GeV$\\
7  & 1 & 0 & 1 & 0.2246E+06 & 4.469  &  .4151 & 4.884   &                    \\
8  & 1 & 0 & 1 & 0.1066E+01 & 3.172  &--.1664 & 3.006   &     \\
9  & 1 & 0 & 1 & 0.2246E+06 & 4.150  &  .5021 & 4.652   &                    \\
10 & 1 & 0 & 1 & 0.2246E+06 & 9.328  &  2.492 & 11.82   &                    \\
\hline
1  & 2 & 0 & 1 & 0.2246E+06 &  4.643  &        &         &     \\
4  & 2 & 0 & 1 & 0.2246E+06 &--2.687  &        &         &     \\
9  & 2 & 0 & 1 & 0.2246E+06 &--2.873  &        &         &     \\
10 & 2 & 0 & 1 & 0.2246E+06 &--3.040  &        &         &     \\
\hline
1  & 3 & 0 & 1 & 0.2246E+06 & .02006 & &       & {\tt ICMP}=1    \\
   &   &   &   & 0.2246E+06 & .02753 &         &         & {\tt ICMP}=2    \\
\hline
1  & 1 & 0 & 2,3 & 0.2246E+06 & 17.82  &        & 17.41   &     \\
2  & 1 & 0 & 2,3 & 0.2246E+06 & 3.748  &        & 4.133   &     \\
4  & 1 & 0 & 2,3 & 0.2246E+06 & 7.195  &        & 7.783   &     \\
7  & 1 & 0 & 2,3 & 0.2246E+06 & 4.337  &        & 4.736   &     \\
\hline
1  & 1 & 1 & 2,3 & 0.2246E+06 & 17.82  &        & 17.41   &          \\
1  & 1 & 2 & 2,3 & 0.2246E+06 & 17.95  &        & 17.41   &          \\
4  & 1 & 1 & 2,3 & 0.2246E+06 & 7.189  &        & 7.783   &          \\
4  & 1 & 2 & 2,3 & 0.2246E+06 & 7.189  &        & 7.783   &          \\
8  & 1 & 1 & 2,3 & 0.1066E+01 & 8.100  &        & 7.933   & {\tt IDIS}=2  \\
   &   &   &     & 0.1066E+01 & 8.070  &        & 7.903   & {\tt IDIS}=1 \\
\hline \hline
\end{tabular}
%
\normalsize
\caption
[
\it
Born cross sections and radiative corrections for $e^+ p$ scattering
for different values of flags {\tt IMEA}, {\tt ICOR}, {\tt IDSP}, {\tt
 IOPT}
]
{\it
Cross sections and radiative corrections for
$e^+ p$ scattering as functions of flags {\tt IMEA}, {\tt ICOR}, {\tt
  IDSP}, {\tt IOPT}.
Parameters:
$x = 10^{-3}, y = 0.1$, $E_e = 26.7~{\rm GeV}$, $E_p = 820.0~{\rm GeV}$,
and
{\tt IVPL} $=1$ (i.e. running $\alpha$).
The parton density parameterization {\tt CTEQ3LO} is chosen together with
the low $Q^2$ damping factor corresponding to {\tt IVAR} $=0$.
Longitudinal structure functions are disregarded.
For the ${\cal O}(\alpha^2)$ corrections $e^+ e^-$ beam conversion
terms were added.
For {\tt IOPT} $=2$ ($3$), the values of $\delta^{(i)}$ are  given in
columns ~$6 (8)$. Note the comments in~\cite{jblla2} for the
branches selected by {\tt IMEA} $=5,6$.
\label{T1}
}
\end{table}


%
\begin{figure}[hb]
\begin{center}
\vspace*{2.3cm}
\mbox{
\epsfysize=20.0cm
\epsffile{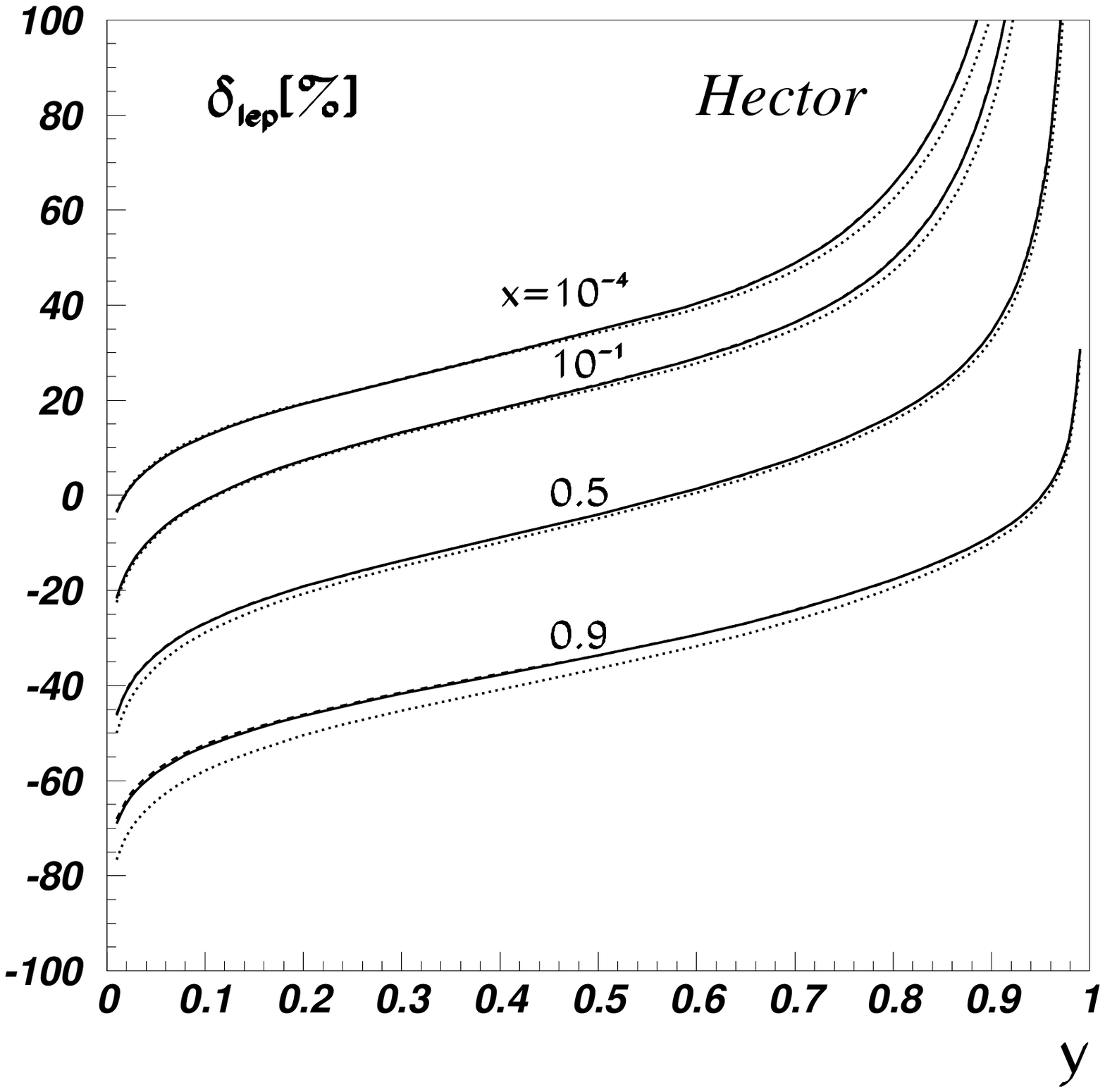}
}
\end{center}
\vspace*{-2.2cm}
\caption
[\it
Radiative corrections in leptonic variables
]
{\it
Radiative corrections in leptonic variables for the default settings
in percent.
Dotted lines: ${\cal O}(\alpha)$, dashed lines: ${\cal O}(\alpha^2)$,
solid lines: in addition soft photon exponentiation.
}
\label{fcc1}
\end{figure}
%

%
\begin{figure}[hb]
\begin{center}
\vspace*{2.3cm}
\mbox{
\epsfysize=20.0cm
\epsffile{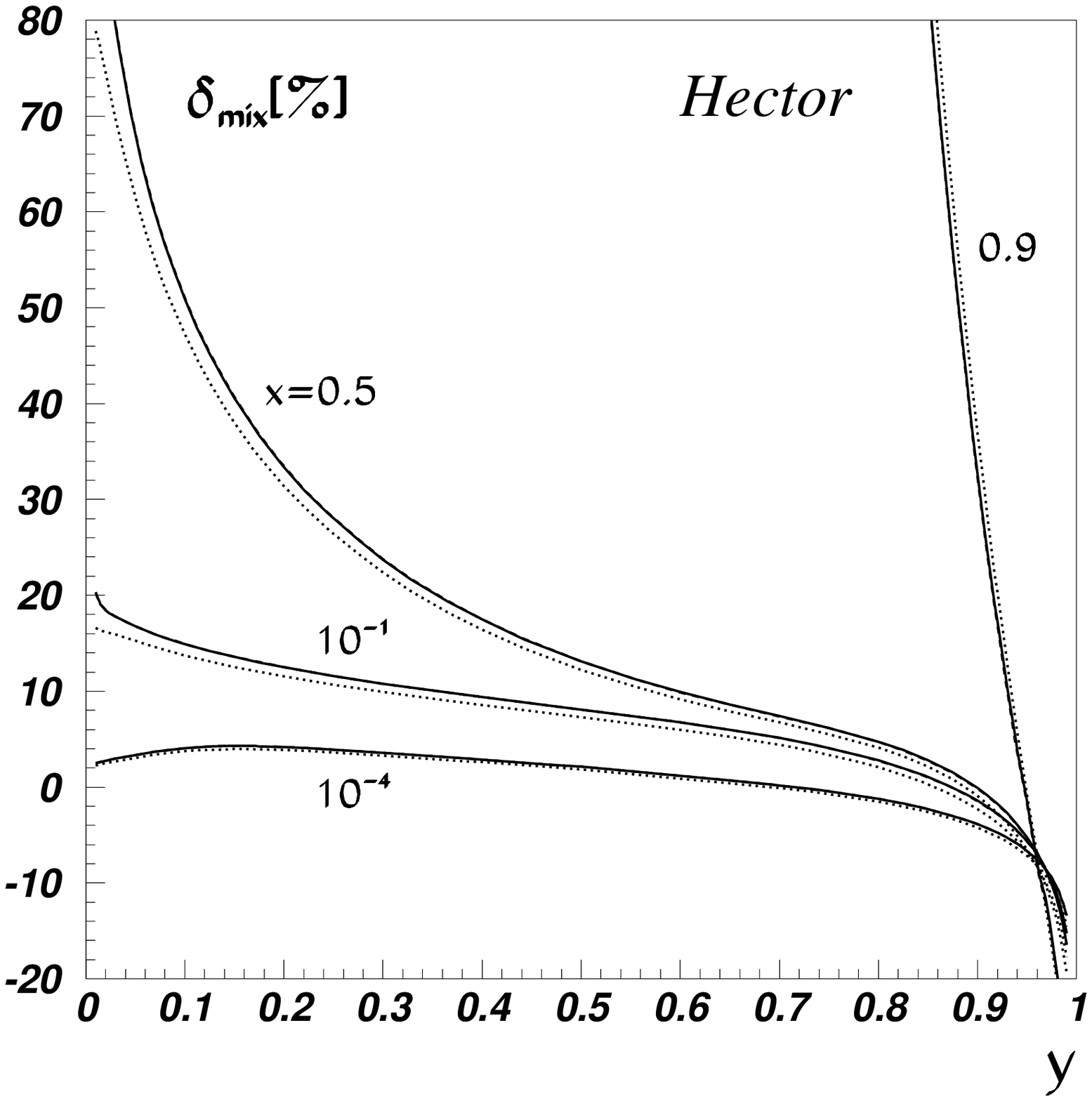}
}
\end{center}
\vspace*{-2.2cm}
\caption
[\it
Radiative corrections in mixed variables
]
{\it
Radiative corrections in mixed variables for the default settings
in percent.
Dotted lines: ${\cal O}(\alpha)$, dashed lines: ${\cal O}(\alpha^2)$,
solid lines: in addition soft photon exponentiation.
}
\label{fcc2}
\end{figure}
%

%
\begin{figure}[hb]
\begin{center}
\vspace*{2.3cm}
\mbox{
\epsfysize=20.0cm
\epsffile{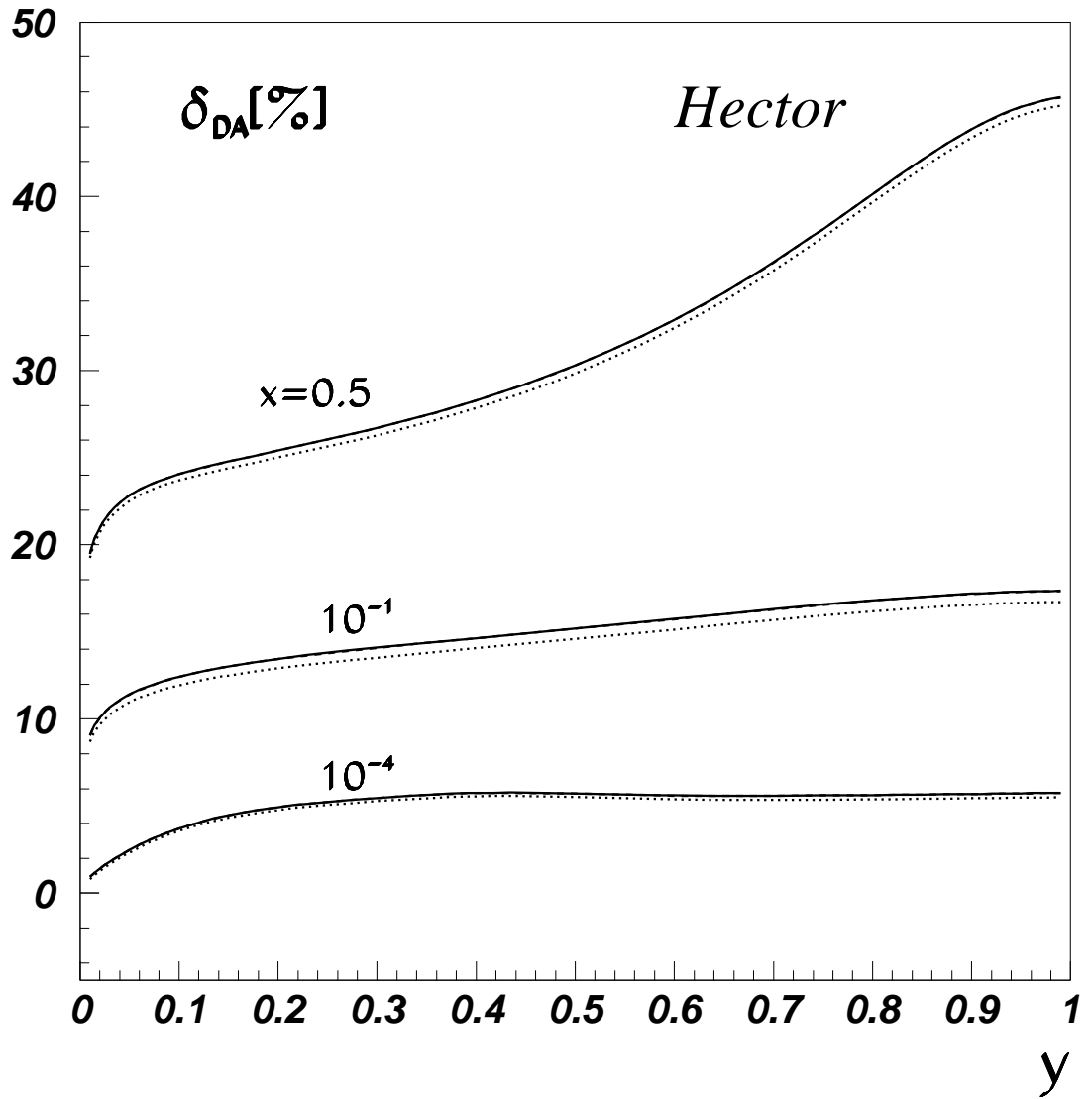}
}
\end{center}
\vspace*{-2.2cm}
\caption
[\it
Radiative corrections in double angle variables
]
{\it
Radiative corrections in double angle variables for the default settings
in percent.
Dotted lines: ${\cal O}(\alpha)$, dashed lines: ${\cal O}(\alpha^2)$,
solid lines: in addition soft photon exponentiation.
}
\label{fcc3}
\end{figure}
%

\clearpage

\section{Output\label{outp}}
\ezero
The code writes a file
{\tt HECTOR.OUT} by default ({\tt IOUT}=0).
The user can organize her/his own
output in the user file {\tt usrout.f}, which is called for
{\tt IOUT}=2. Option {\tt IOUT}=3 pipelines
output to the user file {\tt usrgrf.f}.
Examples for these subroutines and some instructions on their use are
given in the files {\tt usrout.f} and {\tt usrgrf.f}.
Examples of the program output may be found in appendices~C, D. \\[0.15cm]

\noindent
The program writes the following messages into the file {\tt HECTOR.OUT}:
\begin{tabbing}
\= $\bullet$ ~\= program Logo labeled with the version number; \\[.15cm]
\> $\bullet$ \> the data cards which have been read in
                the {\tt FFREAD} format; \\[.15cm]
\> $\bullet$ \> in case of an {\it incompatible\/}
      flag definition the program writes the values of \\
\> \> the corresponding flags into  the file {\tt HECTOR.OUT}
      and terminates thereafter; \\[.15cm]
\> $\bullet$ \>  version number of the
library
{\tt PDFLIB}
                (if activated by the user); \\[.15cm]
\> $\bullet$ \> an information about the chosen program branch; \\[.15cm]
\> $\bullet$ \> type of process and variables selected; \\[.15cm]
\> $\bullet$ \> target type; \\[.15cm]
\> $\bullet$ \> information about the projectile particle: its
charge, type and polarization; \\[.15cm]
\> $\bullet$ \> type the structure functions used; \\[.15cm]
\> $\bullet$ \> choice of the
low $Q^2$ modification used; \\[.15cm]
\> $\bullet$ \> input parameters
and calculated electroweak parameters; \\[.15cm]
\> $\bullet$ \> energy characteristics of the colliding
                particles; \\[.15cm]
\> $\bullet$ \> {\tt ZCUT}
cut parameter (for {\tt IOPT}=1); \\[.15cm]
\> $\bullet$ \> choice of bin variables {\tt VAR1} and {\tt VAR2};
                \\[.15cm]
\> $\bullet$ \> type of grid in each variable; \\[.15cm]
\> $\bullet$ \> number of bins in each variable; \\[.15cm]
\> $\bullet$ \> label of the chosen set of parton densities
                (for {\tt ISTR}=0). \\[.15cm]
\end{tabbing}
%
The results of the
calculation are presented in six columns: values of the first
variable {\tt VAR1}, values of the second variable {\tt VAR2},
the $s$ values in GeV$^2$, Born cross section values in nbarn,
corrected cross section values in nbarn, correction values in $\%$.

\vspace{1cm}
\noindent
{\bf Acknowledgement}~~We would like to thank Paul S\"oding for his
constant support during the performance of the project. A.A., D.B.,
and L.K. would like to thank DESY--Zeuthen for the warm hospitality
extended to them and for financial support. A.A. thanks the
Heisenberg--Landau foundation for financial support. The project was
founded in part by the EC-Network `Capital Humain et Mobilite' under
contract CHRX--CT92--0004. We would like to thank Dick Roberts
for a version of the MRSA distribution in the DIS scheme.

\clearpage

\addcontentsline{toc}{section}{References}

\newpage
\appendix
\def\theequation{\Alph{section}.\arabic{equation}}
\section
{Definition of scaling variables in presence of hard photon emission
\label{appvardef}
}
\ezero
\newcommand {\hxl} {\mbox{${\hat x}_{l}  $}}
\newcommand {\hyh} {\mbox{${\hat y}_{h}  $}}
\newcommand {\hxh} {\mbox{${\hat x}_{h}  $}}
\newcommand {\xesi} {\mbox{$x_{_{\mathrm{e\Sigma}}}$}}
\newcommand {\qesi} {\mbox{$Q^2_{_{\mathrm{e\Sigma}}}$}}
\newcommand {\hxesi} {\mbox{$\hat{x}_{_{\mathrm{e\Sigma}}}$}}
\newcommand {\hyesi} {\mbox{$\hat{y}_{_{\mathrm{e\Sigma}}}$}}
\newcommand {\hqesi} {\mbox{$\hat{Q}^2_{_{\mathrm{e\Sigma}}}$}}
\newcommand {\hxan} {\mbox{${\hat x}_  {\theta y}$}}
\newcommand {\hyan} {\mbox{${\hat y}_  {\theta y}$}}
\newcommand {\hqan} {\mbox{${\hat Q}^2_{\theta y}$}}
\newcommand {\hqjb} {\mbox{$\hat{Q}^2_{_{\mathrm{JB}}}$}}
\newcommand {\hxjb} {\mbox{$\hat{x}_{_{\mathrm{JB}}}$}}
\newcommand {\hyjb} {\mbox{$\hat{y}_{_{\mathrm{JB}}}$}}
\newcommand {\hql} {\mbox{$\hat Q ^2_l$}}
\newcommand {\hym} {\mbox{$\hat y _ m $}}
\newcommand {\hxm} {\mbox{$\hat x _ m $}}
In this appendix, we collect the definitions of all the sets of
kinematical variables, which may be used in \he.
Some motivations for the specific choices may be found
in~\cite{KIN0}-\cite{KIN4}.

\bigskip

\centerline{ \Large \it Leptonic variables:}

\bigskip

\noindent
The scaling variables are defined from leptonic variables:
\ba
 \yl &=& \frac{p_1(k_1-k_2)}{p_1k_1}, \\
\label{lep_1}
 \Ql &=& (k_1-k_2)^2 = -2k_1k_2,      \\
\label{lep_2}
 \xl &=& \frac{\Ql}{S \yl}.
\label{lep_3}
\ea

\bigskip

\centerline{ \Large \it Hadronic variables:
}

\bigskip

\noindent
The scaling variables are defined from hadronic variables using the
4--vector of the current jet:
\ba
 \yh &=& \frac{ p_1(p_2-p_1)}{p_1k_1}, \\
 \qh &=& (p_2-p_1)^2,                  \\
 \xh &=& \frac{\qh}{S\yh  }.
\label{hadvar}
\ea

\bigskip

\centerline{ \Large \it  Jaquet-Blondel variables~\cite{KIN1}:
}

\bigskip

\noindent
Again, the scaling variables are defined from hadronic variables.
Using the transverse hadronic momentum in the laboratory system,
${\vec p}_2^{\perp}$,
instead of $p_2$
and the energy flow in
$z$ direction
in the definition of
the scaling variables\footnote{Note that
in the complete
${\cal O}(\alpha)$ calculations for Jaquet-Blondel and mixed variables
in refs.~\cite{LLAmix} the {\it approximate}
relation $\yjb \approx \yh$ was used which holds iff all masses in the
final state can be neglected with respect to the particle momenta.}
:
\ba
 \yjb &=& \frac{\Sigma}{2 E_e}   ,            \\
\label{scl_jby}
 \qjb &=& \frac{({\vec p}_{2\perp})^2}{1-\yjb},     \\
\label{scl_12}
 \xjb &=& \frac{\qjb}{S \yjb}.
\label{scl_jbx}
\ea
with
\ba
 \Sigma = \sum_h (E_h- p_{h,z}),
\label{sigme}
\ea
and
$E_h, p_{h,x}, p_{h,y}, p_{h,z}$
being the four-momentum components of the hadronic final state.

\bigskip

\centerline{ \Large \it  Mixed variables~\cite{KIN2}:
}

\bigskip

\noindent
The scaling variables are defined from both leptonic and Jaquet-Blondel
variables:
\ba
   \qm  &=& \Ql,    \\
   \ym  &=& \yjb,    \\
   \xm  &=& \frac{\Ql}{S \yjb}.
\ea
For mixed variables, the allowed values of $x$ are not necessarily
restricted to be smaller than one; see for details appendix B.3.1
of~\cite{MI}.

\bigskip

\centerline{ \Large \it  $\Sigma$  method~\cite{KIN4}
}

\bigskip

\noindent
The scaling variables are defined from both leptonic and hadronic
variables as being measured in the laboratory system:

\ba
 \ysi &=& \frac{\Sigma}{\Sigma + E^{'}_l(1-\cos \theta_l)}, \\
\label{ysi}
 \qsi &=& \frac{({\vec k}_{2\perp})^2}{1-\ysi} ,        \\
 \xsi &=& \frac{\qsi}{S \ysi},
\label{xsi}
\ea
The integration boundary $z_0^{\Sigma,f}$ in the case of final state
radiation is the solution of the equation
\ba
x \left [1 - y(1-z_0)\right]^2 = z^3_0,
\label{eqcub}
\ea
\ba
z_0^{\Sigma,f}
= E_1^{1/3}
         + E_2 E_1^{-1/3}+ \frac{1}{3} x y^2,
\label{eqcubs}
\ea
with
 \ba
  E_1 &=& x(E_1^a + E_1^b), \\
  E_1^a &=&
         \frac{1}{2}
     - y \left( 1-\frac{1}{2}y-\frac{1}{27} x^2 y^5 \right)
        +\frac{1}{3} x y^3 (1-y),                          \\
  E_1^b &=& \frac{1}{6} \sqrt{
         9 (1 - 4 y +  y^4)
         +2 y^2 \left[ 27 - 18y - \frac{2}{3}x y^4
         +2 x y \left( \frac{1}{3}- y + y^2 \right)\right]},
                          \\
 E_2 &=&
       \frac{2}{3} x y \left( 1-y+\frac{1}{6} x y^3 \right).
 \ea


\vspace{1.cm}

\bigskip

\newpage
\centerline{ \Large \it  $e\Sigma$  method~\cite{KIN4}:
}

\bigskip

\noindent
The $e\Sigma$  method mixes variables from the set of leptonic
variables and of the $\Sigma$  method:
\ba
 \qesi &=& \ql, \\
 \xesi &=& \xsi    ,
\\
 \yesi &=& \frac{\qesi}{S \xesi}=\frac{\ql}{S \xsi}.
\ea
The integration boundary $z_0^{\Sigma,f}$
is the same as in the $\Sigma$ method.

\bigskip

\centerline{ \Large \it  Double angle method~\cite{KIN3}:
}

\bigskip

\noindent
The scaling variables are defined from both leptonic and hadronic
scattering angles as being measured in the laboratory system.

The definitions of scaling variables in the double angle method
are:
\ba
 \qdo
       &=&
\frac{\displaystyle
          4 E_l^2  \cos^2{\frac{ \theta_l}{2} } }
           {\displaystyle \sin^2{\frac { \theta_l}{2} }
         +          \sin  {\frac { \theta_l}{2} }
                    \cos  {\frac { \theta_l}{2} }
                    \tan  {\frac { \theta_h}{2} }},
\\
\ydo
& = &
     1 - \frac{\displaystyle \sin \frac{\theta_l}{2} }
              {\displaystyle \sin \frac{\theta_l}{2} +
                \cos\frac{\theta_l}{2}
                 \tan\frac{\theta_h}{2}},
\ea
\ba
  \xdo &=& \frac{\qdo   }{ S  \ydo }.
\ea

\bigskip

\centerline{ \Large \it  $\theta y$ method~\cite{jblla2}:
}

\bigskip

\noindent
A quite similar opportunity is to express the cross sections in terms
of $\theta_e$ in the laboratory system and $\yjb$.
The kinematical variables are:
\ba
\qan    &=& 4 E_e^2 (1-\yjb) \frac{1+\cos\theta_l}{ 1-\cos\theta_l} ,
\label{any_2}
\\
 \yan  &=& \yjb ,    \\
\label{any_1}
 \xan  &=& \frac{\qan}{S \yan}.
\label{scl_jbxx}
\ea
%

\clearpage
\section{Table of user options
\label{tuo}
}

\vspace*{-0.5cm}

\begin{table}[h]
\begin{center}


\begin{tabular}{||r|l|c||l|c|r|l||} \hline \hline
  &{{\tt FFREAD} flag}&type& parameter & common     & default &
                   ~references  \\
                      \hline \hline
 1&{\tt EELE}& R  &{\tt EEL}   &{\tt HECPAR}& 30.0  & 4.1
                                                          \\ \hline
 2&{\tt ETAR}& R  &{\tt EPR}   &{\tt HECPAR}& 820.0 & 4.1
                                                           \\ \hline
 3&{\tt GCUT}& R  &{\tt ECUT}  &{\tt HECGMC}& 0.0   & 4.2
                                                          \\ \hline
 4&{\tt HM2C}& R  &{\tt AMF2CT}&{\tt HECHDC}&  0.0  & 4.2
                                                          \\ \hline
 5&{\tt IBCO}& I  &{\tt IBCO}  &{\tt HECGRD}& 1     & 4.5
                                                          \\ \hline
 6&{\tt IBEA}& I  &{\tt IBEAM} &{\tt HECOPT}&  --1  & 4.1
                                                          \\ \hline
 7&{\tt IBIN}& I  &{\tt IBIN}  &{\tt HECBIN}&  0    & 4.5
                                                          \\ \hline
 8&{\tt IBOS}& I  &{\tt ITERM} &{\tt HECCOR}&    4  & 4.3
                                                          \\ \hline
 9&{\tt IBRN}& I  &{\tt IUBRN} &{\tt HECBRN}&    0  & 4.3
                                                          \\ \hline
10&{\tt ICMP}& I  &{\tt ICMPT} &{\tt HECCOR}&    1  & 4.3
                                                          \\ \hline
11&{\tt ICOR}& I  &{\tt ICOR}  &{\tt HECCOR}&    5  & 4.3
                                                          \\ \hline
12&{\tt IDIS}& I  &{\tt IDIS}  &{\tt HECIDS}&    1  & 4.3
                                                          \\ \hline
13&{\tt IDQM}& I  &{\tt IDQM}  &{\tt HECIDS}&    1  & 4.3
                                                          \\ \hline
14&{\tt IDSP}& I  &{\tt IDSP}  &{\tt HECCOR}&    0  & 4.3
                                                          \\ \hline
15&{\tt IEPC}& I  &{\tt IEPC}  &{\tt HECCOR}&    0  & 4.3
                                                          \\ \hline
16&{\tt IEXP}& I  &{\tt IEXP}  &{\tt HECDPI}&    0  & 4.3
                                                          \\ \hline
17&{\tt IGCU}& I  &{\tt IGCUT} &{\tt HECGMC}&    0  & 4.2
                                                          \\ \hline
18&{\tt IGRP}& I  &{\tt NGROUP}&{\tt HECPDF}&    3  & 4.4
                                                          \\ \hline
19&{\tt IHCU}& I  &{\tt IHCUT} &{\tt HECIHC}&    0  & 4.2
                                                          \\ \hline
20&{\tt ILEP}& I  &{\tt ILEPTO}&{\tt HECOPT}&    1  & 4.1
                                                          \\ \hline
21&{\tt ILOW}& I  &{\tt ILOW  }&{\tt HECLOW}&    0  & 4.3
                                                          \\ \hline
22&{\tt IMEA}& I  &{\tt IMEAS} &{\tt HECOPT}&    1  & 4.2
                                                          \\ \hline
23&{\tt IMOD}& I  &{\tt MODEL} &{\tt HECFLG}&    0  & 4.4
                                                          \\ \hline
24&{\tt IOPT}& I  &{\tt IOPT}  &{\tt HECOPT}&    1  & 4.3
                                                          \\ \hline
25&{\tt IORD}& I  &{\tt IORD}  &{\tt HECCOR}&    1  & 4.3
                                                          \\ \hline
26&{\tt IOUT}& I  & {\tt IOUT} &{\tt HECOUP}&    0  & 4.7
                                                          \\ \hline
27&{\tt ISCH}& I  &{\tt ISCHM} &{\tt HECQCD}&    2  & 4.4
                                                          \\ \hline
28&{\tt ISET}& I  &{\tt NSET}  &{\tt HECPDF}&   41  & 4.4
                                                          \\ \hline
29&{\tt ISSE}& I  &{\tt ISSET} &{\tt HECSTF}&    1  & 4.4
                                                          \\ \hline
30&{\tt ISTR}& I  &{\tt ISTR}  &{\tt HECSTF}&    0  & 4.4
                                                          \\ \hline
31&{\tt ITAR}& I  &{\tt ITARG} &{\tt HECOPT}&    1  & 4.1
                                                          \\ \hline
32&{\tt ITER}& I  &{\tt ITERAD}&{\tt HECCOR}&    1  & 4.4
                                                          \\ \hline
33&{\tt ITV1}& I  &{\tt ITV1}  &{\tt HECGRD}&     0 & 4.5
                        \\  \hline  \hline
 \end{tabular}
 \end{center}
 \caption{           \it  \label{usropt}
 User defined flags in the file {\tt HECTOR.INP}}

 \vspace*{-1.cm}

 \end{table}

 \clearpage

 \newpage

\begin{table}[h]
\begin{center}
 \begin{tabular}{||r|l|c||l|c|r|l||} \hline \hline
   &{{\tt FFREAD}~flag }&type& parameter &common      &default~~~~&
                    ~references
                    \\ \hline \hline
34&{\tt ITV2}& I  &{\tt ITV2}  &{\tt HECGRD}&     2 & 4.5
                                                          \\ \hline
35&{\tt IUSR}& I  &{\tt IUSR} &{\tt HECUSR}&    0  & 4.6
                                                           \\ \hline
36&{\tt IVAR}& I  &{\tt IVAR} &{\tt HECFLG}&    0  & 4.4
                                                           \\ \hline
37&{\tt IVPL}& I  &{\tt IVPOL}&{\tt HECVPL}&    1  & 4.3
                                                           \\ \hline
38&{\tt IWEA}& I  &{\tt IWEAK}&{\tt HECFLG}&   0   & 4.3
                                                           \\ \hline
39&{\tt NVA1}& I  &{\tt NV1}  &{\tt HECBIN}&   6   & 4.5
                                                           \\ \hline
40&{\tt NVA2}& I  &{\tt NV2}  &{\tt HECBIN}&  20   & 4.5
                                                           \\ \hline
41&{\tt POLB}& R  &{\tt POL}  &{\tt HECPAR}&  0.0  & 4.1
                                                           \\ \hline
42&{\tt THC1}& R  &{\tt THCUT1} &{\tt HECGMC}& 0.0 & 4.2
                                                           \\ \hline
43&{\tt THC2}& R  &{\tt THCUT2} &{\tt HECGMC}& 0.0 & 4.2
                                                           \\ \hline
44&{\tt VAR1}& R  &{\tt V1(100)}&{\tt HECBIN}& ---&  4.5
                                                           \\ \hline
45&{\tt VAR2}& R  &{\tt V2(100)}&{\tt HECBIN}& ---&  4.5
                                                        \\ \hline
46&{\tt V1MN}& R  &{\tt V1MI} &{\tt HECGRD}& 1.0E--04& 4.5
                                                        \\ \hline
47&{\tt V1MX}& R  &{\tt V1MA} &{\tt HECGRD}&  1.0  & 4.5
                                                        \\ \hline
48&{\tt V2MN}& R  &{\tt V2MI} &{\tt HECGRD}& 1.0E--02& 4.5
                                                        \\ \hline
49&{\tt V2MX}& R  &{\tt V2MA} &{\tt HECGRD}&  1.0  & 4.5
                                                        \\ \hline
50&{\tt Q2CU}& R  &{\tt TCUT} &{\tt HECHDC}&   0.0 & 4.2
                                                        \\ \hline
51&{\tt ZCUT}& R  &{\tt EECU} &{\tt HECPAR}& 35.0  & 4.2
               \\   \hline \hline
\end{tabular}  \\
\vspace{1cm}
 Table~\arabic{table} (continued):
{\it The flags to be set by the user in
   the file\/ {\tt HECTOR.INP}.}
\label{flagstab} \end{center}
\end{table}

The default bins in $x$ are:
\begin{eqnarray}
x & = & 10^{-4},\ 10^{-3},\ 10^{-2},\ 10^{-1},\ 0.5,\ 0.9\ . \label{defx}
\end{eqnarray}
The default values of $y$ are defined in table~7.
They are located at 20 equidistant points in the interval from 0.01 to 1.

\clearpage
\section{Example file {\tt HECTOR.OUT}\label{hector.outs}}
The following sample file {\tt HECTOR.OUT} corresponds to the first
row of Table 3.
\\
\begin{verbatim}

          ------- HECTOR v1.00 -------

1           USER'S DIRECTIVES TO RUN THIS JOB
            ----------------------------------


 ***** DATA CARD CONTENT     LIST
 ***** DATA CARD CONTENT     C.... SETTING OF FLAGS AND PARAMETERS
 ***** DATA CARD CONTENT
 ***** DATA CARD CONTENT     C.... FLAGS
 ***** DATA CARD CONTENT     IOPT    1
 ***** DATA CARD CONTENT     IMEA    1
 ***** DATA CARD CONTENT     IBEA    1
 ***** DATA CARD CONTENT     ILEP    1
 ***** DATA CARD CONTENT     ITAR    1
 ***** DATA CARD CONTENT
 ***** DATA CARD CONTENT     IORD    1
 ***** DATA CARD CONTENT     IBOS    4
 ***** DATA CARD CONTENT     ICOR    1
 ***** DATA CARD CONTENT     ICMP    1
 ***** DATA CARD CONTENT     IEPC    0
 ***** DATA CARD CONTENT     IBRN    1
 ***** DATA CARD CONTENT
 ***** DATA CARD CONTENT     IDSP    0
 ***** DATA CARD CONTENT     IDIS    1
 ***** DATA CARD CONTENT     IDQM    1
 ***** DATA CARD CONTENT     ILOW    0
 ***** DATA CARD CONTENT
 ***** DATA CARD CONTENT     IMOD    0
 ***** DATA CARD CONTENT     ISTR    0
 ***** DATA CARD CONTENT     ISSE    1
 ***** DATA CARD CONTENT     ITER    1
 ***** DATA CARD CONTENT     IVAR    0
 ***** DATA CARD CONTENT     IGRP    3
 ***** DATA CARD CONTENT     ISET    17
 ***** DATA CARD CONTENT     ISCH    0
 ***** DATA CARD CONTENT
 ***** DATA CARD CONTENT     IWEA    0
 ***** DATA CARD CONTENT     IVPL    1
 ***** DATA CARD CONTENT     IEXP    0
 ***** DATA CARD CONTENT
 ***** DATA CARD CONTENT     IHCU    0
 ***** DATA CARD CONTENT     Q2CU    0.0
 ***** DATA CARD CONTENT     HM2C    0.0
 ***** DATA CARD CONTENT
 ***** DATA CARD CONTENT     IGCU    0
 ***** DATA CARD CONTENT     GCUT    0.0
 ***** DATA CARD CONTENT     THC1    0.0
 ***** DATA CARD CONTENT     THC2    0.0
 ***** DATA CARD CONTENT
 ***** DATA CARD CONTENT     C.... PARAMETERS
 ***** DATA CARD CONTENT     POLB    0.0
 ***** DATA CARD CONTENT     EELE    26.70
 ***** DATA CARD CONTENT     ETAR    820.0
 ***** DATA CARD CONTENT     ZCUT     0.00
 ***** DATA CARD CONTENT
 ***** DATA CARD CONTENT     C.... LATTICE
 ***** DATA CARD CONTENT     IBIN    0
 ***** DATA CARD CONTENT     IBCO    1
 ***** DATA CARD CONTENT     ITV1    4
 ***** DATA CARD CONTENT     ITV2    4
 ***** DATA CARD CONTENT     NVA1    1
 ***** DATA CARD CONTENT     NVA2    1
 ***** DATA CARD CONTENT     V1MN    1.E-4
 ***** DATA CARD CONTENT     V1MX    1.0
 ***** DATA CARD CONTENT     V2MN    0.01
 ***** DATA CARD CONTENT     V2MX    0.5
 ***** DATA CARD CONTENT     VAR1    0.001
 ***** DATA CARD CONTENT     VAR2    0.1
 ***** DATA CARD CONTENT
 ***** DATA CARD CONTENT     IOUT    0
 ***** DATA CARD CONTENT
 ***** DATA CARD CONTENT     IUSR    0
 ***** DATA CARD CONTENT
 ***** DATA CARD CONTENT     STOP
BINS:
VAR1(   1)=   .1000E-02
VAR2(   1)=   .1000E+00
 ***** LLA & NTLA BRANCH
 ***** O(ALPHA*L) CORRECTIONS IN LLA
 ***** LEPTON MEASUREMENT, NEUTRAL CURRENT
 ***** CHARGED ANTI-LEPTON SCATTERING
 ***** PROTON TARGET
 ***** LEPTON BEAM POLARIZATION= .0
 ***** L-BEAM: E
 ***** PARTON DESCRIPTION FOR STRUCTURE FUNCTIONS IS USED
 ***** CTEQ3 PDF PARAMETERIZATION, LO
 ***** LOW Q2 MODIFICATION: Q2 < Q2MIN: VOLKONSKY/PROKHOROV DAMPING
 ***** Q2MIN=   4.000  GeV**2
 ***** INPUT ELECTOROWEAK PARAMETERS USDED:
 ***** Z-BOSON     MASS AMZ (GEV) =    91.175
 ***** W-BOSON     MASS AMW (GEV) =    80.399
 ***** HIGSS BOSON MASS AMH (GEV) =   300.000
 ***** TOP QUARK   MASS AMT (GEV) =   180.000
 ***** CALCULATED EW PARAMETERS:
 ***** SIN(THETA_W)   =    .232000
 ***** DELTA_R        =    .036948
 ***** BEAM KINEMATICS *****
 ***** LEPTON BEAM ENERGY (GEV) = 26.70000076293945
 ***** PROTON BEAM ENERGY (GEV) = 820.0
 ***** BINNING IN X & Y
 ***** USER SUPPLIED BINS IN VARIABLE 1
 ***** USER SUPPLIED BINS IN VARIABLE 2
 ***** NUMBER OF BINS IN V1: NVA1= 1
 ***** NUMBER OF BINS IN V2: NVA2= 1
 CTEQ3(LO )
     VAR1       VAR2        S         BORN     SIG_CORR     DELTA
   .100E-02   .100E+00   .876E+05   .225E+06   .254E+06   .130E+02
 \end{verbatim}

\clearpage
\section{Test run output \label{tro}}
\ezero

\vspace{-0.35cm}

 \begin{verbatim}
          ------- HECTOR v1.00 -------

1           USER'S DIRECTIVES TO RUN THIS JOB
            ----------------------------------


 ***** DATA CARD CONTENT     LIST
 ***** DATA CARD CONTENT     C.... SETTING OF FLAGS AND PARAMETERS
 ***** DATA CARD CONTENT
 ***** DATA CARD CONTENT     C.... FLAGS
 ***** DATA CARD CONTENT     IOPT    3
 ***** DATA CARD CONTENT     IMEA    1
 ***** DATA CARD CONTENT     IBEA    1
 ***** DATA CARD CONTENT     ILEP    1
 ***** DATA CARD CONTENT     ITAR    1
 ***** DATA CARD CONTENT
 ***** DATA CARD CONTENT     IORD    3
 ***** DATA CARD CONTENT     IBOS    4
 ***** DATA CARD CONTENT     ICOR    5
 ***** DATA CARD CONTENT     ICMP    1
 ***** DATA CARD CONTENT     IEPC    0
 ***** DATA CARD CONTENT     IBRN    1
 ***** DATA CARD CONTENT
 ***** DATA CARD CONTENT     IDSP    0
 ***** DATA CARD CONTENT     IDIS    1
 ***** DATA CARD CONTENT     IDQM    1
 ***** DATA CARD CONTENT     ILOW    0
 ***** DATA CARD CONTENT
 ***** DATA CARD CONTENT     IMOD    0
 ***** DATA CARD CONTENT     ISTR    0
 ***** DATA CARD CONTENT     ISSE    1
 ***** DATA CARD CONTENT     ITER    1
 ***** DATA CARD CONTENT     IVAR    0
 ***** DATA CARD CONTENT     IGRP    3
 ***** DATA CARD CONTENT     ISET    17
 ***** DATA CARD CONTENT     ISCH    0
 ***** DATA CARD CONTENT
 ***** DATA CARD CONTENT     IWEA    0
 ***** DATA CARD CONTENT     IVPL    1
 ***** DATA CARD CONTENT     IEXP    1
 ***** DATA CARD CONTENT
 ***** DATA CARD CONTENT     IHCU    0
 ***** DATA CARD CONTENT     Q2CU    0.0
 ***** DATA CARD CONTENT     HM2C    0.0
 ***** DATA CARD CONTENT
 ***** DATA CARD CONTENT     IGCU    0
 ***** DATA CARD CONTENT     GCUT    0.0
 ***** DATA CARD CONTENT     THC1    0.0
 ***** DATA CARD CONTENT     THC2    0.0
 ***** DATA CARD CONTENT
 ***** DATA CARD CONTENT     C.... PARAMETERS
 ***** DATA CARD CONTENT     POLB    0.0
 ***** DATA CARD CONTENT     EELE    26.70
 ***** DATA CARD CONTENT     ETAR    820.0
 ***** DATA CARD CONTENT     ZCUT     0.00
 ***** DATA CARD CONTENT
 ***** DATA CARD CONTENT     C.... LATTICE
 ***** DATA CARD CONTENT     IBIN    0
 ***** DATA CARD CONTENT     IBCO    1
 ***** DATA CARD CONTENT     ITV1    4
 ***** DATA CARD CONTENT     ITV2    4
 ***** DATA CARD CONTENT     NVA1    1
 ***** DATA CARD CONTENT     NVA2    1
 ***** DATA CARD CONTENT     V1MN    1.E-4
 ***** DATA CARD CONTENT     V1MX    1.0
 ***** DATA CARD CONTENT     V2MN    0.01
 ***** DATA CARD CONTENT     V2MX    0.5
 ***** DATA CARD CONTENT     VAR1    0.0001  0.01  0.1  0.5  0.90
 ***** DATA CARD CONTENT     VAR2    0.01    0.1   0.5  0.9  0.99
 ***** DATA CARD CONTENT
 ***** DATA CARD CONTENT     IOUT    1
 ***** DATA CARD CONTENT
 ***** DATA CARD CONTENT     IUSR    0
 ***** DATA CARD CONTENT
 ***** DATA CARD CONTENT     STOP
BINS:
VAR1(   1)=  0.1000E-03
VAR1(   2)=  0.1000E-01
VAR1(   3)=  0.1000E+00
VAR1(   4)=  0.5000E+00
VAR1(   5)=  0.9000E+00
VAR2(   1)=  0.1000E-01
VAR2(   2)=  0.1000E+00
VAR2(   3)=  0.5000E+00
VAR2(   4)=  0.9000E+00
VAR2(   5)=  0.9900E+00
 ***** O(ALPHA) COMPLETE CORRECTIONS SUPPLEMENTED
       BY LLA TERMS IN HIGHER ORDERS
 ***** LEPTON MEASUREMENT, NEUTRAL CURRENT
 ***** CHARGED ANTI-LEPTON SCATTERING
 ***** PROTON TARGET
 ***** LEPTON BEAM POLARIZATION=  0.0000000000000000E+00
 ***** L-BEAM: E
 ***** PARTON DESCRIPTION FOR STRUCTURE FUNCTIONS IS USED
 ***** CTEQ3 PDF PARAMETERIZATION, LO
 ***** LOW Q2 MODIFICATION: Q2 < Q2MIN: VOLKONSKY/PROKHOROV DAMPING
 ***** Q2MIN=   4.000  GeV**2
 ***** INPUT ELECTOROWEAK PARAMETERS USDED:
 ***** Z-BOSON     MASS AMZ (GEV) =    91.175
 ***** W-BOSON     MASS AMW (GEV) =    80.399
 ***** HIGSS BOSON MASS AMH (GEV) =   300.000
 ***** TOP QUARK   MASS AMT (GEV) =   180.000
 ***** CALCULATED EW PARAMETERS:
 ***** SIN(THETA_W)   =   0.232000
 ***** DELTA_R        =   0.036948
 ***** BEAM KINEMATICS *****
 ***** LEPTON BEAM ENERGY (GEV) =   26.70000076293945
 ***** PROTON BEAM ENERGY (GEV) =   820.0000000000000
 ***** BINNING IN X & Y
 ***** USER SUPPLIED BINS IN VARIABLE 1
 ***** USER SUPPLIED BINS IN VARIABLE 2
 ***** NUMBER OF BINS IN V1: NVA1=           5
 ***** NUMBER OF BINS IN V2: NVA2=           5
 CTEQ3(LO )
     VAR1       VAR2        S         BORN     SIG_CORR     DELTA
  0.100E-03  0.100E-01  0.876E+05  0.113E+10  0.109E+10 -0.353E+01
  0.100E-03  0.100E+00  0.876E+05  0.382E+08  0.429E+08  0.124E+02
  0.100E-03  0.500E+00  0.876E+05  0.113E+07  0.153E+07  0.348E+02
  0.100E-03  0.900E+00  0.876E+05  0.324E+06  0.684E+06  0.111E+03
  0.100E-03  0.990E+00  0.876E+05  0.271E+06  0.176E+07  0.549E+03
  0.100E-01  0.100E-01  0.876E+05  0.137E+06  0.134E+06 -0.233E+01
  0.100E-01  0.100E+00  0.876E+05  0.175E+04  0.200E+04  0.139E+02
  0.100E-01  0.500E+00  0.876E+05  0.573E+02  0.785E+02  0.369E+02
  0.100E-01  0.900E+00  0.876E+05  0.150E+02  0.343E+02  0.129E+03
  0.100E-01  0.990E+00  0.876E+05  0.123E+02  0.229E+03  0.175E+04
  0.100E+00  0.100E-01  0.876E+05  0.120E+04  0.941E+03 -0.216E+02
  0.100E+00  0.100E+00  0.876E+05  0.110E+02  0.109E+02 -0.971E+00
  0.100E+00  0.500E+00  0.876E+05  0.283E+00  0.348E+00  0.232E+02
  0.100E+00  0.900E+00  0.876E+05  0.611E-01  0.115E+00  0.878E+02
  0.100E+00  0.990E+00  0.876E+05  0.494E-01  0.594E+00  0.110E+04
  0.500E+00  0.100E-01  0.876E+05  0.956E+01  0.514E+01 -0.463E+02
  0.500E+00  0.100E+00  0.876E+05  0.717E-01  0.524E-01 -0.269E+02
  0.500E+00  0.500E+00  0.876E+05  0.126E-02  0.121E-02 -0.398E+01
  0.500E+00  0.900E+00  0.876E+05  0.149E-03  0.200E-03  0.344E+02
  0.500E+00  0.990E+00  0.876E+05  0.112E-03  0.471E-03  0.319E+03
  0.900E+00  0.100E-01  0.876E+05  0.661E-02  0.204E-02 -0.691E+02
  0.900E+00  0.100E+00  0.876E+05  0.387E-04  0.183E-04 -0.528E+02
  0.900E+00  0.500E+00  0.876E+05  0.546E-06  0.362E-06 -0.337E+02
  0.900E+00  0.900E+00  0.876E+05  0.540E-07  0.493E-07 -0.857E+01
  0.900E+00  0.990E+00  0.876E+05  0.400E-07  0.523E-07  0.307E+02
\end{verbatim}

\vspace{-1cm}

\end{document}